\documentclass[%
aps,
prx,%
amsmath,amssymb,showpacs,
reprint,%
floatfix,
twocolumn,
nofootinbib,
longbibliography
]{revtex4-1}

\usepackage[disable]{todonotes}
\usepackage{amsmath,amsfonts,amssymb}
\usepackage{cases} 
\usepackage{xfrac}

\newcommand{\ket}[1]{|#1\rangle}                   
\newcommand{\bra}[1]{\langle #1|}                  
\newcommand{\braket}[2]{\langle #1 | #2 \rangle}   
\def\doubleunderline#1{\underline{\underline{#1}}} 

\allowdisplaybreaks[3]

\begin{document}
	
	\title{Ab initio few-mode theory for quantum potential scattering problems}
	
	\author{Dominik \surname{Lentrodt}}
	\email[]{dominik.lentrodt@mpi-hd.mpg.de}
	
	\author{J\"org \surname{Evers}}
	\email[]{joerg.evers@mpi-hd.mpg.de}
	
	\affiliation{Max-Planck-Institut f\"ur Kernphysik, Saupfercheckweg 1, 69117 Heidelberg, Germany}
	
	
	\begin{abstract}
	Few-mode models have been a cornerstone of the theoretical work in quantum optics, with the famous single-mode Jaynes-Cummings model being only the most prominent example. In this work, we develop ab initio few-mode theory, a framework connecting few-mode system-bath models to ab initio theory. We first present a method to derive exact few-mode Hamiltonians for non-interacting quantum potential scattering problems and demonstrate how to rigorously reconstruct the scattering matrix from such few-mode Hamiltonians. We show that upon inclusion of a background scattering contribution, an ab initio version of the well known input-output formalism is equivalent to standard scattering theory. On the basis of these exact results for non-interacting systems, we construct an effective few-mode expansion scheme for interacting theories, which allows to extract the relevant degrees of freedom from a continuum in an open quantum system. As a whole, our results demonstrate that few-mode as well as input-output models can be extended to a general class of problems, and open up the associated toolbox to be applied to various platforms and extreme regimes. We outline differences of the ab initio results to standard model assumptions, which may lead to qualitatively different effects in certain regimes. The formalism is exemplified in various simple physical scenarios. In the process we provide proof-of-concept of the method, demonstrate important properties of the expansion scheme, and exemplify new features in extreme regimes.
	\end{abstract}
	
	\maketitle
	
	\section{Introduction}
	
	Scattering theory is a major tool in a variety of platforms. However, particularly for quantum dynamical systems, solving the scattering problem is often difficult, not least due to the infinitely many degrees of freedom provided by the scattering continuum. Consequently, it is a crucial task to reduce the complexity of the theoretical description by extracting the relevant degrees of freedom of the system. In practice, these often turn out to be only few, especially when the system features resonances or long-lived decaying states \cite{Kukulin1989,Cohen-Tannoudji1998b}, as is the case in various platforms of quantum dynamics. To name a few examples, electronic transport in mesoscopic physics \cite{Datta1995,Blanter2000,Rotter2017} and resonances in atomic \cite{Burke1965,Smith1966} as well as nuclear \cite{Mahaux1969,Mitchell2010} physics can often be interpreted as particles scattering on a Schr\"odinger potential, while light scattering in cavity QED \cite{Berman1994,Haroche2013,Ritsch2013}, photonics \cite{Joannopoulos2008,Rotter2017}, and many other optical platforms is governed by Maxwell's equations.

	In quantum optics, this idea of few relevant modes has been implemented in a famous model known as the input-output formalism~\cite{YurkeDenker1984,Collett1984,Gardiner1985}. It is based on a system-bath Hamiltonian where a few modes characterizing the system's dynamics are coupled to an external continuum. The few-mode character of this model enables a variety of approximations and as a result system-bath methods form the cornerstone for a large bulk of theoretical work \cite{Gardiner2004,Carmichael1999}, and an impressive toolbox has been developed to apply the input-output formalism to various problems and physical situations, including cavity QED \cite{Gardiner2004,Yurke2004}, quantum networks \cite{Zhang2013,Reiserer2015} and photon transport \cite{Xu2015,Caneva2015,Xu2017}. It further allows to connect the scattering properties of such systems to well studied few-mode models for light-matter interaction, such as the single-mode Jaynes-Cummings model \cite{Jaynes1963a} and its generalizations, including the Rabi model \cite{Rabi1936,Braak2011,Rossatto2017}, the Dicke model \cite{Dicke1954a,Kilin1980,Kirton2018}, and many more.
	
	\begin{figure*}[!htbp]
		\includegraphics[width=1.0\textwidth]{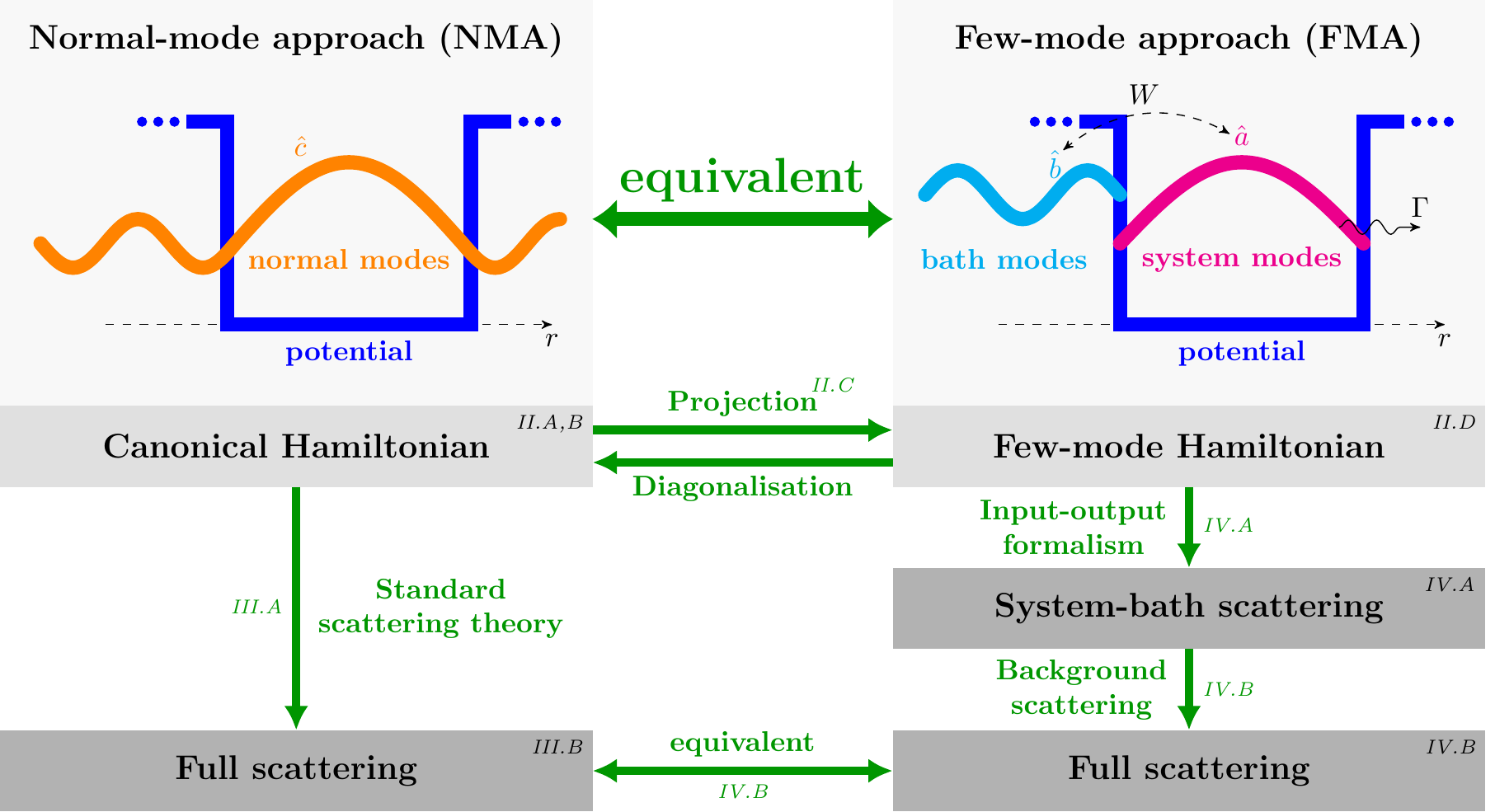}
		\caption{Schematic of the theoretical connections on non-interacting theories presented in this paper. The left hand side represents the normal-mode approach (NMA) to quantum potential scattering, where one can rigorously obtain a Hamiltonian from canonical quantization. The latter is conveniently expressed in terms of normal mode operators $\hat{c}$ (top left picture). The right hand side represents the few-mode approach (FMA), usually employed in terms of phenomenological models. There, a discrete set of system modes is coupled to a continuum of bath modes (top right picture), corresponding to operators $\hat{a}$ and $\hat{b}$, respectively, with coupling constant $W$ and loss rate $\Gamma$. We show that the Hamiltonians on each side can be connected by a basis transformation. A common method to calculate scattering observables in the FMA, known as the input-output formalism, can further be connected to standard scattering theory by inclusion of a background scattering contribution. Having shown that the Hamiltonians as well as the methods for calculating scattering observables can be rigorously connected, allows the normal-mode and the few-mode approaches to quantized potential scattering theory to be regarded as equivalent.\label{fig::schematic}}
	\end{figure*}
	
	However, despite their success, there are several open questions related to input-output models. In many cases, the input-output formalism is applied phenomenologically \cite{Gardiner2004}, that is the structure of its Hamiltonian is assumed and its parameters are fitted to data. For good cavities or more generally isolated resonances, this approach is natural, since one would not expect a weakly leaky system to differ grossly from a completely closed system. However, the applicability of input-output theory has been debated in the bad cavity and overlapping modes regimes \cite{Barnett1988a,Dutra2000,Dutra2001}, for systems with absorption \cite{Khanbekyan2005}, as well as more recently in the ultra-strong and deep-strong coupling regimes \cite{Bamba2014,FriskKockum2019}.
	Besides these fundamental concerns, due to the unknown origin of the Hamiltonian there is often no systematic way to calculate the phenomenological coupling and decay rates, which inhibits design possibilities. Additionally, it is unclear under what circumstances the method is appropriate, hindering applications in more general scattering theory settings beyond quantum optics \cite{Search2002,Gardiner2004b}, which have been sparse so far.

	A number of ab initio methods have been developed to address these issues, and much progress has been made pursuing multiple avenues, such as macroscopic QED \cite{Knoell1987,Dutra2004,Khanbekyan2005,Vogel2006,Scheel2008,Esfandiarpour2018}, modes-of-the-universe \cite{Lang1973a,Cerjan2016,Glauber1991,Scully1997}, local density of states \cite{Dung2000,Krimer2014} as well as pseudo-modes \cite{Garraway1997a,Garraway1997b} approaches for quantum optics. For general wave mechanics and scattering theory, alternative ways to rigorously extract the relevant dynamics have been investigated, including different types of quasi-modes \cite{Gamow1928,Fox1961,Ching1998,Dalton1999a,Lamprecht1999,Dutra2000,Tureci2005,Kristensen2014,Alpeggiani2017,Hughes2018,Lalanne2018}, the related constant flux states \cite{Tureci2008a,Krimer2014,Cerjan2016,Malekakhlagh2016}, temporal coupled-mode theory \cite{Haus1984_BOOK,Fan2003}, and various methods from the theory of chaotic scattering \cite{Mitchell2010,Dittes2000,Rotter2009,Zaitsev2010}, all of which have found multiple applications. While these approaches do not raise the concerns of few-mode Hamiltonians, they only rarely connect to the large toolbox available in few-mode input-output theory, and are consequently often limited in other ways. A major step forward has been the ab initio derivation of a system-bath Hamiltonian with infinite number of system modes for Maxwell's equations by Viviescas\&Hackenbroich \cite{Viviescas2003}. This motivates the question whether such a connection to ab initio methods could also be established for few-mode theory, and how to rigorously reconstruct the scattering information from such Hamiltonians using input-output methods.

	Here, we develop \textit{ab initio few-mode theory}, providing a rigorous foundation for established models and extending the reach of associated methods to extreme regimes.

	As a first and founding set of results, we derive an exact link between standard scattering theory, few-mode Hamiltonians, and the input-output formalism for quantum potential scattering systems without interactions. The result is based on and extends methods from system-bath theory in quantum optics \cite{Viviescas2003}, scattering theory in quantum chemistry \cite{Domcke1983}, and quantum field theory \cite{Glauber1991}, with the goal to make input-output methods a general and rigorous tool for second quantized scattering problems. We find crucial differences between the ab initio approach and common model assumptions, such as frequency dependent couplings and cross-mode decay terms, as well as a background scattering contribution, all of which are significant particularly in the overlapping modes regime and may cause qualitatively new effects. We emphasize that despite these differences, our ab initio version of the input-output formalism does not increase the theoretical complexity of the problem compared to phenomenological models, only the coupling constants have to initially be calculated from the scattering geometry to obtain the Hamiltonian. The latter further offers design opportunities.

	In a second step and based on the exact results for non-interacting systems, we develop an effective few-mode expansion scheme for interacting quantum potential scattering problems, where the concept of extracting a few relevant degrees of freedom becomes a powerful tool. In this context, ab initio few-mode theory extends and provides a number of advantages to phenomenological few-mode theory, including famous field-matter interaction models such as the Jaynes-Cummings model \cite{Jaynes1963a} and others mentioned above. Firstly, the non-interacting system is now always treated exactly, such that the advantages of the exact results in the non-interacting part of the paper are inherited. Secondly, a systematic few-mode expansion scheme for the interacting dynamics can now be constructed, which disentangles various approximations. Thirdly, our method directly connects to the toolbox of phenomenological few-mode theory, such that frequently used techniques do not have to be abandoned. Lastly, our formalism extends the reach of few-mode theory in general, making models such as the open Jaynes-Cummings model applicable in more extreme regimes and for different physical systems. Each of the advantages is demonstrated using representative examples from the field of light-matter interactions.

	In combination, our work connects phenomenological models in cavity QED to ab initio quantization, and shows that the input-output formalism can be applied in highly open systems such as the overlapping modes regime \cite{Petermann1979,Hackenbroich2002,Heeg2015c,Zhong2017,Dittes2000}, non-Hermitian photonics \cite{El-Ganainy2018,Miri2019,Ozdemir2019} or other platforms featuring significant leakage \cite{Stockman2018,Fernandez-Dominguez2018}, and regimes of extreme light-matter coupling, such as the ultra-strong \cite{Carusotto2013,Forn-Diaz2019,FriskKockum2019} or multi-mode strong coupling regime \cite{Krimer2014,Sundaresan2015}.

	\begin{figure}[t]
		\includegraphics[scale=1.0]{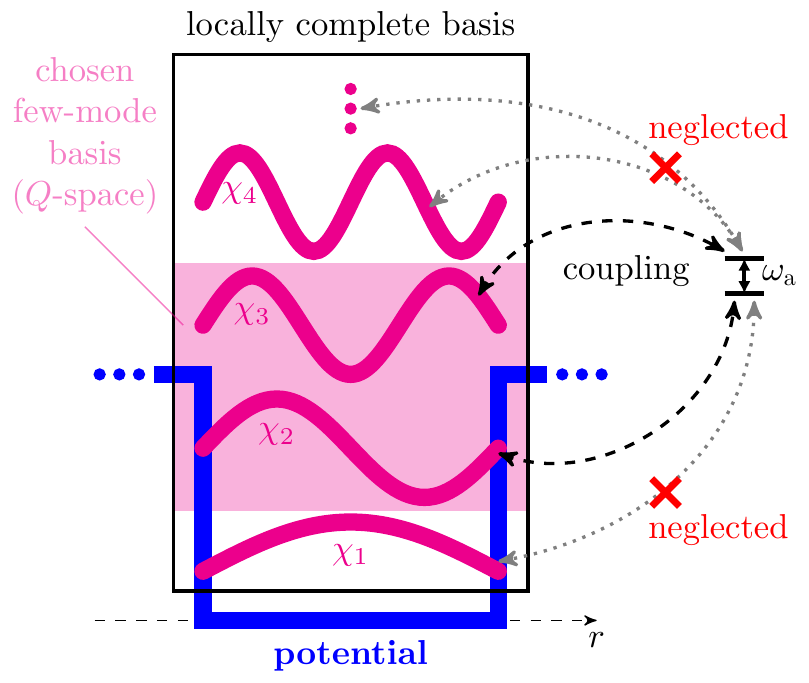}
		\caption{
			Schematic illustrating the part of the paper on interacting systems, in particular the few-mode approximation and constructive approach to choosing few-mode bases for an ab initio effective few-mode expansion. 
			After introducing system modes and bath modes (see also Fig.~\ref{fig::schematic}), the few-mode approximation consists of neglecting the interaction of the interacting subsystem (e.g.~a two-level atom located inside the cavity with transition frequency $\omega_\textrm{a}$) with the bath modes. In the figure, an example of a two-mode basis is shown as the states inside the magenta shaded box.
			Ideally, the set of system modes is chosen exploiting physical insight into the system under study, to facilitate the modeling of the system with as few modes as possible. In the absence of any prior knowledge, a constructive approach can be used to determine a few-mode basis. For this, a locally complete basis (states inside black box) is found as solutions to the Dirichlet boundary value problem in the potential region, which in general contains infinitely many modes. A few-mode basis  is then given by a subset of the locally complete basis. Varying the number of modes in the few-mode basis and performing the few-mode approximation in each case yields a systematic expansion scheme. \label{fig::schematic_modeSelection}}
	\end{figure}
	
	Beyond cavity QED, our results show the equivalence between the input-output formalism and standard scattering theory, paving the way for the application of simple system-bath models to more general quantum scattering problems. The latter promotes existing methods from wave scattering theory as they are used for example in chaotic scattering \cite{Beenakker1997}, nuclear physics \cite{Mitchell2010}, mesoscopic physics \cite{Rotter2017,Datta1995} and non-Hermitian systems \cite{Dittes2000,Moiseyev2011,Rotter2009} to the second quantized level \cite{Hackenbroich2002}. From this perspective our method may advance the exchange of methods and concepts \cite{Rotter2017} between currently separated fields.

	Fig.~\ref{fig::schematic} provides an overview of the first set of results on non-interacting systems presented in this paper and explains its structure.
	The left hand side represents established ab initio methods, for example based on the canonical quantization of a wave equation. In this paper, we consider the Schr\"odinger equation and a special case of Maxwell's equation for a dielectric medium as particular examples of quantum scattering problems. Fig.~\ref{fig::schematic} depicts the more general principle illustrated by a model potential (blue) with a schematic normal mode (orange). The normal mode basis is convenient, since it diagonalizes the Hamiltonian, which is obtained from the canonical quantization procedure. In Sec.~\ref{sec::schroed_CanQuant}, this approach is reviewed for the Schr\"odinger equation. The normal modes then obtain associated operators $\hat{c}$ (Sec.~\ref{sec::nmbfs}). The equations of motion for these operators can be solved using standard scattering theory, to obtain the scattering matrix (Sec.~\ref{sec::scatt}). Throughout the paper, we will denote this approach as the {\it normal mode approach} (NMA). 
	On the right hand side, the {\it few-mode approach} (FMA) is depicted, on which we focus here. It is usually employed in the form of phenomenological models, featuring a small number of discrete system modes coupled to an external bath with coupling constant $W$ and complex energy shift / loss rate $\Gamma$. The input-output formalism is then used to calculate the scattering between the bath modes via the system modes.

	Fig.~\ref{fig::schematic_modeSelection} provides an overview of the results on interacting systems and in particular illustrates the concept of effective few-mode expansions. In this part, we use a paradigmatic system from the theory of light-matter interactions, namely a two-level atom inside a cavity, as an example.

    The paper is organized as follows. As our first result, we project the full problem into a system-bath representation in Sec.~\ref{sec::schroed_systBathBasis} and use it to derive an ab initio few-mode Hamiltonian for the Schr\"odinger field in Sec.~\ref{sec::schroed}. As our main result for non-interacting systems, in Sec.~\ref{sec::FMscatt}, we rigorously reconstruct the full scattering matrix from the ab initio few-mode Hamiltonian obtained in Sec.~\ref{sec::schroed} using a suitable input-output formalism. We in particular show in Sec.~\ref{sec::scattEquiv} that the equivalence to the full scattering solution obtained from the NMA can only be established if a so-called background scattering term is included, which translates the bath modes scattering on the system into the asymptotically free modes.
	Our results thus not only connect the Hamiltonians on each side, which govern the dynamical equations of the system, but also the methods for computing scattering observables. This promotes the FMA and the input-output formalism to a rigorous theory and allows the two pictures to be used as equivalent approaches, which each have their advantages in practical situations.
	In Sec.~\ref{sec::Maxwell}, we present corresponding results for the dielectric Maxwell equations, which form the basis for major fields of application of input-output models such as cavity QED. These results are brought into a practical context in Sec.~\ref{sec::models} by comparing to what is usually done in corresponding phenomenological approaches. 
	For illustration and proof-of-concept purposes, Sec.~\ref{sec::exSyst} discusses a Fabry-Perot cavity with variable mirror quality and a double barrier tunneling potential as example systems to illustrate the results on non-interacting systems.
	Finally, in Sec.~\ref{sec::eff} the formalism for interacting quantum systems is developed. We first describe how ab initio few-mode theory allows to construct a systematic effective few-mode expansion to approximate the interacting system. We then outline the advantages of the method, which are inherited from the exact description of the non-interacting system. We demonstrate each advantage individually by explicit calculations for example systems.
	In Sec.~\ref{sec::outlook}, we discuss possible applications and generalizations of the formalism in detail, before we conclude in Sec.~\ref{sec::conclusions}. The appendices give details on the formalism.

	\section{Ab initio few-mode Hamiltonians}\label{sec::schroed}
	In order to link the FMA to the NMA, we begin by establishing a direct connection between the typical Hamiltonians in the two fields (see Section~\ref{sec::schroed} labels in Fig.~\ref{fig::schematic}). On the NMA side, this is a diagonal normal modes Hamiltonian which can be obtained from the canonical quantization of a wave equation~\cite{CohenTannoudji1997,Imamoglu1999}. On the FMA side, a \textit{system} and a \textit{bath} appear as coupled degrees of freedom~\cite{Gardiner2004} (see Fig.~\ref{fig::schematic}). Via a suitable basis transformation \cite{Domcke1983,Viviescas2003}, we show that the two descriptions are equivalent for an arbitrary number of system modes. Based on this equivalence, we promote the few-mode input-output model to an ab initio theory in Sec.~\ref{sec::FMscatt}.
	
	Our technique can be applied to a general class of wave equations. In this Section, we demonstrate its working principle on the Schr\"odinger equation
	\begin{align}\label{equ::schrodinger_eom}
	H \psi(r,t) = i \frac{\partial}{\partial t} \psi(r,t)\,,
	\end{align}
	where $\psi(r,t)$ is the wave function, $H = H_0 + V(r)$ is the first quantized Hamiltonian, $V(r)$ is a real-valued potential that vanishes at large $|r|$ and $H_0=K=-\frac{1}{2}\frac{\partial^2}{\partial r^2}$ is the free kinetic energy operator. For simplicity, we work with $\hbar=m=1$ and restrict ourselves to one dimension, the technique is however not limited to this setting. 
	
	\subsection{Canonical quantization}\label{sec::schroed_CanQuant}
	Second quantization (see Appendix~\ref{sec::app_schroed_CanQuant} for details) of Eq.~\eqref{equ::schrodinger_eom} yields the Hamiltonian
	\begin{align}\label{equ::schrodinger_hamiltonian}
	\hat{H}  =  \int dr \: \hat{\psi}^\dag(r,t)\, H \,\hat{\psi}(r,t)\,,
	\end{align}
	where $\hat{\psi}(r,t)$, $\hat{\psi}^\dag(r,t)$ are now operators with bosonic commutation relations
	\begin{align}\label{equ::schrodinger_fieldCommutator}
		[\hat{\psi}(r,t),\,\hat{\psi}^\dag(r',t)] = \delta(r-r')\,.
	\end{align}
	
	\subsection{Normal mode basis \& Fock space}\label{sec::nmbfs}
	It is useful to write the second quantized Hamiltonian in terms of normal mode creation and annihilation operators. To this end the field operator can be expanded in a normal mode basis
	\begin{align}\label{equ::schrodinger_fieldOpModeExpansion}
	\hat{\psi}(r,t) = \sum_m \int dE(k) \: \phi_m(r,k)\:\hat{c}_m(k,t)\,.
	\end{align}
	Here, the normal mode $\phi_m(r,k)$ is defined as an eigenstate of the time-independent Schr\"odinger equation
	\begin{align}\label{equ::schrodinger_timeIndep}
	H \, \phi_m(r,k) = E(k)\, \phi_m(r,k)
	\end{align}
	with energy $E(k)$ and further quantum numbers denoted by the index $m$.

	With appropriate mode normalization (see Appendix~\ref{sec::app_modeNorm}) the second quantized Hamiltonian is
	\begin{align}\label{equ::schrodinger_H_modes}
	\hat{H} = \sum_m \int dE(k) \: E(k)\: \hat{c}^\dag_m(k,t) \hat{c}^{\mathstrut}_m(k,t)\,.
	\end{align}
	The normal mode operators $\hat{c}_m(k,t)$ satisfy the canonical ladder operator commutation relations, for example,
	\begin{align}\label{equ::schrodinger_ladderCommutator}
	[\hat{c}^{\mathstrut}_m(k,t),\,\hat{c}^\dag_{m'}(k',t)] = \delta^{\mathstrut}_{mm'} \: \delta(E(k)-E(k'))\,.
	\end{align}
	We note that in the normal mode basis, the Hamiltonian is diagonal. The normal modes generally form a continuum, since they include scattering states, and are also known as modes-of-the-universe in the context of electromagnetic radiation~\cite{Lang1973a}.
	
	\subsection{System-and-bath representation}\label{sec::schroed_systBathBasis}
	To obtain a system-bath representation of the Hamiltonian~\cite{Viviescas2003}, we would like to split the normal mode operators into a discrete set of system operators $\hat{a}_\lambda$ and a continuum of bath operators $\hat{b}_{m}(k)$ via a basis transformation of the form
	\begin{align}\label{equ::schrodinger_cExpansion}
	\hat{c}_m(k) &= \sum_{\lambda \in \Lambda_Q}\:  \alpha^*_{\lambda m}(k)\:  \hat{a}^{\mathstrut}_\lambda \nonumber \\
	&+ \sum_{m'} \int dE(k') \:  \beta^*_{m m'}(k, k')\: \hat{b}^{\mathstrut}_{m'}(k')\,,
	\end{align}
	where $\alpha^*_{\lambda m}(k)$ and $\beta^*_{m m'}(k, k')$ are expansion coefficients. This separation of the Hilbert space into two parts gives a Hamiltonian with couplings between the system and the bath modes, thus a non-diagonal Hamiltonian. A similar basis transformation with infinite number of system modes has been obtained by Viviescas\&Hackenbroich \cite{Viviescas2003}. Our method extends their approach, such that the discrete set of system modes denoted by $\Lambda_Q$ can be chosen to contain only few or even a single mode, and does not need to span a region in position space as a basis. This way effective few-mode theories capturing the relevant resonant dynamics can be formulated (see Sec.~\ref{sec::eff} for details).
	
	However, constructing such a few-mode basis is non-trivial. For Eq.~\eqref{equ::schrodinger_cExpansion} to be a consistent basis transformation, the system and bath together have to span the original Hilbert space. To connect to quantum noise theory~\cite{Gardiner2004}, we would also like $\hat{a}_\lambda$ and $\hat{b}_{m}(k)$ to be bosonic operators, which places a restriction on their commutation relations, and the Hamiltonian to be of so-called Gardiner-Collett form \cite{Gardiner1985,Gardiner2004}. In the following, we show that all of these conditions can be ensured by using Feshbach projections \cite{Feshbach1958,*Feshbach1962,*Feshbach1967,Domcke1983} to select a certain set of system states corresponding to the second quantized system operators $\hat{a}_\lambda$.
	
	\subsubsection{Feshbach projection for states}\label{sec::FeshStates}
	The idea of the Feshbach projection formalism~\cite{Feshbach1958,*Feshbach1962,*Feshbach1967} is to reformulate the Schr\"odinger equation Eq.~\eqref{equ::schrodinger_eom}, which describes the wave propagation in the full Hilbert space, in terms of wave equations in two subspaces, which are then coupled to each other. In this spirit we follow Domcke \cite{Domcke1983} to first express the eigenstates of the Schr\"odinger equation in terms of the subspace eigenstates.
	
	We start by defining projection operators $Q, P$ such that
	\begin{equation}\label{equ::proj_projOp_props}
	P^2 = P,~Q^2 = Q,~P+Q=1\,.
	\end{equation}
	$Q$ will correspond to the system subspace and $P$ to the bath subspace, which together span the full Hilbert space. However, we note that in the few-mode case, $Q$ and $P$ itself generally do not correspond to disjunct regions in position space. Specifically, the $Q$-space projector is defined by choosing a set of system modes $\Lambda_Q = \{\ket{\chi_\lambda}\}$, which are discrete normalized states that span the $Q$-space such that
	\begin{equation}\label{equ::proj_QProjector}
	Q = \sum_{\lambda \in \Lambda_Q} \ket{\chi_\lambda} \bra{\chi_\lambda}\,.
	\end{equation}
	We further require\footnote[3]{This requirement is imposed to obtain a Hamiltonian in the second quantized case, where the system states do not couple to each other directly. It does not restrict the generality since $Q$ and $H_{QQ}$ commute.} that these states be eigenstates of the projected $Q$-space Hamiltonian $H_{QQ} = QHQ$, that is
	\begin{equation}\label{equ::proj_Q_statesDef}
	H_{QQ}\,\ket{\chi_\lambda} = E_\lambda\, \ket{\chi_\lambda}\,.
	\end{equation}
	Analogously, we can define the bath modes $\ket{\tilde{\psi}_{m}(k)}$ as eigenstates of the $P$-space Hamiltonian
	\begin{equation}\label{equ::proj_P_statesDef}
		H_{PP}\,\ket{\tilde{\psi}_{m}(k)} = E(k)\,\ket{\tilde{\psi}_{m}(k)}\,.
	\end{equation}
	These states form a continuum and can only be determined uniquely after choosing appropriate boundary conditions \cite{Newton1982,Domcke1983}, which will become relevant in the context of scattering in Sec.~\ref{sec::scatt}.
	
	We note that the hermicity of the subspace Hamiltonians implies certain orthogonality conditions for their eigenstates (see Appendix \ref{sec::app_modeNorm} for details), which will become relevant in the context of quantization in Sec.~\ref{sec::FeshOps}.
	
	We can now write the eigenstates in full space as an expansion over the subspace eigenstates
	\begin{align}\label{equ::proj_modeSep}
		\ket{\phi_m(k)}  &=  Q\: \ket{\phi_m(k)} + P\: \ket{\phi_m(k)}
		\\
		&= \sum_{\lambda \in \Lambda_Q} \: \alpha_{\lambda m}(k)\: \ket{\chi_\lambda}  \nonumber
		\\
		& + \sum_{m'}\int dE(k')\:\beta^{\mathstrut}_{m m'}(k, k')\: \ket{\tilde{\psi}_{m'}^{\mathstrut}(k')}\,. \label{equ::proj_modeSep_b}
	\end{align}
	This can be interpreted as a system-bath expansion of the normal mode states. Importantly, the coefficients
	\begin{subequations}\label{equ::schrodinger_coefficients_systBath} 
	\begin{align}
		\alpha_{\lambda m}(k)&=\braket{\chi_\lambda}{\phi_m(k)}\,,
		\\
		\beta_{m m'}(k, k')&=\braket{\tilde{\psi}_{m'}(k')}{\phi_m(k)}
	\end{align}
	\end{subequations}
	can be calculated without direct knowledge of the normal mode functions $\phi_m(r,k)$ by so-called separable expansions (see Appendix \ref{appSub} for details), which can have computational advantages \cite{Domcke1983}.
	
	\subsubsection{Feshbach projection for operators}\label{sec::FeshOps}
	The separation of the dynamics into two coupled subspaces can alternatively be formulated in Fock space by introducing operators $\hat{a}_\lambda$ and $\hat{b}_{m}(k)$ corresponding to the system modes $\ket{\chi_\lambda}$ and bath modes $\ket{\tilde{\psi}_{m}(k)}$, respectively. It can be shown (see Appendix \ref{sec::app_SBoperators}) that analogously to Eq.~\eqref{equ::proj_modeSep_b}, the normal mode operators relate to these system-bath operators via
	\begin{align}\label{equ::schrodinger_cExpansion2}
	\hat{c}_m(k) =& \sum_{\lambda \in \Lambda_Q} \: \alpha^{*\mathstrut}_{\lambda m}(k) \:  \hat{a}^{\mathstrut}_\lambda\nonumber \\
	& + \sum_{m'} \int dE(k') \: \beta^{*\mathstrut}_{m m'}(k, k')\: \hat{b}^{\mathstrut}_{m'}(k')\,,
	\end{align}
	which is the operator system-bath expansion Eq.~\eqref{equ::schrodinger_cExpansion}, with the coefficients now given by Eqs.~\eqref{equ::schrodinger_coefficients_systBath}. In addition, the operators $\hat{a}_\lambda$ and $\hat{b}_{m}(k)$ fulfill the desired commutation relations \cite{Viviescas2003} (see Appendix \ref{sec::app_SBoperators} for details), that is they are each bosonic degrees of freedom and the system commutes with the bath. It has previously been unclear whether the latter holds in the bad cavity regime and alternative models have been suggested \cite{Dutra2001}. Now, the condition can be ensured constructively using the Feshbach projection method, even in the few-mode case. 
	
	\subsection{Ab initio few-mode Hamiltonian}\label{sec::schroed_GC}
	Applying the system-bath expansion Eq.~\eqref{equ::schrodinger_cExpansion2} to the second quantized Hamiltonian Eq.~\eqref{equ::schrodinger_H_modes} and using Appendices \ref{sec::app_SBoperators}, \ref{app_coeffIds} we obtain
	\begin{align}\label{equ::GC_H}
	\hat{H}  = & \sum_{\lambda \in \Lambda_Q} E^{\mathstrut}_\lambda \hat{a}^\dag_\lambda \hat{a}^{\mathstrut}_\lambda + \sum_m \int dE(k)\: E(k) \: \hat{b}^\dag_m(k) \hat{b}^{\mathstrut}_m(k) \nonumber \\
	& + \sum_{\lambda \in \Lambda_Q} \sum_m \int dE(k) \: \left [W^{\mathstrut}_{\lambda m}(k)\: \hat{a}^\dag_\lambda \hat{b}^{\mathstrut}_m(k) + \mathrm{h.c.} \right]\,, 
	\end{align}
	with the coupling constants
	\begin{equation}\label{equ::GCH_couplings}
	W^{\mathstrut}_{\lambda m}(k) := \bra{\vphantom{\hat{k}^-}\chi^{\mathstrut}_\lambda} H \ket{\tilde{\psi}_{m}^{\mathstrut}(k)}\,.
	\end{equation}
	We have thus derived an ab initio few-mode Hamiltonian of Gardiner-Collett form for the Schr\"odinger equation. We note that the few-mode Hamiltonian exactly captures the system's dynamics, equivalently to the Hamiltonian in its normal mode representation Eq.~\eqref{equ::schrodinger_H_modes}, even though the system modes are discrete and their number is finite. This feature opens new theoretical possibilities when interactions such as atoms are present inside the cavity, as we investigate in detail in Sec.~\ref{sec::eff}.
	
	We note that Eq.~\eqref{equ::GC_H} generalizes the Hamiltonian derived by Viviescas\&Hackenbroich \cite{Viviescas2003} from an infinite to an arbitrary number of system modes and to a general class of wave equations. More importantly, as we will show in the following sections, an ab initio input-output formalism can now be used to reconstruct the scattering information, which for Viviescas\&Hackenbroich's Hamiltonian \cite{Viviescas2003} is hindered by the appearance of divergent series in the infinite mode case (see Appendix \ref{sec::app_ViviescasScattering}). Due to the non-trivial behavior of this limit, which has already been noted in \cite{Domcke1983}, the few-mode Hamiltonians proposed here are better suited to achieve this task.
	
	For completeness, we further note that the inverse of the presented basis transformation constitutes a Fano diagonalization \cite{Fano1961} of the system-bath Hamiltonian. A similar basis transformation has been investigated in \cite{Dalton2001} in relation to pseudo-modes theory \cite{Garraway1997a,Garraway1997b}, and in an early paper \cite{Barnett1988a} considering an approximate treatment.
	
	\section{Quantum potential scattering}\label{sec::scatt}
	In practice, system-bath Hamiltonians are most commonly used as {phenomenological models for quantum mechanical systems~\cite{Gardiner2004,Yurke2004}. Their great value arises since scattering observables can be calculated using the famous input-output formalism~\cite{Gardiner2004,Yurke2004}, which is a standard tool  in quantum optics. Despite its success, the input-output formalism only addresses the scattering problem from the perspective of a model Hamiltonian, which the inventors called a ``simplified representation of reality''~\cite{Gardiner2004}.
	In the previous section, we showed how to rigorously derive few-mode system-bath Hamiltonians from canonical quantization, and thereby eliminated the need for the ad hoc assumption of a model Hamiltonian. With this ab initio version of the Hamiltonian at hand, we now have the tools to connect the input-output formalism to scattering theory.
	
	To set a foundation for comparison, in this section, we first derive scattering theory results in the first and second quantized setting as a reference (see also Sec.~\ref{sec::scatt} labels in Fig.~\ref{fig::schematic}).
	
	\subsection{First quantized potential scattering theory}
	
	\subsubsection{Standard scattering theory}
	For a wave equation such as the time-dependent Schr\"odinger equation Eq.~\eqref{equ::schrodinger_eom}, the scattering problem is given by the question of how an incoming wave-packet defined in the infinite past evolves into an outgoing wave-packet in the infinite future \cite{Newton1982}. For elastic scattering this information can be encoded in the on-shell scattering matrix $S_{mm'}(k)$, which is defined by the linear relation between states $\ket{\phi^{(+)}_m(k)}$ and $\ket{\phi^{(-)}_m(k)}$ \cite{Newton1982}
	\begin{equation}\label{equ::potScatt_defS_linSup}
		\ket{\phi^{(+)}_m(k)} = \sum_{m'} \: \ket{\phi^{(-)}_{m'}(k)} \: S^{\mathstrut}_{m'm}(k)\,.
	\end{equation}
	The states $\ket{\phi^{(\pm)}_m(k)}$ are the normal modes defined in Eq.~\eqref{equ::schrodinger_timeIndep} as eigenstates of the Hamiltonian. The $(\pm)$ corresponds to a choice of boundary conditions. As usual in scattering theory, $(+)$ is the state with a controlled incoming free state, and $(-)$ is the state with a controlled outgoing free state \cite{Newton1982}.
	
	Another useful scattering quantity is the transition operator $T$ defined by
	\begin{equation}\label{equ::potScatt_T_full_expansion}
		\ket{\phi^{(+)}_m(k)} = \ket{k^{\mathstrut}_m} + G_0^{(+)}\, T \,\ket{k^{\mathstrut}_m}
	\end{equation}
	where $G_0^{(+)}$ is the free propagator given via the free Hamiltonian $H_0$ as
	\begin{equation}\label{equ::potScatt_freePropagator}
		G_0^{(+)}=\left(E(k) - H_0 + i\eta\right)^{-1}\,,
	\end{equation}
	and $\ket{k_m}$ is an eigenstate of $H_0$ with $H_0\ket{k_m}=E(k)\ket{k_m}$.
	The operator $T$ thus quantifies transitions between a full eigenstate and a free eigenstate. It is linked to the on-shell scattering matrix defined above via~\cite{Newton1982}
	\begin{equation}
		S_{mm'}(k)  = \delta_{mm'} - 2\pi i \, T_{mm'}(k)
	\end{equation}
	with $ T_{mm'}(k) = \bra{k_m} T \ket{k_{m'}}$.
	
	The scattering properties can thus be obtained by solving the eigenproblem for the full Hamiltonian and computing their transition probabilities to freely propagating states.
	
	\subsubsection{Potential scattering via projection operators}\label{sec::potScatt_Domcke}
		\begin{figure}[t]
			\includegraphics[width=\columnwidth]{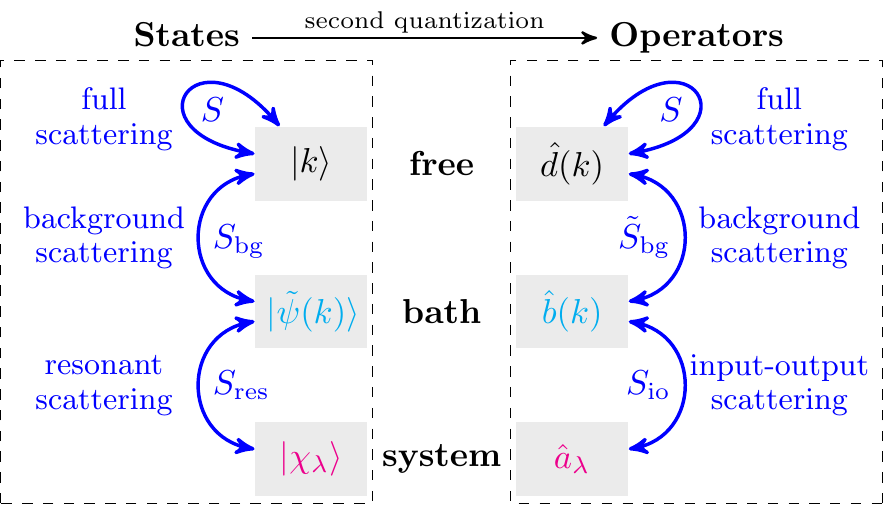}
			\caption{Schematic of the different degrees of freedom and scattering connections in the Feshbach projection formalism for potential scattering problems. The full scattering matrix $S$ can be used to relate asymptotically free states or operators to each other. The background scattering $S_\mathrm{bg}$ arises from a basis transformation of the free states into the bath states. The bath states scatter via $S_{\textrm{res}}$ on the system states, which span part of the region where the scattering potential $V(r)$ is non-zero. Similarly on the operator level, the scattering between bath operators that are coupled to the system is given by the input-output scattering matrix $S_\mathrm{io}$. To obtain the full scattering matrix, a background scattering contribution $\tilde{S}_\mathrm{bg}$ has to be applied. }\label{fig::stateScatt}
		\end{figure}
	Domcke showed \cite{Domcke1983} that instead of using the eigenstates in full space, the scattering matrix can also be calculated from the system and bath states that we used in Sec.~\ref{sec::schroed}.  Details on the calculation are summarized in Appendix \ref{sec::app_DomckeScatt}. Here we will focus on the definitions and interpretation of the results relevant to our work. The relation between the different states and scattering matrices used below is illustrated in the left part of Fig.~\ref{fig::stateScatt}.
	
	We first define a transition operator $T_{\textrm{res}}$ by considering the bath modes as ``free'' states. Analogously to Eq.~\eqref{equ::potScatt_T_full_expansion}, omitting matrix subscripts for brevity, we can write
	\begin{equation}
		P\ket{\phi^{(+)}(k)} = \ket{\tilde{\psi}^{(+)}(k)} + \tilde{G}^{(+)} \, T^{\mathstrut}_{\textrm{res}} \, \ket{\tilde{\psi}^{(+)}(k)}\,,
	\end{equation}
	where $\tilde{G}^{(+)}$ is the Green function for $P$-space propagation
	\begin{equation}
		\tilde{G}^{(+)} = \left(E(k) - H_{PP} + i\epsilon\right)^{-1}\,.
	\end{equation}
	We can then quantify the scattering between bath states by a scattering matrix
	\begin{equation}
		S_{\textrm{res}}(k) \equiv \mathbb{I} - 2\pi i\: T_{\textrm{res}}(k)\,,
	\end{equation}
	where $T_{\textrm{res}}(k)$ is the matrix element of $T_{\textrm{res}}$ on the basis of retarded bath states.
	
	However the bath states are not necessarily free states. Therefore there is a residual scattering contained in the asymptotic structure of the bath states, which can be described by a transition operator for transitions from a bath state to a free state
	\begin{equation}
		\ket{\tilde{\psi}^{(+)}(k)} = \ket{k} + G_0^{(+)}\, T^{\mathstrut}_{\textrm{bg}}\, \ket{k}\,.
	\end{equation}
	The background scattering matrix $S_\textrm{bg}$ is  again defined as the corresponding on-shell scattering matrix
	\begin{equation}
		S_{\textrm{bg}}(k) \equiv \mathbb{I} - 2\pi i \,T_{\textrm{bg}}(k)\,,
	\end{equation}
	where $T_{\textrm{bg}}(k)$ is the matrix element of $T_{\textrm{bg}}$ on the basis of free states. The effect of $S_{\textrm{bg}}(k)$ can thus be interpreted as an asymptotic basis transformation between bath states and free states.
	
	The full scattering matrix $S$ is then decomposed into the resonant scattering matrix $S_\textrm{res}$ and the background scattering matrix $S_{\textrm{bg}}$ via~\cite{Domcke1983} 
	\begin{align}
		S(k) &= S_{\textrm{bg}}(k) \,S_{\textrm{res}}(k)\,. \label{equ::Domcke_S}
	\end{align}In terms of the system and bath states these matrices read (see Appendix \ref{sec::app_DomckeScatt})
	\begin{align}
		S_{\textrm{res}}(k) &= \mathbb{I} - 2\pi i\: \bra{\tilde{\psi}^{(+)}(k)} H_{PQ}  G_{QQ} H_{QP}\ket{\tilde{\psi}^{(+)}(k)} \,,\label{equ::Domcke_Sres}
		\\
		S_{\textrm{bg}}(k) &= \mathbb{I} - 2\pi i \: \bra{k} (H_{PP}-K) \ket{\tilde{\psi}^{(+)}(k)}\,. \label{equ::Domcke_Sbg}
	\end{align}
	We note that unlike in the quasi-modes approach \cite{Ching1998,Kristensen2014,Alpeggiani2017,Lalanne2018}, the ``resonant'' part in the Feshbach projection formalism does not necessarily correspond to the resonances of the wave equation, that is the poles of the scattering matrix. However by choosing the system states appropriately, certain resonances can be selected, such that their poles appear in $S_\mathrm{res}$, and the remaining poles appear in $S_\mathrm{bg}$. This behavior has been investigated partially in \cite{Domcke1983} and we demonstrate its significance for extracting few-mode dynamics in Sec.~\ref{sec::exSyst}. In the context of interacting theories the concept further becomes a powerful tool to construct effective few-mode expansions, which we show in Sec.~\ref{sec::eff}. 
	
	We further note that from the viewpoint of the entire scattering problem, both $S_\mathrm{res}$ and $S_\mathrm{bg}$ are unphysical on their own, since their properties depend on the arbitrary choice of the  system states. However, they individually may provide accurate approximations of the full scattering matrix in the vicinity of their corresponding resonances (see also Sec.~\ref{sec::exSyst}), such that the choice of system states becomes a resource allowing the extraction of relevant properties of the whole system.

	\subsection{Second quantized potential scattering theory}\label{sec::potScatt_opScatt}
	In the second quantized setting, one investigates the dynamics of operators defined by the Hamiltonian and its corresponding Heisenberg equations of motion. That is, the quantization procedure promotes the wave equation to a non-relativistic quantum field theory, such that correlation functions can be computed and interactions can be considered.
	
	For potential scattering we can define asymptotically free operators by expanding the quantum field in a free mode basis instead of in its normal mode basis. If $\phi_m^{(\textrm{free})}(r,k) = \braket{r}{k_m^{\mathstrut}}$ are the field distribution of the free eigenstates,
	then the free state expansion of the field operator reads
	\begin{align}\label{equ::potScatt_freeFieldExpansion}
		\hat{\psi}(r,t) = \sum_m \int dE(k) \: \phi^{(\textrm{free})}_{m}(r,k)\: \hat{d}_m(k,t)\,,
	\end{align}
	where $\hat{d}_m(k,t)$ are the free bosonic operators satisfying canonical commutation relations.
	
	One can solve the Heisenberg equations of motion for these operators (see Appendix \ref{sec::app_opScatt} for details) to obtain a scattering relation
	\begin{equation}\label{equ::potScatt_free_IO}
		\hat{d}^{(\textrm{out})}_{m}(k) = \sum_{m'}S^{\mathstrut}_{mm'}(k)\: \hat{d}^{(\textrm{in})}_{m'}(k)\,,
	\end{equation}
	where the asymptotically free in [out] operators are interaction picture operators in the infinite past [future], that are defined via adiabatically switching on [off] of the potential in the corresponding time limits (see Appendix \ref{sec::app_opScatt} for details).
	
	In the case of potential scattering, the operator scattering matrix can be shown  to be exactly the first quantized scattering matrix Eq.~\eqref{equ::potScatt_defS_linSup}~\cite{Glauber1991}. This correspondence between the solution to the wave equation and its second quantized analogue is also required for consistency, since on average the result from the wave equation should be obtained, that is $\langle\hat{d}_{\textrm{out}}(k)\rangle = S\langle\hat{d}_{\textrm{in}}(k)\rangle$.
	
	For clarity, we emphasize that the scattering matrices employed here relate different asymptotic \textit{operators}. The relation of this formulation to scattering between initial and final states of the quantum field has, for example, been noted in \cite{Fan2010,Xu2015,Trivedi2018} in the context of few-photon transport.
	
	\section{Few-mode scattering}\label{sec::FMscatt}
	We now show how to rigorously reconstruct the full scattering information from the ab initio few-mode Hamiltonian derived in Sec.~\ref{sec::schroed} using the input-output formalism. We further show the equivalence of the input-output formalism result to that of standard scattering theory (see Sec.~\ref{sec::FMscatt} labels in Fig~\ref{fig::schematic}). The applicability of the input-output formalism is thus not limited to the good cavity regime, but applies to a general class of quantum scattering problems and in extreme regimes.
	
	\subsection{Ab initio input-output formalism}\label{sec::scatt_IO}
	
	We now apply the input-output formalism \cite{Gardiner1985, Gardiner2004,Viviescas2003} to our ab initio few-mode Hamiltonian Eq.~\eqref{equ::GC_H}. This constitutes solving the Heisenberg equations of motion for the Hamiltonian Eq.~\eqref{equ::GC_H}, which are
	\begin{align}\label{equ::IO_Heisenberg1}
		\frac{d}{dt}\hat{a}_\lambda(t) 
		=& -iE_\lambda \,\hat{a}_\lambda(t) \nonumber
		\\
		& -i\sum_m \int dE(k) \: W_{\lambda m}(k)\, \hat{b}_m(k,t)\,,
	\end{align}
	\begin{align}\label{equ::IO_Heisenberg2}
		\frac{d}{dt}\hat{b}_m(k,t) 
		= -iE(k) \, \hat{b}_m(k,t) - i\sum_{\lambda\in\Lambda_Q} W^*_{\lambda m}(k)\: \hat{a}_\lambda(t)\,. 
	\end{align}
	We can solve Eq.~\eqref{equ::IO_Heisenberg2} formally in terms of the initial time $t_0$ and final time $t_1$ as
	\begin{align}\label{equ::IO_formalT0}
		\hat{b}_m(k,t) &= e^{-iE(k)(t-t_0)} \:\hat{b}_m(k,t_0) \nonumber
		\\
		& - i\sum_{\lambda\in\Lambda_Q} W_{\lambda m}^*(k) \int_{t_0}^{t} dt' e^{-iE(k) (t-t')} \:\hat{a}_\lambda(t') 
	\end{align}
	and
	\begin{align}\label{equ::IO_formalT1}
		\hat{b}_m(k,t) &= e^{-iE(k)(t-t_1)} \:\hat{b}_m(k,t_1) \nonumber
		\\
		& + i\sum_{\lambda\in\Lambda_Q} W_{\lambda m}^*(k) \int_{t}^{t_1} dt' e^{-iE(k) (t-t')} \:\hat{a}_\lambda(t')\,, 
	\end{align}
	respectively.
	As usual in quantum noise theory \cite{Gardiner1985,Viviescas2003} and in analogy with the quantum field theory definition (see Sec.~\ref{sec::potScatt_opScatt}) we define the in- and out- operators
	\begin{subequations}
	\begin{align}
		\hat{b}^{\textrm{(in)}}_m(k) &= e^{iE(k)t_0} \:\hat{b}_m(k,t_0)\,,\\
		\hat{b}^{\textrm{(out)}}_m(k) &= e^{iE(k)t_1} \:\hat{b}_m(k,t_1)\,,
	\end{align}
	\end{subequations}
	respectively.
	Taking initial [final] times to negative [positive] infinity gives the input-output relation
	\begin{align}\label{IO}
		\hat{b}^{\textrm{(out)}}_m(k) - \hat{b}^{\textrm{(in)}}_m(k) = -i \sum_{\lambda\in\Lambda_Q} W^{*\mathstrut}_{\lambda m}(k) \:\hat{a}^{\mathstrut}_\lambda(k)\,,
	\end{align}
	where the Fourier transform of $\hat{a}_\lambda(t)$ is defined by
	\begin{equation}\label{equ::IO_FTdef}
		\hat{a}_\lambda(k) = \int_{-\infty}^{\infty} dt' e^{iE(k)t'} \:\hat{a}_\lambda(t')\,.
	\end{equation}	
	Substituting the formal solution Eq.~\eqref{equ::IO_formalT0} into Eq.~\eqref{equ::IO_Heisenberg1} and inverting the resulting matrix equation gives
	\begin{equation}\label{equ::IO_modeSol}
		\hat{a}^{\mathstrut}_\lambda(k) = 2\pi \sum_{\lambda'\in\Lambda_Q} \sum_m \:D^{-1}_{\lambda \lambda'}(k)^{\mathstrut}\: W^{\mathstrut}_{\lambda' m}(k) \:\hat{b}_m^{\textrm{(in)}\mathstrut}(k)\,,
	\end{equation}
	where we defined $D^{-1}$ as the inverse of the matrix of
	\begin{equation}\label{equ::IO_Dmatrix}
		D_{\lambda \lambda'}(k) = (E(k)-E_\lambda)\delta_{\lambda \lambda'} + \Gamma_{\lambda \lambda'}(k).
	\end{equation}
	The decay matrix (see also Fig.~\ref{fig::schematic}) is given by
	\begin{align}\label{equ::IO_decayMatrix}
	\Gamma_{\lambda \lambda'}(k)
	&= - \sum_m \int dE(k') \frac{W^{\mathstrut}_{\lambda m}(k') W^*_{\lambda'm}(k')}{E(k)-E(k') + i\epsilon}\\
	&=: -\Delta_{\lambda \lambda'}(k) + i\gamma_{\lambda \lambda'}(k)\,,
	\end{align}
	where we have defined the real and imaginary parts of $\Gamma_{\lambda \lambda'}(k)$ as $\Delta_{\lambda \lambda'}(k)$ and $\gamma_{\lambda \lambda'}(k)$. In the latter equation, the limit $\epsilon\rightarrow 0^+$ is implied. For $\lambda \neq \lambda'$, the complex decay matrix $\Gamma_{\lambda \lambda'}(k)$ describes couplings between the system modes, whereas the diagonal parts correspond to frequency shifts $\Delta_{\lambda \lambda}$ and loss rates $\gamma_{\lambda \lambda}$.
	
	We note that to obtain this expression, the Fourier transform integrals have been regularized (see Appendix \ref{sec::app_FourReg} for details), analogously to what is usually done in time-independent scattering theory \cite{Newton1982}. We further note that a Markov approximation is not necessary in this derivation \cite{Viviescas2003}.
	
	Upon substitution of Eq.~\eqref{equ::IO_modeSol} into Eq.~\eqref{IO} we can read off the scattering matrix
	\begin{equation}\label{equ::IO_ViviescasScattering}
		S^{\mathstrut}_{\textrm{io}}(k) = \delta_{mm'} - 2\pi i \sum_{\lambda, \lambda'} W^*_{\lambda m}(k) D^{-1}_{\lambda \lambda'}(k) W^{\mathstrut}_{\lambda' m'}(k),
	\end{equation}
	such that
	\begin{equation}\label{equ::IO_bath_inputOutput}
	\hat{b}^{\textrm{(out)}}(k) = S^{\mathstrut}_{\textrm{io}}(k) \:\hat{b}^{\textrm{(in)}}(k).
	\end{equation}
	The subscript `io' stands for `input-output' to indicate that this scattering matrix has been obtained by solving the quantum statistical operator equations of motion of the ab initio few-mode Hamiltonian using the input-output formalism of quantum noise theory \cite{Gardiner1985,Gardiner2004}.
	
	\subsection{Equivalence to standard scattering theory}\label{sec::scattEquiv}

	We now show that the above calculation is equivalent to the full quantum scattering calculation, only expressed in a different basis. The relation is best understood by analogy to the state case (see Fig. \ref{fig::stateScatt}).
	
	Firstly we recognize that, using the definition of the coupling constants Eq.~\eqref{equ::GCH_couplings} as well as the completeness relations of the subspace eigenstates and Eq.~\eqref{equ::proj_Q_statesDef}, the decay matrix Eq.~\eqref{equ::IO_decayMatrix} can be written as
	\begin{align}
		\Gamma_{\lambda \lambda'}(k)
		= -\bra{\chi_\lambda}H_{QP}\tilde{G}^{(+)}H_{PQ}\ket{\chi_{\lambda'}}\,.
	\end{align}
	We have now chosen the bath states fulfilling retarded boundary conditions $\ket{\tilde{\psi}^{(+)}(k)}$ \cite{Newton1982,Domcke1983}, since by writing Eq.~\eqref{equ::IO_modeSol} in terms of the incoming operator, we have decided to solve an initial value scattering problem.
	
	From Eq.~\eqref{equ::IO_Dmatrix}, the $D$-matrix therefore consists of the matrix elements
	\begin{align}\label{Viviescas_D}
		D_{\lambda \lambda'}(k)  
		= \bra{\chi_\lambda}E(k) - H_{QQ} - H_{QP} \tilde{G}^{(+)} H_{PQ} \ket{\chi_{\lambda'}}\,.
	\end{align}
	Noting that the effective $Q$-space Hamiltonian is
	\begin{align}\label{Viviescas_Heff_Q}
	H_{\textrm{eff}} = H_{QQ} + H_{QP} \tilde{G}^{(+)} H_{PQ}\,,
	\end{align}
	we see that the inverse of the $D$-matrix coincides with the matrix elements
	\begin{equation}
		D^{-1}_{\lambda \lambda'}(k) = \bra{\chi_\lambda}G_{QQ}\ket{\chi_{\lambda'}}
	\end{equation}
	of the $Q$-space propagator $G_{QQ} = \left[E(k) - H_{\textrm{eff}}\right]^{-1}$.
	
	Substituting into Eq.~\eqref{equ::IO_ViviescasScattering}, again using the definition of the coupling constants Eq.~\eqref{equ::GCH_couplings} and the completeness relations of the subspace eigenstates, 
	we find that
	\begin{align}
		S^{\mathstrut}_{\textrm{io}}(k) &= \mathbb{I} - 2\pi i \: \bra{\tilde{\psi}^{(+)}(k)} H_{PQ}G_{QQ}H_{QP} \ket{\tilde{\psi}^{(+)}(k)} \nonumber
		\\
		&= S^{\mathstrut}_{\textrm{res}}(k)\,.
	\end{align}
	Thus the expression for the input-output scattering matrix $S^{\mathstrut}_{\textrm{io}}(k)$ coincides with the scattering matrix $S^{\mathstrut}_{\textrm{res}}(k)$ in Eq.~(\ref{equ::Domcke_Sres}) obtained from potential scattering theory using the Feshbach projection formalism~\cite{Domcke1983}.
	
	From our interpretation of the resonant scattering matrix in Sec.~\ref{sec::potScatt_Domcke}, it is to be expected that $S^{\mathstrut}_{\textrm{io}}(k)$ is not the full scattering matrix. The ab initio few-mode Hamiltonian Eq.~\eqref{equ::GC_H} only contains information about the dynamics of the system and bath modes, which interact via the coupling terms. Despite capturing these dynamics exactly, it does not contain information about the structure of the bath modes. In addition, the bath operators are not asymptotically free. Therefore analogously to the first quantized potential scattering case in Sec.~\ref{sec::potScatt_Domcke}, an asymptotic basis transformation is needed to translate from the bath operators in Eq.~\eqref{equ::IO_bath_inputOutput} to the asymptotically free operators in Eq.~\eqref{equ::potScatt_free_IO}, as schematically shown in Fig.~\ref{fig::stateScatt}. We further know from Eq.~(\ref{equ::Domcke_S}) that this transformation can be expressed as the background scattering matrix $S_{\textrm{bg}}(k)$. Therefore, the full scattering matrix can be calculated from the input-output result by
	\begin{equation}
		S(k) = S_{\textrm{bg}}(k)S_{\textrm{io}}(k)\,.
	\end{equation}
	To summarize, the background scattering contribution translates the bath mode scattering from the input-output formalism into free-state scattering as usually observed in spectroscopic experiments.
	
	We have thus clarified the relation of our ab initio FMA to the NMA and conventional quantum scattering theory. The two approaches are equivalent if care is taken to compute the scattering between asymptotically free operators in both cases. Figures~\ref{fig::schematic} and \ref{fig::stateScatt} illustrate the equivalence and the relation between the different operators.

	\section{Application to Maxwell's equations}\label{sec::Maxwell}
	While so far we have presented the construction of ab initio few-mode Hamiltonians on the example of the Schr\"odinger equation, our technique is in fact quite general. The essential requirements are that the Hilbert space of the quantum system can be separated into two orthogonal subspaces and that each subspace is spanned by a set of orthonormal modes. In the Schr\"odinger case, these conditions were ensured by the hermicity of the corresponding operators. One can thus envision an application of the formalism to a variety of quantized scattering problems.
	One such problem with practical relevance in quantum optics and cavity QED is the scattering of light from dielectric materials, described by Maxwell's equations. Since this field is a main application of system-bath theory and the input-output formalism as a phenomenological model, the question arises if our ab initio FMA can be applied to this setting as well.
	
	In the following, we analyze this question for the simplest possible case of a linear, isotropic, non-absorbing dielectric medium in one dimension, with only a single polarization considered. We show that within the rotating wave approximation (RWA), the correspondence between the input-output formalism and the potential scattering approach can be established.

	Our assumptions allow us to write the wave equation for a component $A(r,t)$ of the vector potential as~\cite{Viviescas2004}
	\begin{equation}\label{equ::max_scalarEOM}
		\frac{\partial^2}{\partial r^2}A(r,t) = \varepsilon(r) \frac{\partial^2}{\partial t^2}A(r,t)\,,
	\end{equation}
	where $\varepsilon(r)$ is the dielectric function and again $c=1$. The applicability of this scalar Helmholtz equation to physical scenarios has been discussed in \cite{Rotter2017}.
	This problem is closely related to our treatment of the Schr\"odinger equation, since the corresponding time-independent equation for the normal modes $f_m(r,k)$ \cite{Viviescas2003},
	\begin{equation}\label{equ::max_timeIndep}
		\frac{\partial^2}{\partial r^2} f_m(r,\omega) + \varepsilon(r) \omega^2 f_m(r,\omega) = 0\,,
	\end{equation}
	can be written as a Schr\"odinger equation with an energy-dependent potential \cite{Ching1998,Rotter2017}
	\begin{equation}\label{equ::kDep_potential}
		\tilde{V}(r,\omega) = \frac{1-\varepsilon(r)}{2}\: \omega^2\,.
	\end{equation}
	The normal modes of this wave equation are still orthogonal, but under a modified inner-product \cite{Viviescas2003,Rotter2017}
	\begin{align}\label{equ::kDep_innerProduct}
		\braket{x}{y} = \int dr\:  \varepsilon(r)\, x^*(r)\, y(r)\,.
	\end{align}
	
	The Maxwell wave equation can be quantized canonically \cite{Glauber1991} (see Appendix \ref{sec::app_maxwell_quant} for details), similarly to the Schr\"odinger case in Sec.~\ref{sec::schroed_CanQuant}. However due to the double time-derivative, the Hamiltonian now contains coordinate operators $\hat{q}$ and momentum operators $\hat{p}$ \cite{Glauber1991,Viviescas2003}, such that the corresponding commutation relations differ \cite{Glauber1991,Viviescas2003}.
	
	The separation into system and bath operators via a Feshbach projection can also be performed analogously to the Schr\"odinger case (see Appendix \ref{sec::app_maxwell_Feshbach} for details). The resulting few-mode Hamiltonian is of the form \cite{Viviescas2003}
	\begin{align}
	\hat{H} =& \sum_{\lambda\in \Lambda_Q} \omega_{\lambda}^{\mathstrut} \hat{a}_\lambda^\dag \hat{a}^{\mathstrut}_\lambda + \sum_m \int d\omega \: \omega \, \hat{b}_m^\dagger(\omega) \hat{b}^{\mathstrut}_m(\omega) \nonumber
	\\
	&+ \sum_{\lambda,m} \int d\omega \,\left [\mathcal{W}^{\mathstrut}_{\lambda m}(\omega) \: \hat{a}^\dagger_\lambda\, \hat{b}^{\mathstrut}_m(\omega)  \right. \nonumber \\
	&\left. \quad\quad\quad\quad~+ \mathcal{V}_{\lambda m}(\omega) \: \hat{a}_\lambda \,\hat{b}_m(\omega) + h.c. \right ] \,. \label{equ::max_GCH}
	\end{align}
	We note the appearance of counter-rotating terms in the system-bath coupling \cite{Viviescas2003}, which are also a result of the second time-derivative in the Maxwell wave equation.
	
	\subsection{Scattering in the rotating wave approximation}\label{sec::Maxwell_rot}
	We proceed with the analysis by applying the rotating wave approximation, which simplifies the Hamiltonian Eq.~\eqref{equ::max_GCH} to
	\begin{align}
	\hat{H}^{\mathstrut}_\textrm{rot} =& \sum_{\lambda\in \Lambda_Q} \omega^{\mathstrut}_\lambda \hat{a}_\lambda^\dag \hat{a}^{\mathstrut}_\lambda + \sum_m \int d\omega\: \omega \,\hat{b}_m^\dagger(\omega) \hat{b}^{\mathstrut}_m(\omega) \nonumber
	\\
	&+ \sum_{\lambda,m} \int d\omega \: \left [\mathcal{W}^{\mathstrut}_{\lambda m}(\omega)\: \hat{a}^\dagger_\lambda \,\hat{b}^{\mathstrut}_m(\omega) + h.c.\right ]\,. \label{equ::max_GCH_rot}
	\end{align}
	One can solve the equations of motion for this Hamiltonian analogously to Sec.~\ref{sec::scatt_IO}. The resulting scattering matrix is
	\begin{equation}\label{equ::max_scattIO}
	S^{\textrm{(rot)}}_{\textrm{io}}(\omega) = \delta_{mm'} - 2\pi i \sum_{\lambda, \lambda'} \mathcal{W}^*_{\lambda m}(\omega) \mathcal{D}^{-1}_{\lambda \lambda'}(\omega) \mathcal{W}^{\mathstrut}_{\lambda' m'}(\omega)\,,
	\end{equation}
	with
	\begin{equation}\label{equ::max_Dmatrix_rot}
	\mathcal{D}_{\lambda \lambda'}(\omega)^{\mathstrut} = (\omega-\omega_\lambda)\delta_{\lambda \lambda'} + \Gamma'_{\lambda \lambda'}(\omega)
	\end{equation}
	and
	\begin{align}\label{equ::max_decayMatrix_rot}
	\Gamma'_{\lambda \lambda'}(\omega)
	&= - \sum_m \int d\omega' \: \frac{\mathcal{W}^{\mathstrut}_{\lambda m}(\omega') \, \mathcal{W}^*_{\lambda'm}(\omega')}{\omega-\omega' + i\epsilon}\,.
	\end{align}
	In order to compare to scattering theory, we substitute the definition of the coupling constants $\mathcal{W}$ and translate to the Schr\"odinger normalization and energy labeling by (see Appendix~\ref{sec::app_maxwell_quant})
	\begin{equation}
		\mathcal{W}_{\lambda m}(\omega) = \frac{1}{2\sqrt{\omega_\lambda \omega}} \tilde{W}_{\lambda m}(\omega) = \frac{1}{\sqrt[4]{2E_\lambda}} W_{\lambda m}(k)\,,
	\end{equation}
	where $W_{\lambda m}(k)$ are the coupling constants corresponding to the scattering normalization.
	The scattering matrix then reads
	\begin{equation}\label{equ::max_scattIO_2}
		S^{\textrm{(rot)}}_{\textrm{io}}(k) = \delta_{mm'} - 2\pi i \sum_{\lambda, \lambda'} W^*_{\lambda m}(k) (D_\textrm{rot}^{-1}(k))^{\mathstrut}_{\lambda \lambda'} W^{\mathstrut}_{\lambda' m'}(k)\,,
	\end{equation}
	with
	\begin{equation}\label{equ::max_Dmatrix_rot2}
	(D_\textrm{rot}(k))^{\mathstrut}_{\lambda \lambda'} = 2\sqrt{E_\lambda}(\sqrt{E(k)}-\sqrt{E_\lambda})\delta^{\mathstrut}_{\lambda \lambda'} + \Gamma^\textrm{(rot)}_{\lambda \lambda'}(k)\nonumber
	\end{equation}
	and
	\begin{align}\label{equ::max_decayMatrix_rot2}
	\Gamma^\textrm{(rot)}_{\lambda \lambda'}(k)
	= -  \sum_m \int \frac{dE(k')}{2\sqrt{E(k')}} \frac{W^{\mathstrut}_{\lambda m}(k') W^*_{\lambda'm}(k')}{\sqrt{E(k)}-\sqrt{E(k')} + i\epsilon}\,.
	\end{align}
	We now see that these integrals are different to the ones encountered in scattering theory, due to the square-rooted energy dependence. However, since these expressions were derived under the assumption that the rotating wave approximation holds, we should also approximate $2\sqrt{E_\lambda} \approx \sqrt{E_\lambda} + \sqrt{E(k)}$ and $2\sqrt{E(k')} \approx \sqrt{E(k)} + \sqrt{E(k')}$ in the relevant energy ranges of above expressions. Substitution of these approximations shows that
	\begin{equation}
	D_\textrm{rot}(k) \approx D(k)\,,
	\end{equation}
	such that from comparing Eq.~\eqref{equ::max_scattIO_2} with Eq.~\eqref{equ::IO_ViviescasScattering} we get
	\begin{equation}
	S^{\textrm{(rot)}}_{\textrm{io}}(k) \approx S^{\mathstrut}_{\textrm{res}}(k)\,.
	\end{equation}
	This means that if the rotating wave approximation applies and is carried through consistently, the correspondence between the input-output operator scattering and the resonant state scattering matrix still holds. We note that it is in fact crucial to perform the above second step within the rotating wave approximation, in order to obtain the correct pole structure of the system propagator yielding a converging multi-mode expansion (see also Sec.~\ref{sec::eff} and Appendix \ref{sec::app_AnalytConvergence}).
	
	We further note that a similar correspondence can be established within the slowly-varying envelope approximation as outlined in Appendix~\ref{sec::app_maxwell_scatt}, an approximation which only modifies the time-dependence of the system and still yields the exact steady-state response.
	
	As a result, we find that within these approximations, our formalism can be applied straightforwardly to the scalar Helmholtz wave equation in the same way as for the Schr\"odinger equation, if a modified inner product and an energy-dependent potential are considered.
	
	\subsection{Scattering beyond the rotating wave approximation}\label{sec::Maxwell_beyondRot}
	Going beyond these approximations, we note that the input-output formalism does not require neglecting the counter-rotating terms \cite{Ciuti2006}. Without RWA, an additional linear equation for the conjugated operators has to be considered, which couples to the original equations via the counter-rotating terms. The input-output calculation can thus in principle be performed analogously.
	
	From the discussion in Sec.~\ref{sec::scatt} and Fig.~\ref{fig::stateScatt} it is clear that this will yield an input-output scattering matrix describing scattering between bath operators, which has to be multiplied by a background term to obtain the full scattering between asymptotically free operators. 
	The key difficulty now is to relate the contour integrals appearing in the operator scattering calculation (such as Eq.~\eqref{equ::max_decayMatrix_rot2}) to the matrix elements in the state scattering calculation (such as Eq.~\eqref{equ::Domcke_Sres}). In the case of the Schr\"odinger equation, a correspondence between the state scattering and the operator scattering has been shown in Sec.~\ref{sec::scattEquiv}, using the relation of the contour integrals to the bath Green function. In the Maxwell case, this correspondence is obscured due to the rooted energy dependence in the contour integrals. The origin of this can be understood since for Maxwell's equations, the field satisfying the wave equation has mixed operator contributions $A(r,t) \sim \hat{b} \tilde{\psi} + \hat{b}^\dagger \tilde{\psi}^*$, while for the Schr\"odinger equation $\psi(r,t) \sim \hat{b} \tilde{\psi}$.
	We note that conceptually the lack of such a correspondence makes no difference and the input-output scattering matrix can still be calculated if the contour integrals are evaluated correctly. Only now it is not clear if $S_{\textrm{io}}(k) = S_{\textrm{res}}(k)$ can be invoked to simplify the calculation. 
	
	As a result, we conclude that even beyond the rotating wave approximation our formalism can be applied to calculate ab initio input-output scattering matrices, however the precise form of the corresponding background scattering matrix on the operator scattering level remains to be determined (see also Fig.~\ref{fig::stateScatt}).
	
	\section{Practical aspects}\label{sec::models}

	Before turning to an example calculation, we conclude our analysis with practical remarks, in particular focusing on applications in cavity QED. Applying the ab initio FMA discussed here in essence entails two parts. The first part is the calculation of the quantum optical parameters and coupling constants entering the Hamiltonian and the input-output relations. The second part is the solution of the equations of motion resulting from the Hamiltonian. Regarding the second part, it is important to note that the Hamiltonian and the input-output relations obtained from our FMA are quite similar in structure to that of the well-established phenomenological models. This is of great advantage, since it means that the solution methods  established for phenomenological models can also be applied to our approach, once the coupling constants are evaluated.
	
	Nevertheless, there are certain differences to standard phenomenological models, which we discuss in the following. The model input-output relation is usually written in the form \cite{Gardiner2004}
	\begin{equation}\label{equ::modelComp_modelIO_time}
		\hat{b}^{(\textrm{out})}(t) - \hat{b}^{(\textrm{in})}(t) = -i\sum_{\lambda}\sqrt{\kappa_\lambda} \: \hat{a}_\lambda(t)\,,
	\end{equation}
	or alternatively in terms of the corresponding Fourier transforms
	\begin{equation}\label{equ::modelComp_modelIO_spectral}
		\hat{b}^{(\textrm{out})}(\omega) - \hat{b}^{(\textrm{in})}(\omega) = -i\sum_{\lambda}\sqrt{\kappa_\lambda}\: \hat{a}_\lambda(\omega)\,,
	\end{equation}
	from which a spectrum can be computed. Here, $\kappa_\lambda$ is the coupling constant between the cavity mode $\lambda$ and the external bath mode considered. 
	
	The corresponding input-output relation derived within our approach reads  (compare Eq.~\eqref{IO})
	\begin{equation}\label{equ::modelComp_abInitioIO_time}
		\hat{b}^{\textrm{(out)}}_m(\omega) - \hat{b}^{\textrm{(in)}}_m(\omega) = -i \sum_{\lambda} \mathcal{W}^{*\mathstrut}_{\lambda m}(\omega)\: \hat{a}^{\mathstrut}_\lambda(\omega)\,.
	\end{equation}
	This expression is similar in structure to Eq.~\eqref{equ::modelComp_modelIO_spectral}, only now the cavity-bath coupling is frequency dependent. It is important to note that this frequency dependence also includes the possibility that the couplings change considerably within the spectral width of a single resonance, which cannot be captured by fitting a phenomenological Lorentzian mode to the response of the system.  An example for this will be shown in Sec.~\ref{sec::exSyst}.
	
	Next, we turn to the equations of motion for the cavity modes. Including a loss constant $\gamma$, a typical equation of motion within a phenomenological model reads 
	\begin{equation}\label{equ::modelComp_modelLangevin2_time}
		\frac{d}{dt}\hat{a}_\lambda(t) = -i\omega_\lambda\: \hat{a}_\lambda(t) - i\sqrt{\kappa_\lambda} \: \hat{b}^{(\textrm{in})}(t) - \gamma_\lambda \: \hat{a}_\lambda(t)\,.
	\end{equation}
	This can again be expressed in Fourier space as
	\begin{equation}\label{equ::modelComp_modelLangevin2_spectral}
		-i\omega\hat{a}_\lambda(\omega) = -i\omega_\lambda\: \hat{a}_\lambda(\omega) - i\sqrt{\kappa_\lambda} \:\hat{b}^{(\textrm{in})}(\omega) - \gamma_\lambda\: \hat{a}_\lambda(\omega)\,,
	\end{equation}
	so that spectroscopic quantities such as reflection or transmission spectra can be obtained by substituting Eq.~\eqref{equ::modelComp_modelLangevin2_spectral} into Eq.~\eqref{equ::modelComp_modelIO_spectral}. When atoms or other quantum systems are present inside the cavity, additional terms are added to describe cavity-atom interactions (see also Sec.~\ref{sec::eff}).
	
	The corresponding  Langevin equation in our ab initio few-mode theory reads (compare Eqs.~(\ref{equ::IO_modeSol}, \ref{equ::max_decayMatrix_rot}))
	\begin{align}
		-i\omega\hat{a}^{\mathstrut}_\lambda(\omega) &= -i\omega^{\mathstrut}_\lambda \:\hat{a}^{\mathstrut}_\lambda(\omega) - 2\pi i \sum_m \mathcal{W}^{\mathstrut}_{\lambda m}(\omega)\:  \hat{b}^{(\textrm{in})}_m(\omega) \nonumber
		\\
		&- \sum_{\lambda'}[\gamma^{\mathstrut}_{\lambda \lambda'}(\omega) + i\Delta^{\mathstrut}_{\lambda \lambda'}(\omega)]\: \hat{a}^{\mathstrut}_{\lambda'}(\omega)\,. \label{equ::modelComp_abinitioLangevin}
	\end{align}
	Comparing this with Eq.~\eqref{equ::modelComp_modelLangevin2_spectral}, we again find frequency dependent decay and coupling constants. Additionally, next to the loss rates $\gamma_{\lambda \lambda}$, an imaginary contribution $\Delta_{\lambda \lambda}$ appears, which induces a frequency shift. Furthermore, both the loss and the frequency shift parameters are now matrices, such that cross-mode coupling terms with $\lambda\neq\lambda'$ are present. Such cross-mode terms bear the potential for qualitatively different phenomena, for example, spontaneously generated coherences~\cite{Ficek_Swain,Kiffner_Vacuum_Processes,Heeg2013a}.
	
	Also the frequency dependence of the coupling constants may lead to qualitative differences to phenomenological models, since in the time-domain, it implies  non-Markovian dynamics. For example, the input-ouput relation in the time domain can be obtained by Fourier transforming Eq.~\eqref{equ::modelComp_abInitioIO_time} and reads~\cite{Viviescas2003}
	\begin{equation}
		\hat{b}^{\textrm{(out)}}_m(t) - \hat{b}^{\textrm{(in)}}_m(t) = -i \sum_{\lambda} \mathcal{W}^{*\mathstrut}_{\lambda m}(t) * \hat{a}^{\mathstrut}_\lambda(t)\,,
	\end{equation}
	where $\mathcal{W}^{*\mathstrut}_{\lambda m}(t) = \int \frac{d\omega}{2\pi}\: e^{-i\omega t} \mathcal{W}^{*\mathstrut}_{\lambda m}(\omega)$ and $*$ denotes a convolution. The output field thus depends on the history of the cavity mode operators. A similar connection is obtained when writing the Langevin equation in the time domain. We note that such non-Markovian input-output relations have been studied in detail in \cite{Diosi2012,Zhang2013}.
	
	We therefore see that our ab initio few-mode theory can be employed as a tool to calculate cavity spectra analogously to the phenomenological approach, and the computational simplicity of the phenomenological models is not destroyed by the ab initio method. In particular for spectral observables, including frequency dependent couplings does not incur significant additional complexity. The main task to apply the formalism will thus lie in calculating the frequency dependent coupling and decay constants from the cavity geometry by employing the projection operator equations in Sec.~\ref{sec::schroed}. After this calculation, the complete tool box of the input-output formalism and system-bath theory can be applied and the various approximation schemes that are available for few-mode systems can be employed. For details on how these statements generalize in the presence of interactions refer to Sec.~\ref{sec::eff}.
	
	\begin{figure}
		\includegraphics[width=\columnwidth]{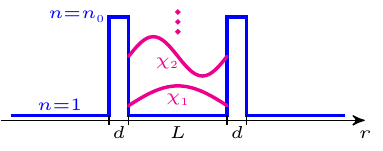}
		\caption{Model potential with two barriers. In the Maxwell case, this corresponds to the Ley-Loudon model for a two-sided Fabry-Perot cavity  \cite{Ley1987}, and the solid blue curve shows the spatial refractive index distribution. For simplicity, in the calculation, the thin-mirror limit $d\rightarrow0$ is considered, with  $n_0\rightarrow \infty$ such that $\eta=n_0^2 d$ remains finite \cite{Ley1987,Viviescas2004}. In the cavity, the first two perfect-cavity modes $\chi_1, \chi_2$ are shown as magenta curves. For the Schr\"odinger case, the solid blue curve indicates the potential energy, which defines a tunneling problem. }\label{fig::maxwell_fabryPerot}
	\end{figure}
	
	\begin{figure*}
		\includegraphics[scale=1.0]{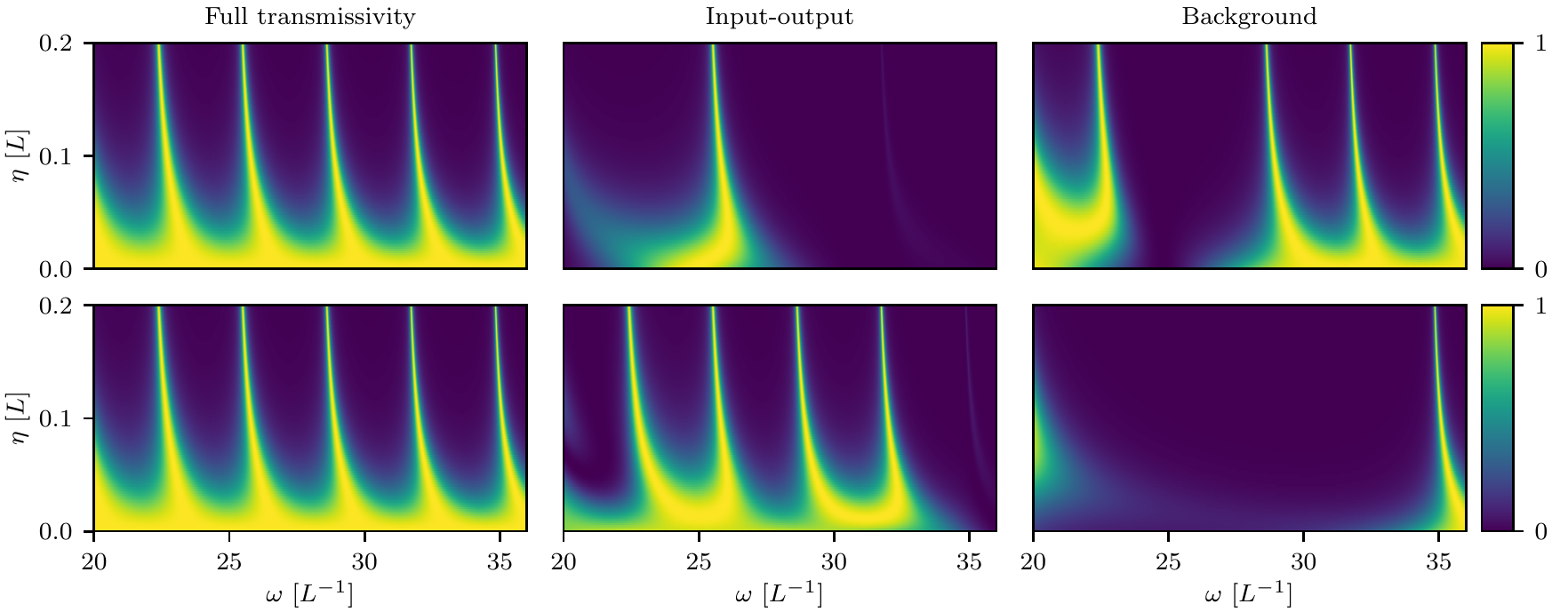}
		\caption{\label{fig::maxwell_doubleDelta2D}Transmission spectra for a Fabry-Perot cavity as a function of the mirror quality $\eta$. The top row shows the case in which the system comprises the single mode $\lambda=8$. The bottom row shows corresponding results with the system consisting of the modes $\lambda\in \{7,8,9,10\}$. In both cases, the left column illustrates the full transmissivity of the system. The middle and right columns show the input-output ($S_{\mathrm{io}}$) and the background ($S_{\mathrm{bg}}$) contributions, respectively. The full result can either be obtained from standard methods such as a transfer matrix formalism also known as Parratt's formalism \cite{Parratt1954,Rohlsberger2005}, or as a product of the input-output and background scattering matrices.}
	\end{figure*}

	\section{Example: Double barrier potential}\label{sec::exSyst}
	
	To illustrate our formalism for non-interacting theories and as a proof-of-concept, we perform explicit calculations for the example of a one-dimensional potential featuring two barriers, see Fig.~\ref{fig::maxwell_fabryPerot}. Because our derivation in the Maxwell case works analogously to the Schr\"odinger case, it is tempting to assume that the two wave equations will give similar results. Below, we show that this is not the case, because they lead to different potentials in the respective Hamiltonians, and thus to different scattering properties. In each case, we demonstrate how our few-mode formalism enables the extraction of relevant resonant dynamics.

	\subsection{Maxwell case: Fabry-Perot cavity}\label{sec::maxwell_FabryPerot}
	In the Maxwell case, the two-barrier potential is realized using a spatially varying index-of-refraction distribution, and corresponds to a two-sided Fabry-Perot cavity with a semi-transparent mirror at each end. For simplicity, we consider the thin-mirror limit $d\rightarrow0$  with  $n_0\rightarrow \infty$ such that $\eta=n_0^2 d$ remains finite, which is known as the Ley-Loudon model \cite{Ley1987,Viviescas2004}.  This model is one of the simplest cavity geometries with tunable sharp resonances. The mirror quality can be characterized by $\eta = n_0^2d$, which relates to the energy dependent mirror reflectivity via $r(\omega) = {i\omega\eta}/(2-i\omega\eta)$ \cite{Ley1987,Viviescas2004}. 
	Within this model, the potential in the Maxwell case thus becomes
	\begin{equation}\label{eq:potmaxwell}
		\tilde{V}(r,\omega) = -\big[\eta_1 \delta(r-L/2) + \eta_2 \delta(r+L/2)\big] \omega^2/2 \,.
	\end{equation}
	For this system a natural choice of cavity modes are the ``perfect cavity modes'', that is eigenstates in the cavity region with Dirichlet boundary conditions at the mirrors given by
	\begin{equation}\label{eq::perfect-modes}
		\chi_\lambda(r) = \sqrt{\frac{2}{L}} \sin(\omega_\lambda r), \quad 0\leq r \leq L.
	\end{equation}
	The eigenfrequencies are $\omega_\lambda = \lambda \pi/L$ and $L$ is the cavity length.
	
	\begin{figure*}
		\includegraphics[scale=1.0]{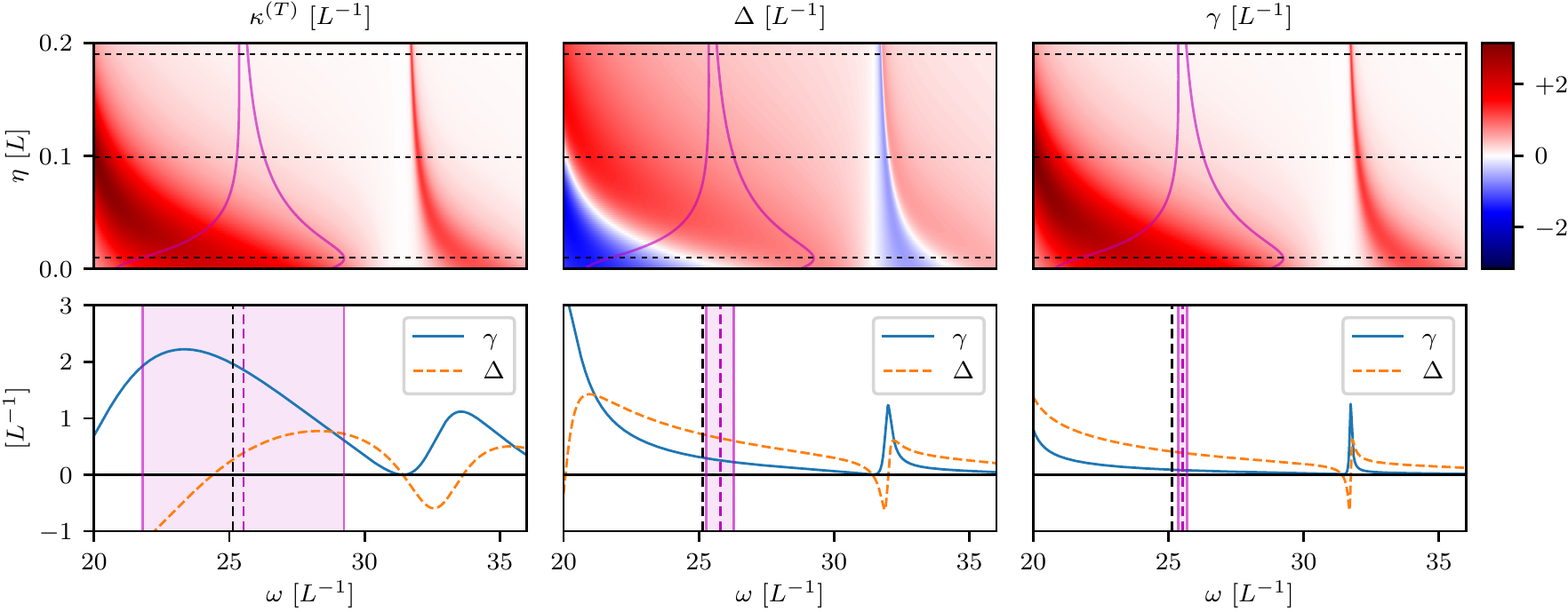}
		\caption{\label{fig::maxwell_doubleDelta2D_couplings}Quantum optical parameters calculated via the ab initio few-mode theory. The upper row shows the transmission coupling strength $\kappa^{(T)} = 2\pi |\mathcal{W}|^2$, the mode frequency shift $\Delta$ and the resonance width $\gamma$ as a function of frequency and mirror quality $\eta$. The parameters are as in the upper row of Fig.~\ref{fig::maxwell_doubleDelta2D}, with the system comprising the single mode $\lambda = 8$. The magenta curves indicate the width of the resonance as function of $\eta$. 
		The lower row shows cuts through the upper panels at fixed $\eta=0.01, 0.1, 0.19$ (left to right, corresponding to the transition from a bad to a good cavity) indicated as dotted lines in the upper panel.  The respective widths of the modes are indicated as shaded magenta regions, defined as twice the value of $\gamma$. The vertical dashed lines indicate the bare center frequency of the mode (black) as well as the actual center frequency (magenta).}
	\end{figure*}
	
	Based on these states, we numerically evaluate the input-output scattering matrix $S_{\mathrm{io}}$ and the corresponding background scattering matrix $S_{\mathrm{bg}}$ in the rotating wave approximation. Due to the cavity being open on both sides, this is a two channel problem featuring transmission as well as reflection. Each part in the relation $S = S_{\mathrm{bg}}S_{\mathrm{io}}$ thus is a $2\times2$ matrix. 
	
	In Fig.~\ref{fig::maxwell_doubleDelta2D}, we show transmission spectra for the cavity as a function of the mirror quality, and compare it to the individual resonant input-output ($S_{\mathrm{io}}$) and background ($S_{\mathrm{bg}}$) contributions. In all cases, the full transmissivity coincides with the product of the resonant and the background contributions, as has been shown in Sec.~\ref{sec::scattEquiv}. The upper row illustrates the case in which the system space comprises a single mode with $\lambda = 8$. The lower row shows corresponding results with four resonant modes as the system part ($\lambda \in \{7,8,9,10\}$).
	As expected, for a good cavity with high $\eta$, well-resolved transmission resonances are obtained, which naturally split into the resonant and the background contributions. Each mode that is included in the few-mode Hamiltonian removes a resonance peak from the background and adds it to the input-output scattering matrix. This means that in the vicinity of the included resonances, one can expect that  the input-output result alone gives a good representation of the scattering behavior. 
	But towards the bad-cavity limit ($\eta \to 0$), the modes start to overlap, and the separation into resonant and background part becomes non-trivial. As a result, the background part is crucial, and more modes are required for the input-output matrix to capture the resonance behavior in the same frequency range. Also, the position of the mode resonance systematically shifts with the quality factor $\eta$, which is a consequence of the imaginary contribution $\delta$ found in the ab initio equations. Furthermore, the resonant modes become asymmetric with respect to their central frequencies, and are no longer of Lorentzian shape. This asymmetry can be understood since the width of the resonances decreases for this cavity with increasing energy. As a result there is more overlap of any particular resonance with its lower energy neighbor than with its higher energy neighbor, which also leads to the formation of two distinct pairs of modes in the case of multiple system modes in the lower row of Fig.~\ref{fig::maxwell_doubleDelta2D}.
	
	Next, we study the quantum optical parameters extracted from our ab initio approach. Fig.~\ref{fig::maxwell_doubleDelta2D_couplings} shows the transmission coupling strength $\kappa^{(T)}$ entering the input-output relation, the mode frequency shift $\Delta$ and the decay rate $\gamma$ as a function of frequency and mirror quality $\eta$. All plots correspond to the upper panel of Fig.~\ref{fig::maxwell_doubleDelta2D}, with a single mode as the system subspace. In the upper panels of Fig.~\ref{fig::maxwell_doubleDelta2D_couplings}, the solid purple curve indicates the spectral width of the mode as function of $\eta$. The lower panels show three cuts through the plots in the upper panel, for different values of $\eta$. In these lower panels, the purple shaded area indicates the spectral width of the mode, which grows towards lower $\eta$. As expected, for a high-quality cavity, the system parameters calculated using the ab initio method are approximately constant over the spectral width of the resonance. Thus we again find that a phenomenological approach with constant parameters is well-suited to model the cavity dynamics. However, towards the bad-cavity limit, the system parameters significantly change within the spectral width of the mode, rendering a modeling using fixed phenomenological rates difficult. Finally, the vertical lines in the lower panel indicate the difference of the actual mode frequency from the ``bare'' mode frequency, that is the effect of the imaginary part $\Delta$.
	
	From these results, we conclude that our formalism can indeed be used to extract the resonant dynamics of the system, by choosing the relevant modes that participate in the dynamics. We further conclude that the input-output formalism is not limited to the good cavity regime, however has to be applied with care when the cavity features overlapping modes, since background scattering and frequency dependence of the quantum parameters become sizable and cannot be neglected.
	
	\begin{figure}
		\includegraphics[width=\columnwidth]{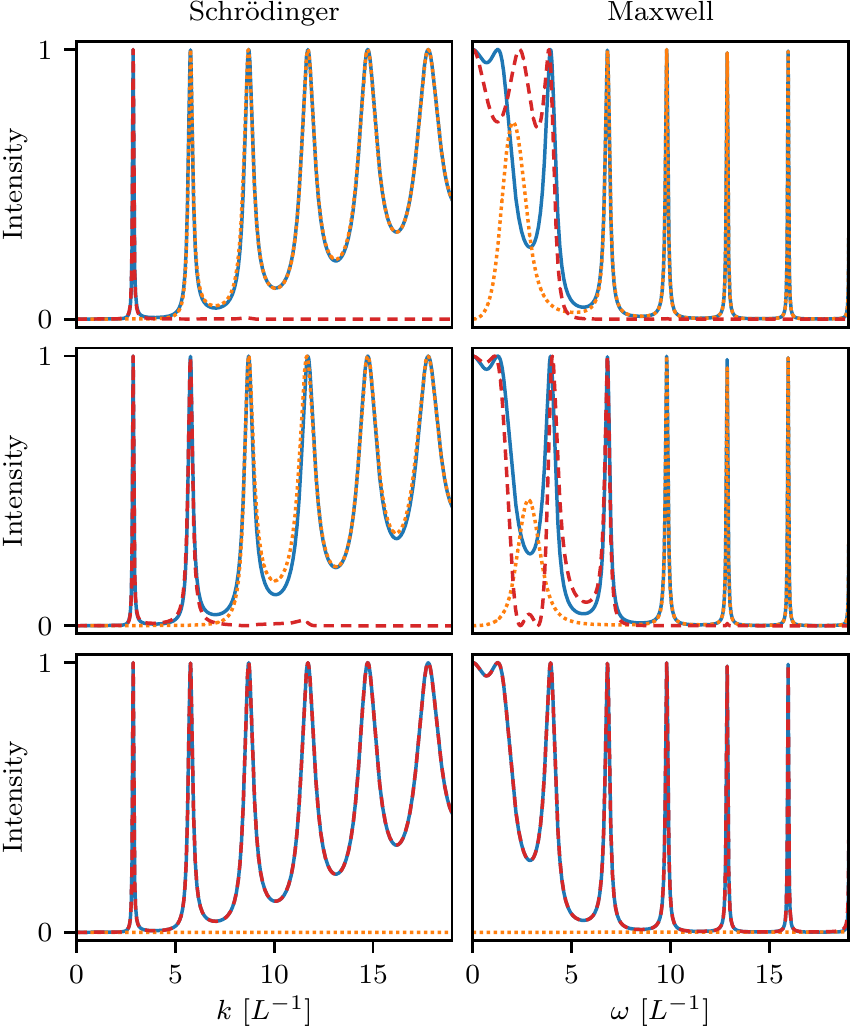}
		\caption{Transmission spectra for a Schr\"odinger (left column) and a Maxwell (right column) Fabry-Perot cavity, with different sets of system modes (top: $\lambda= 1$, middle: $\lambda\in \{1,2\}$, bottom: $\lambda\in \{1,2,\dots,100\}$). In each case, the full (solid blue), input-output (dashed red) and background (dotted orange) transmissivity are shown. For both wave equations, each added system mode transfers a resonance peak from the background to the input-output contribution, such that in the many mode case featuring 100 modes, the input-output result alone agrees with the full transmissivity (bottom panels). For sharp resonance peaks, the input-output result captures the behavior in the relevant energy range, if the corresponding modes are included (top and middle left panels). For overlapping resonances, the background contribution is crucial even in the vicinity of the resonance peak (top and middle right panels).}\label{fig::SchroedMax}
	\end{figure}
	
	\subsection{Schr\"odinger case: Tunneling problem}

	In the Schr\"odinger case, the double-barrier potential structure shown in Fig.~\ref{fig::maxwell_fabryPerot} defines a tunneling problem, and can be written as
	\begin{equation}
		V(r) = \xi_1 \delta(r-L/2) + \xi_2 \delta(r+L/2)\,. \label{eq:potschroed}
	\end{equation}
	We note that this potential has prefactors independent of the energy, while the corresponding potential Eq.~(\ref{eq:potmaxwell}) for the Maxwell case is proportional to $\omega^2$, and thus energy dependent. This gives rise to crucial differences between the Schr\"odinger and the Maxwell wave equation, which we discuss below. 
	
	Fig.~\ref{fig::SchroedMax} compares the transmissivity in the Schr\"odinger case and the Maxwell case, 
	for the parameters $\xi_1=\xi_2=10 L^{-1}$ and $\eta_1=\eta_2=0.5 L$. The three rows correspond to a system space containing one mode (top row: $\lambda = 1$), two modes (middle row, $\lambda \in \{1,2\}$), or the many-mode limit (bottom row, $\lambda \in \{1, \dots, 100\}$). 
	
	The Schr\"odinger transmissivity features sharp resonances at low energies, which can be understood by noting that at low energies it is less likely for a particle to tunnel through or overcome the confining barriers (see Fig.~\ref{fig::maxwell_fabryPerot}). With increasing energy, these resonances become broader and start to overlap. Furthermore, the baseline of the transmissivity resonances rises with increasing energy. 
	
	In the Maxwell case, the transmissivity spectrum at low energies is entirely different. This is due to the prefactor $\omega^2$ in the potential, which vanishes at low energies. As a consequence, the modes become broader and the baseline of the transmissivity resonances raises towards lower energies. In contrast, towards higher energies, the potential $\sim \omega^2$ is highly confining and features sharp resonances. On a qualitative level, the frequency dependence of the Maxwell resonances thus appears reversed as compared to the Schr\"odinger case.
	
	Next, we investigate the behavior of the few-mode input-output results further in both cases, by comparing the input-output and background transmissivity separately for different system mode numbers (see Fig.~\ref{fig::SchroedMax}). As expected, we observe that for each additional system mode, a resonance peak gets transferred from the background to the input-output spectrum. For the case where a single mode with $\lambda = 1$ is included, the Schr\"odinger and Maxwell equations show very different behavior. In the Schr\"odinger case the corresponding resonance is sharp and isolated, such that the input-output transmissivity reproduces the full result in the energy range of the resonance peak, even without having to include the background contribution. In the Maxwell case, however, these modes are broad and overlap, such that the background contribution is crucial. It is important to note that this difference is a consequence of the $\omega^2$-dependence of the Maxwell potential, and not of the single-mode approximation. This can be seen from the top panel of Fig.~\ref{fig::maxwell_doubleDelta2D}, where the single mode $\lambda = 8$ is well-represented by the input-output part alone for the Maxwell case. As a result of the $\omega^2$ dependence, the ``perfect'' system modes Eq.~(\ref{eq::perfect-modes}) for barriers of infinite height do not represent the $\lambda = 1$ case of shallow potential barriers well.
	
	We further note that the transmissivity maxima in the Maxwell case of Fig.~\ref{fig::SchroedMax} lie between the ones for the Schr\"odinger equation, despite the identical geometry. On the level of wave equations, this can also be explained by the energy dependence of the potential causing the complex poles of the scattering matrix to shift. In the quantized few-mode Hamiltonian approach, the shift can alternatively be understood as radiative corrections to the bare system states, which we chose to be the perfect cavity states  Eq.~(\ref{eq::perfect-modes}). These corrections arise from the system-bath coupling and are expressed as the complex decay matrix. The shifting effect can thus also be seen in Fig.~\ref{fig::maxwell_doubleDelta2D_couplings}, where the mode frequency shift $\Delta$ remains larger than the mode width for large $\eta$.

	\section{Interacting quantum systems}\label{sec::eff}
	In the previous sections, we have shown how to derive ab initio few-mode Hamiltonians for quantum potential scattering problems and how the full scattering information can be reconstructed from such Hamiltonians using the input-output formalism. We have further demonstrated that by choosing certain states in the few-mode basis, the corresponding spectral resonance peaks can be extracted.
	
	This idea of extracting important degrees of freedom is at the heart of few-mode theory. The concept also naturally leads to a crucial approximation when considering interacting systems, such as atoms coupling to the quantized field, which are often theoretically intractable in their full complexity. The \textit{few-mode approximation} allows to boil down the field continuum to a few relevant degrees of freedom that dominate the interacting dynamics, by neglecting the interaction with other irrelevant modes (see Fig.~\ref{fig::schematic_modeSelection}). Our ab initio few-mode theory now enables this approximation to be performed rigorously, provides new insight on its range of validity and gives practical advantages for its application. The main step behind this progress is the possibility of choosing the system states at will while still treating the free system exactly, such that one can focus on approximating the interaction.
	
	We note that the few-mode approximation has already been employed extensively in the study of cavity QED \cite{Berman1994,Haroche2013,Ritsch2013} and related subjects by using phenomenological few-mode Hamiltonians. Importantly, a large bulk of theoretical tools has been developed to solve and understand the resulting dynamical equations \cite{Gardiner2004,Breuer2002_BOOK,Carmichael2008}, which have found applications in a broad quantum optics context, also beyond cavity QED. However, these approaches inherit the limitations of phenomenological few-mode approaches discussed in the previous sections.
	
	In this section, we show how our ab initio few-mode theory can be applied to interacting quantum systems, providing a number of advantages to phenomenological few-mode theory. Firstly, in ab initio few-mode theory, the empty cavity or potential is treated exactly no matter which system modes are chosen such that the interacting case inherits the advantages from the non-interacting one. Secondly, a systematic effective few-mode expansion scheme can now be constructed where only the interaction is approximated. Thirdly, an important aspect of our method is that it connects to the toolbox of phenomenological few-mode theory, such that frequently used techniques do not have to be abandoned.
	Lastly, this extends the range of few-mode theory to extreme parameter regimes, such as highly open and multi-mode systems, where previously mentioned aspects of ab initio few-mode theory, such as frequency-dependent couplings and background scattering, can be crucial.
	
	In the following, we first outline the construction of ab initio few-mode theory for interacting systems. We then discuss each of the advantages of ab initio few-mode theory mentioned above, and demonstrate them using representative examples.
	
	\subsection{Effective few-mode expansions}\label{sec::int_lightMatter}
	We outline the construction of ab initio few-mode theory using a paradigmatic model from the field of light-matter interactions: a two-level atom in a cavity.
	
	For clarity and consistency with previous sections, the term `system' is reserved for the cavity in the following, and not used to describe the atom, which is referred to as `atom/interaction'.
	
	\subsubsection{Interaction Hamiltonian}	
	The Hamiltonian for a Maxwell field interacting with a single two-level atom is \cite{Scully1997}
	\begin{equation}
		\hat{H} = \hat{H}_\mathrm{field} + \hat{H}_\mathrm{atom} + \hat{H}_\mathrm{int}.
	\end{equation}
	Here, $\hat{H}_\mathrm{field}$ is given by the quantization of the dielectric wave equation from Sec.~\ref{sec::Maxwell}, and can be expressed in the usual normal mode basis by Eq.~\eqref{equ::max_NormH} or equivalently in a few-mode system-bath basis by Eq.~\eqref{equ::max_GCH}. For a two-level system, the atomic Hamiltonian is given by $\hat{H}_\mathrm{atom} = \frac{\omega_\textrm{a}}{2} \hat{\sigma}^z$, where $\omega_\textrm{a}$ is the transition frequency and $\hat{\sigma}^{x,y,z,+,-}$ are the Pauli operators. The interaction Hamiltonian can be obtained by the minimal coupling substitution \cite{Scully1997,Cohen-Tannoudji1998b}, and in the dipole approximation can be written as \cite{Scully1997,Cohen-Tannoudji1998b}
	\begin{equation}\label{equ::int_interactionHamiltonian}
		\hat{H}_\mathrm{int} = -i\omega_\textrm{a}  (d\hat{\sigma}^+ - d^*\hat{\sigma}^-) A(r_a),
	\end{equation}
	where $d$ and $r_a$ are the transition dipole moment and the position of the atom, respectively. Consistently with previous notation, we set $\hbar = m_e=1$. We note that following the minimal coupling prescription, we use the $p \cdot A$ interaction term here \cite{Scully1997,CohenTannoudji1997}, since the canonical quantization scheme that we employed works in the Coulomb gauge \cite{Glauber1991}, and as a result our system-bath Hamiltonians are also in this gauge. We further neglect the $A^2$ term in the interaction. This treatment is known to cause problems in the ultra-strong or deep-strong coupling regimes \cite{FriskKockum2019,Forn-Diaz2019}, whose resolution has been discussed elsewhere (see, for example, \cite{DeLiberato2014,GarciaRipoll2015,Malekakhlagh2016b,Malekakhlagh2017,DeBernardis2018,DiStefano2019,Ridolfo2013}). For our purposes, this approach suffices and we also perform the rotating wave approximation in the light-matter coupling. We note that polarization is already absent, since we considered a scalar version of the Maxwell wave equation in Sec.~\ref{sec::Maxwell}. As before, we employ these assumptions for simplicity, in order to demonstrate the central ideas of ab initio effective few-mode theories. We expect, however, that the method can be extended to a broad class of Hamiltonians (see Sec.~\ref{sec::outlook}), since it only relies on the few-mode concept and the previously constructed basis transformation for the field, which is exact.
	
	The crucial advantage of the few-mode approach arises when we express the field in terms of a mode expansion. In the standard normal modes basis, the expansion results in an interaction Hamiltonian where the atom couples to a continuum of modes (see Eq.~\eqref{equ::max_E_normLadderExpansion}).
	In the few-mode basis, the expansion Eq.~\eqref{equ::max_E-expansion_ladderOps} gives an alternative representation of the interaction Hamiltonian
	\begin{align}\label{eq::H-alternative}
		\hat{H}_\mathrm{int} &= \hat{H}_\textrm{atom-cavity} + \hat{H}_\textrm{atom-bath} \,,
	\end{align}
	with \cite{Viviescas2003,Krimer2014}
	\begin{subequations}
	\begin{align}
		\hat{H}_\textrm{atom-cavity} &= \sum_\lambda g^{\mathstrut}_\lambda \hat{\sigma}^+ \hat{a}^{\mathstrut}_\lambda  + h.c. \,,\\
		\hat{H}_\textrm{atom-bath} &= \sum_m \int d\omega \tilde{g}^{\mathstrut}_m(\omega) \hat{\sigma}^+\hat{b}_m(\omega) + h.c.\,,
	\end{align}
	\end{subequations}
	where the atom-cavity and atom-bath coupling constants are defined analogously to the normal mode case, that is
	\begin{subequations}
	\begin{align}
		g_\lambda &= -id\omega_\textrm{a}\,\sqrt{\frac{1}{2\omega_\lambda}} \chi_\lambda(r_a)\,,  \\
		\tilde{g}_m(\omega) &= -id\omega_\textrm{a}\,\sqrt{\frac{1}{2\omega}} \tilde{\psi}_m(r_a, \omega)\,. 
	\end{align}
	\end{subequations}
	As shown in the previous sections on non-interacting problems, the ab initio few-mode approach allows one to choose the system modes freely without having to approximate the field Hamiltonian $\hat{H}_\mathrm{field}$.
	This enables a systematic few-mode approximation scheme for the interacting theory, which we discuss in detail in the next sections.
	
	\subsubsection{Few-mode expansion scheme}\label{sec::FMexpansion}
	
	We see from Eq.~(\ref{eq::H-alternative}) that in the system-bath basis, the atom couples to the discrete system (cavity) modes as well as to a continuum of bath modes. While this Hamiltonian has been obtained from the normal modes Hamiltonian without further approximations, there is no clear advantage to the normal modes formulation yet, because the Hamiltonian still involves a continuum part. 	
	The \textit{few-mode approximation} consists of only including the atom-cavity interaction, 
	\begin{equation}\label{eq::FMapprox}
		\hat{H}_\mathrm{int} \approx \hat{H}_\textrm{atom-cavity}\,,
	\end{equation}
	such that the continuum part is neglected, where the cavity part includes the chosen system modes. If applicable, this approximation is tremendously useful, since it vastly reduces the complexity of the coupling and the dimension of the coupled system. Phenomenological few-mode theory, encompassing famous models such as the Jaynes-Cummings model \cite{Jaynes1963a}, the Rabi model \cite{Rabi1936,Braak2011} and the Dicke model \cite{Dicke1954a,Kilin1980}, is based on this approximation. Indeed the above interaction Hamiltonian is exactly of the form of a multi-mode Jaynes-Cummings model, emphasizing the close connection between phenomenological and ab initio few-mode theory. However, we found in the previous sections that phenomenological few-mode theory may lead to incorrect predictions already in the non-interacting case, depending on the system and regime under study. 
	
	The key advantage of our ab initio few-mode theory as compared to phenomenological approaches is that the non-interacting system is treated exactly. As a result, we can choose any set of system modes to describe the cavity alone, without affecting the non-interacting part. This allows us to disentangle the few-mode approximation from approximative treatments of the cavity openness.
	
	The few-mode expansion scheme then comprises a systematic variation of the number of system modes, such that the predictions of the approximate interaction Hamiltonian Eq.~(\ref{eq::FMapprox}) converge to the exact results.
	
	\subsubsection{Choice of few-mode basis}\label{sec::FMchoice}
	From the previous section it is clear that the choice of the few-mode basis is important, and we will find below that it in particular affects the rate of convergence as a function of the number of included system modes. Usually, prior knowledge about the system under study can be used to guide the choice of relevant system modes. In general, this constitutes an optimization problem, where the task is to find the minimal and optimal set of modes with respect to an optimization criterion. What constitutes a good set of relevant modes may also depend on what further approximations one would like to make. For example, if one wants to derive a Markovian master equation by tracing out the bath modes, one should try to limit the frequency dependence of the coupling coefficients (see also Sec.~\ref{sec::nonlinear}).
	
	In the absence of any prior knowledge, a constructive approach can be used that allows one to obtain a systematic expansion in the number of included modes. These modes may not be the optimally relevant few-mode basis, but they still provide a non-perturbative series expansion for observables. The method relies on the insight that for strongly confining systems, the perfect cavity eigenstates provide a good few-mode basis. A natural approach, even for weakly confining systems, is thus to solve the Dirichlet boundary value problem in the region of the cavity potential, giving a complete basis set in the region where the atom is located, as illustrated in Fig.~\ref{fig::schematic_modeSelection}. The few-mode basis is obtained by choosing a subset of these states, according to the energy scales set by the atom inside the cavity. The number of modes can then be varied systematically, and in the limit of infinitely many modes, where the few-mode basis becomes complete in the interaction region, should converge to the full solution of the problem (see Sec.~\ref{sec::int_convergence} for a detailed investigation of convergence).
	
	For completeness, we note that the selection of a confinement region with boundary conditions is reminiscent of R-matrix theory \cite{Wigner1947,Lane1958,Burke1977,Descouvemont2010}, a first quantized approach to describe atomic, molecular and nuclear scattering properties, as well as the related exterior complex scaling method \cite{Simon1979,Reed1982,Scrinzi2010} in general resonance theory. In relation to shifting environment degrees of freedom of an open quantum system to obtain Markovian master equations we note a recent and very general result \cite{Tamascelli2018}, generalizing the pseudo-mode approach \cite{Garraway1997a,Garraway1997b,Dalton2001,Mazzola2009} for the spin-boson model.
	
	\subsubsection{Few-mode equations of motion}\label{sec::app_fewEOM}
	From the effective few-mode Hamiltonian Eq.~(\ref{eq::FMapprox}), one can derive Heisenberg-Langevin equations of motion, analogously to what has been done in Sec.~\ref{sec::FMscatt} for the free system. The equations of motion for the atomic operators read
	\begin{subequations}\label{eq::effFM}
	\begin{align}
		\dot{\hat{\sigma}}^+(t) &= i\omega^{\mathstrut}_\textrm{a} \hat{\sigma}^+(t) - i \hat{\sigma}^z(t)\sum_{\lambda} \hat{a}^\dagger_\lambda(t) g^*_\lambda, \label{equ::eff_atomEOM1}
		\\
		\dot{\hat{\sigma}}^-(t) &= -i\omega^{\mathstrut}_\textrm{a} \hat{\sigma}^-(t) + i \hat{\sigma}^z(t)\sum_{\lambda} \hat{a}^{\mathstrut}_\lambda(t) g^{\mathstrut}_\lambda,
		\\
		\dot{\hat{\sigma}}^z(t) &= -2i \hat{\sigma}^+(t) \sum_{\lambda} \hat{a}^{\mathstrut}_\lambda(t) g^{\mathstrut}_\lambda + 2i \hat{\sigma}^-(t) \sum_{\lambda} \hat{a}^{\dagger}_\lambda(t) g^*_\lambda\,.\label{equ::eff_atomEOM3}
	\end{align}
	\end{subequations}
	We note that we have not considered additional loss channels here, such as absorption or other electronic processes in the atom. We further note that in dimensions higher than one, it may be advantageous to trace out some of the bath modes and include them as a direct decay term in the Langevin equations. This can be useful in describing, for example, radiative losses to the side of a Fabry-Perot cavity.
	
	The input-output relation only depends on the system-bath Hamiltonian and hence stays unmodified by the coupling to the atom. Again performing the rotating wave approximation also in the system-bath coupling, we obtain
	\begin{equation}
	\hat{b}^{\mathrm{(out)}}_m(\omega) - \hat{b}^{\mathrm{(in)}}_m(\omega) = -i\sum_\lambda \mathcal{W}^*_{\lambda m}(\omega) \hat{a}^{\mathstrut}_\lambda(\omega).\label{equ::eff_ioEOM}
	\end{equation}
	For the cavity operators, the equations of motion are most easily written in Fourier space analogously to Eq.~\eqref{equ::IO_modeSol} as
	\begin{align}
	\hat{a}_\lambda(\omega) = \sum_{\lambda'} \mathcal{D}^{-1}_{\lambda \lambda'}(\omega) \big[&2\pi \sum_m \mathcal{W}^{\mathstrut}_{\lambda'm}(\omega) \hat{b}^{\mathrm{(in)}}_m(\omega) \nonumber
	\\
	&+ g^*_{\lambda'} \hat{\sigma}^-(\omega)\big]. \label{equ::eff_cavityEOM}
	\end{align}
	We see that by use of the input-output formalism and Heisenberg-Langevin equations, the bath dynamics are completely described by the input-output relation and the driving term in the cavity equation of motion. Therefore the coupled atom-continuum system has been transformed into a driven dissipative few-mode system.

	\subsection{Ab initio few-mode theory for interacting systems in the linear regime}\label{sec::eff_advantages}
	
	In the following, we demonstrate some specific advantages of ab initio few-mode theory mentioned above using a variety of practically relevant examples. In particular, we study the systematic few-mode expansion scheme for problems involving interactions that is offered by ab initio few-mode theory. To this end, we focus on the linear limit of the interacting system, which allows us to systematically investigate various features of the expansion scheme. The non-linear regime will be discussed in Sec.~\ref{sec::nonlinear}.
	
	\subsubsection{Scattering in the linear regime}\label{sec::app_fewEOM_linScatt}
	It is well known that for linear systems, the input-output relations can be solved analytically without further approximations \cite{Walls2008}. However, in obtaining the input-output relation, a Markov approximation \cite{Walls2008} or an approximate extension of frequency integrals \cite{Viviescas2003} is usually performed. Non-Markovian input-output theory \cite{Diosi2012,Zhang2013} has been developed on the basis of phenomenological few-mode Hamiltonians. In our approach, neither of these approximations nor the assumption of a model Hamiltonian \cite{Gardiner1985,Gardiner2004} are necessary.
	
	Consequently, the linear regime is an ideal candidate to demonstrate the advantages of ab initio few-mode theory.
	
	The example of a two-level atom considered above is non-linear in general, but becomes linear in the weak excitation limit, where $\hat{\sigma}^z(t) \approx -1$, which physically corresponds to a weak field driving the atomic ground state, and is a frequently used approximation in quantum optics \cite{Waks2006,Fan2010}. An alternative way of performing the weak excitation approximation is a Holstein-Primakoff transformation \cite{DeLiberato2014,DeBernardis2018}.
	
	\begin{figure*}
		\includegraphics[scale=1.0]{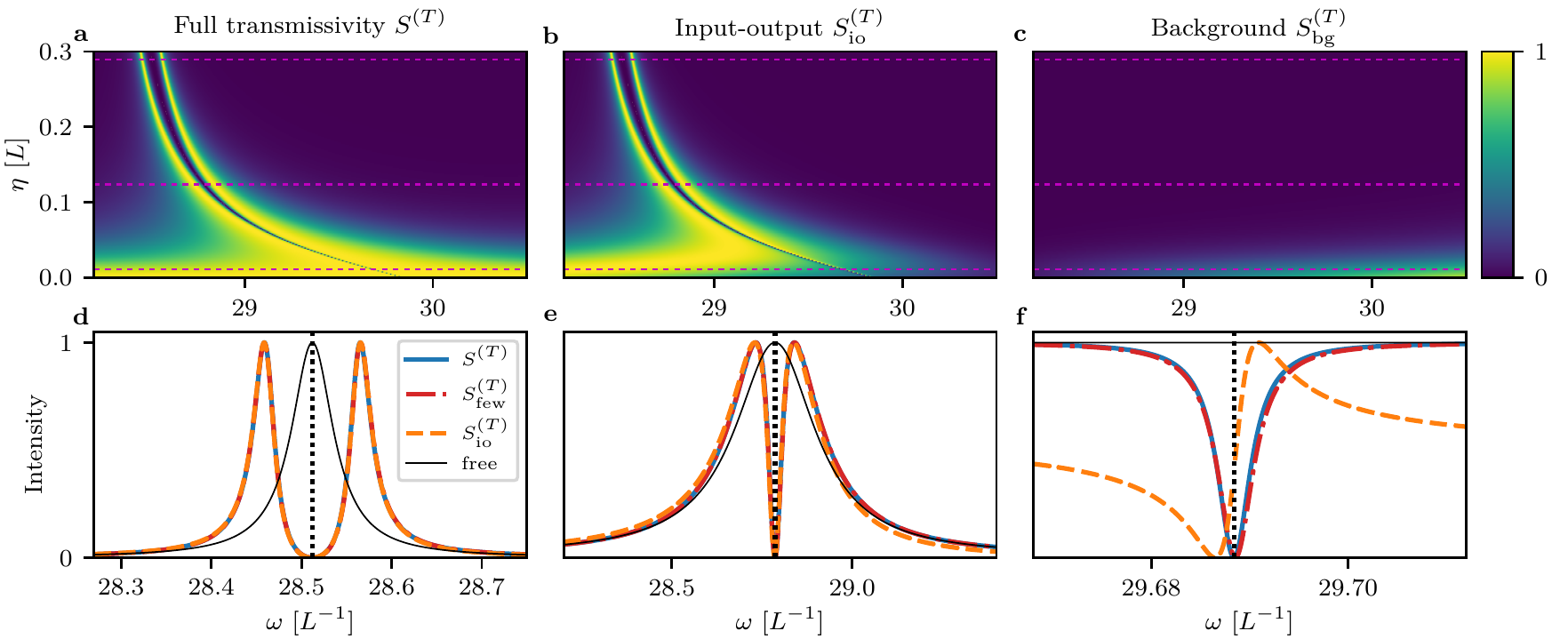}
		\caption{Linear transmission spectra of a coupled atom-cavity system. The cavity is chosen as in Fig.~\ref{fig::maxwell_doubleDelta2D}, and the transition frequency of the atom with $d=0.01$ has been chosen resonant with the 9th empty cavity mode at each mirror quality $\eta$. 
		(a) shows the full transmissivity calculated using linear dispersion theory as a reference. (b) shows the  input-output part without the background contribution which is shown in (c), each calculated with three system modes ($\Lambda_Q = \{\chi_7, \chi_9, \chi_{11}\}$). 
		Panels (d-f) show slices at $\eta=0.289,~0.124,~0.011$, respectively. They correspond to a transition from strong coupling (d) via weak coupling (e) to a regime with negligible cavity confinement (f).
		In the entire range, the ab initio few-mode result ($S_\textrm{few}=S_\textrm{bg}S_\textrm{io}$, red dash-dotted line) agrees well with the full result from linear dispersion theory ($S$, blue solid line), with good convergence already found using a single mode $\chi_9$ for (d) and (e), and using three modes $\Lambda_Q$ for (f). 
		In the weak and strong coupling regimes, the input-output term alone is sufficient to model the interacting system (d,e). But in the regime of strongly overlapping modes and weak confinement, background scattering plays a crucial role (c), such that input-output scattering alone gives a vastly different line shape from the full result (f).\label{fig::purcell_etaScan}} 
	\end{figure*}

	In this limit, the above equations can be solved straightforwardly to give, switching from index to vector-matrix notation,
	\begin{equation}
	\underline{\hat{b}}^{\mathrm{(out)}}(\omega) = \doubleunderline{S}_\mathrm{io}(\omega) \underline{\hat{b}}^{\mathrm{(in)}}(\omega),
	\end{equation}
	with the operator scattering matrix
	\begin{align}
	\doubleunderline{S}_{\mathrm{io}}(\omega) &= \doubleunderline{\mathbb{I}} - 2\pi i \doubleunderline{\mathcal{W}}^\dagger(\omega) \big[\doubleunderline{\mathcal{D}}(\omega) - \frac{1}{\omega-\omega_\textrm{a}}\underline{g}^* \underline{g}^T\big]^{-1} \doubleunderline{\mathcal{W}}(\omega)\nonumber
	\\
	=& \doubleunderline{S}^{\mathrm{(free)}}_{\mathrm{io}}(\omega) - 2\pi i \frac{\doubleunderline{\mathcal{W}}^\dagger(\omega) \doubleunderline{\mathcal{D}}^{-1}(\omega) \underline{g}^* \underline{g}^T \doubleunderline{\mathcal{D}}^{-1}(\omega) \doubleunderline{\mathcal{W}}(\omega)}{\omega-\omega_\textrm{a} - \underline{g}^T \doubleunderline{\mathcal{D}}^{-1}(\omega) \underline{g}^*}. \label{equ::int_linearScatteringIO}
	\end{align}
	The second formula is particularly useful since one can read off the complex level shift $\underline{g}^T \doubleunderline{\mathcal{D}}^{-1}(\omega) \underline{g}^*$ and thus extract the Purcell enhanced line width of the atom
	\begin{equation}
		\gamma_\textrm{S} = -\mathrm{Im}[\underline{g}^T \doubleunderline{\mathcal{D}}^{-1}(\omega) \underline{g}^*],
	\end{equation}
	as well as its cavity modified Lamb shift
	\begin{equation}
		\delta_\textrm{LS} = \mathrm{Re}[\underline{g}^T \doubleunderline{\mathcal{D}}^{-1}(\omega) \underline{g}^*].
	\end{equation}
	These two quantities can thus be directly computed from the cavity geometry using ab initio few-mode theory.
	
	We also see that the effective few-mode theory gives an expansion of the scattering matrix as a sum over the quantum optical coupling constants. As expected, the input-output scattering matrix reduces to the free case Eq.~\eqref{equ::max_scattIO} in the limit $\underline{g}\to0$, where the expression is exact up to the rotating wave approximation in the system-bath coupling. For the interacting case, one can systematically include more cavity modes in the projector basis and observe the series' convergence in the many modes limit, where the few-mode basis, if chosen correctly, approaches a complete set in the region of the atom (see Fig.~\ref{fig::schematic_modeSelection}). The expansion is non-perturbative in the sense that it is not limited to weak atom-mode coupling $\underline{g}$, however due to the rotating wave approximation the above expression does not apply in the ultra-strong coupling regime. Inclusion of the counter-rotating terms would, however, essentially result in additional linearly coupled equations (see also Sec.~\ref{sec::Maxwell_beyondRot}), which can be solved analogously in the linear regime, as has been shown in detail in \cite{Ciuti2006}.
	
	To obtain the full scattering matrix between the observable asymptotically free operators, we have to account for the background scattering contribution again. Since it is only responsible for translating the bath operators into asymptotically free operators, the background scattering is independent of the matter coupling and can be computed as in the free theory.

	\subsubsection{Transition from strong coupling to free space}
	In Fig.~\ref{fig::purcell_etaScan}, we show linear transmission spectra for the Fabry-Perot cavity that has also been investigated in Fig.~\ref{fig::maxwell_doubleDelta2D}, but now containing a single atom at its center, with a dipole moment of $d=0.01$. The linear regime of this interacting system is ideal to demonstrate the advantages of ab initio few-mode theory, since the resulting equations of motion (\ref{eq::effFM}) can be solved without a Markov or semi-classical approximation, as shown in Sec.~\ref{sec::app_fewEOM_linScatt}. Thus the effect of frequency dependent system parameters can be investigated. Additionally, linear dispersion theory \cite{Born1980,Zhu1990} (see Appendix \ref{sec::app_LinDisp}) can be used as a benchmark for comparison in the linear regime. We note that the results in Fig.~\ref{fig::purcell_etaScan} have been obtained using the constructive approach to choosing a system basis (see Section \ref{sec::FMchoice}), without assuming any prior knowledge about the system.
	
	The transmission spectra as a function of the mirror quality $\eta$ show a transition from the strong coupling regime at high $\eta$, via the usual weak coupling regime at intermediate $\eta$, to a regime where the resonances overlap significantly until the situation approaches a weakly confined regime at low $\eta$, essentially corresponding to free space. The lower panels show slices of the two dimensional spectra in each of these regimes.
	To explore the potential of the ab initio few mode approach, we compare linear dispersion theory as a reference ($S$), the results obtained neglecting the background contribution ($S_\textrm{io}$), as well as the full ab initio few-mode result including the background contribution ($S_\textrm{few}=S_\textrm{bg}S_\textrm{io}$).

	We see that in the strong and weak coupling regime, all three approaches agree very well. In both cases, we found  a single mode to be sufficient for good agreement. This is also illustrated in more detail in panels (d) and (e). However in the overlapping modes regime and at weak confinement, the situation is quite different. Panel (f) clearly shows that excluding the background contribution leads to qualitatively wrong predictions. For example, while $S_{\rm{io}}$ without the background contribution predicts an asymmetric Fano-like line shape, the full result including the background contribution remains Lorentzian. Consequently, phenomenological input-output theory fails in this regime, since the background and resonant scattering contributions are not distinguished in these models. Thus, the novel aspects of ab initio few-mode theory come into play and it is crucial that the empty cavity is treated exactly due to the strong mode overlap and absence of isolated resonances. 
		
	We further note that the ab initio few-mode approach has advantages already in the usual strong and weak coupling regime, even though phenomenological input-output theory is sufficient for a quantitative treatment there. Firstly, a rigorous foundation of the method is given. This also has the practical consequence that the quantum optical coupling constants can now be calculated from the cavity geometry, instead of being obtained by a fitting procedure. The latter may lead to theoretical design opportunities for quantum optical properties of complex structures. Secondly, it can already be seen that the toolbox of interacting few-mode and input-output theory can be applied straightforwardly without additional complexity. After the ab initio Hamiltonian of the non-interacting cavity is obtained, the calculation follows standard methods with minor adjustments (see Sec.~\ref{sec::app_fewEOM}).

	\subsubsection{Convergence of the few-mode expansion }\label{sec::int_convergence}
	\begin{figure}[!htbp]
		\includegraphics[width=\columnwidth]{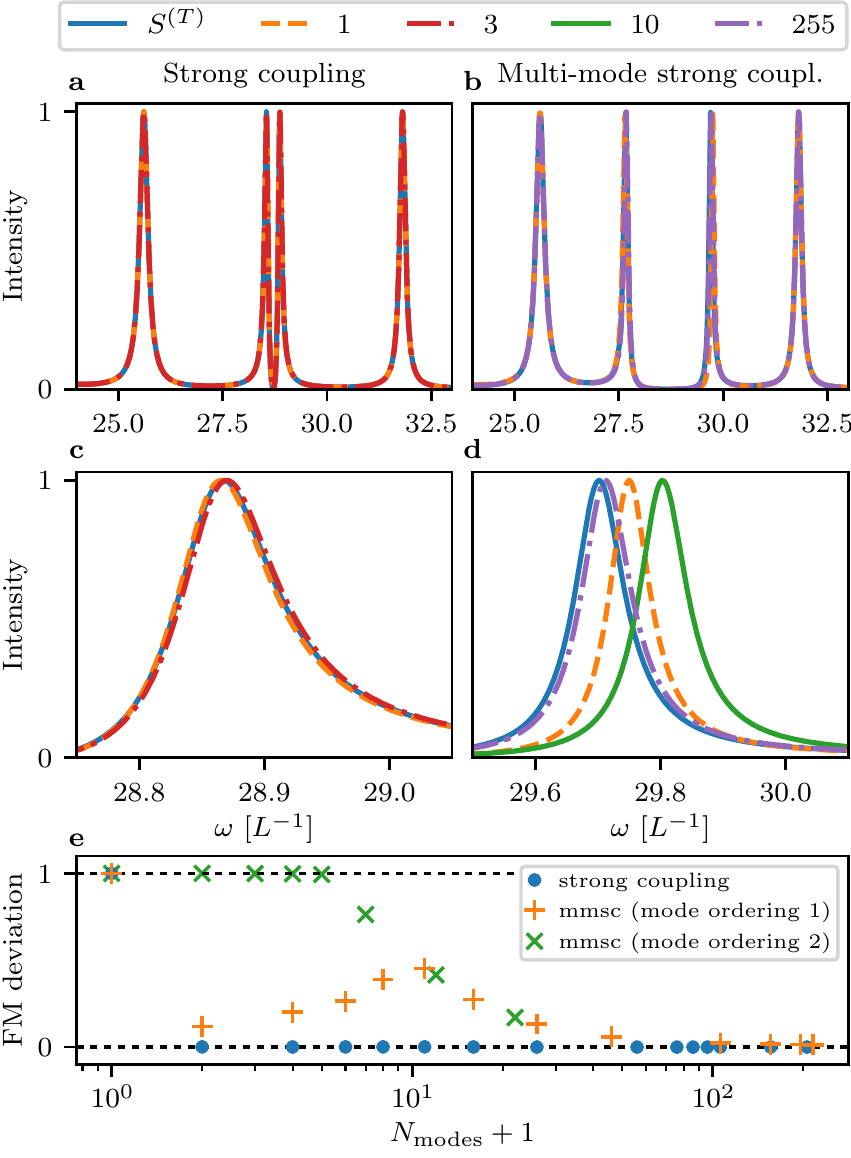}
		\caption{
			Convergence behavior of the few-mode expansion in the strong coupling (a,c) and multi-mode strong coupling (b,d) regime. Panels (a-d) show few-mode spectra at different numbers of system modes ($N_{\rm{modes}} \in \{1, 3, 10, 255\}$, see legend). The solid blue line shows linear dispersion spectra as a benchmark ($S^{(T)}$, solid blue). In the strong coupling case, the convergence is fast, with a single mode being sufficient (c), as expected. At multi-mode strong coupling, the qualitative behavior is already captured in a single mode description, but for quantitative agreement more than 100 modes of the generic mode basis are necessary (d).
			The convergence in the two cases can be quantified by the few-mode (FM) deviation Eq.~\eqref{eq::fm-deviation}. Panel (e) shows this quantity as a function of the mode number ($N_{\rm{modes}}+1$). The strong coupling case (blue dots) shows the expected behavior. In the multi-mode strong coupling case (mmsc, orange crosses, green hatches), the convergence is much slower, and for low mode numbers, the convergence depends on the order in which the modes are added to the few-mode system space, as explained in the main text.
		}\label{fig::maxwell_convergence_new}
	\end{figure}	
	
	As with any series expansion, an important requirement for the effective few-mode expansion is that it should converge as the number of system modes increases. Demonstrating the convergence of the few-mode expansion is particularly important as multi-mode light-matter coupling models are notorious for their divergent behavior (see, for example, \cite{Krimer2014,Gely2017,SanchezMunoz2018,Malekakhlagh2017}).
	
	In Fig.~\ref{fig::maxwell_convergence_new}, we numerically investigate the dependence of few-mode scattering observables on the number of system modes in the strong coupling and the multi-mode strong coupling regime. We again use the cavity geometry from Fig.~\ref{fig::purcell_etaScan}, with $\eta=0.15$ and $d=0.03$ (strong coupling) or $d=0.2$ (multi-mode strong coupling), and $\omega_\textrm{a}=28.71$ resonant with the ninth cavity mode $\chi_9$. The atom is placed at the center of the cavity, such that only odd Dirichlet modes contribute to the interaction.
	
	The few-mode basis is chosen as described in Sec.~\ref{sec::FMexpansion} by solving the Dirichlet boundary value problem. The single mode model contains the dominant ninth mode. We then label each few-mode basis in terms of a mode number $N_\textrm{modes}$ as follows. We first add the odd modes in steps of two in decreasing order of dominance. A mode number of $N_\textrm{modes}=3$ then corresponds to $\Lambda_Q \in \{\chi_7, \chi_9, \chi_{11}\}$,  $N_\textrm{modes}=5$ to $\Lambda_Q \in \{\chi_5, \chi_7, \chi_9, \chi_{11}, \chi_{13}\}$ and so on. Since there are no lower lying modes than $\chi_1$, once it is included, we add the remaining higher lying odd modes in steps of one, such that for example $N_\textrm{modes}=10$ corresponds to $\Lambda_Q \in \{\chi_1, \chi_3,\dots \chi_{15}, \chi_{17}\}$.
	
	Panels (a,c) in Fig.~\ref{fig::maxwell_convergence_new} demonstrate that a single mode is already sufficient in the strong coupling regime of this isolated resonance cavity, as is expected from phenomenological few-mode theory. Panels (b,d) show that in the multi-mode strong coupling regime, the convergence is much slower for the chosen generic mode bases. While a single mode can already reproduce the qualitative features of the spectrum, in order to reach quantitative agreement a relatively large number of modes is required. It is further seen that when including more modes symmetrically around the dominant one (see labeling order described above), the spectral peak first shifts away from its final position before it starts to converge, a feature we will explain below.
	
	Panel (e) quantifies the deviation of ab initio few-mode theory from the benchmark provided by linear dispersion theory as a function of the mode number $N_{\rm{modes}}$. Note that the $x$-axis shows $N_{\rm{modes}}+1$ to allow for a logarithmic representation. The few-mode deviation is defined by 
	\begin{align}\label{eq::fm-deviation}
	  \Delta_{\textrm{few}}=\frac{\sum_\omega |S^{(T)}_\textrm{few}(\omega)-S^{(T)}(\omega)|^2}{\sum_\omega |S^{(T)}_0(\omega)-S^{(T)}(\omega)|^2}\,,
	\end{align}
	where the frequency axis is evaluated on a grid and the superscript $(T)$ indicates the transmission element of the scattering matrix. $S^{(T)}_0(\omega)$ is the zero modes few-mode theory coinciding with the transmission spectrum of the empty cavity. This quantity represents a phase-sensitive $\chi^2$-like deviation metric that is normalized to the zero system modes case. The fast convergence in the strong coupling regime is evident from the sharp deviation drop from zero to a single mode. In the multi-mode strong coupling case, a much slower decline is observed in the generic mode basis, reaching reasonable convergence only at more than 100 modes with the few-mode deviation still decreasing. 
	
	\begin{figure}
		\includegraphics[width=\columnwidth]{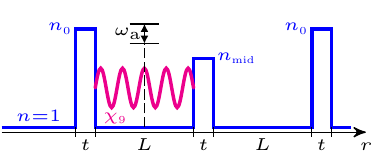}
		\caption{One-dimensional double cavity potential. The Q-space basis modes are chosen as solutions of the Dirichlet problem in the left cavity (the figure shows mode $\chi_9$ in magenta as an example). This choice of basis appears particularly suitable for describing local light-matter interactions in the left cavity, but not in the right cavity. An atom is placed at the center of the left cavity, chosen resonant with the ninth cavity mode.}\label{fig::maxwell_doubleFabryPerot}
	\end{figure}	
	\begin{figure}
		\includegraphics[scale=1.0]{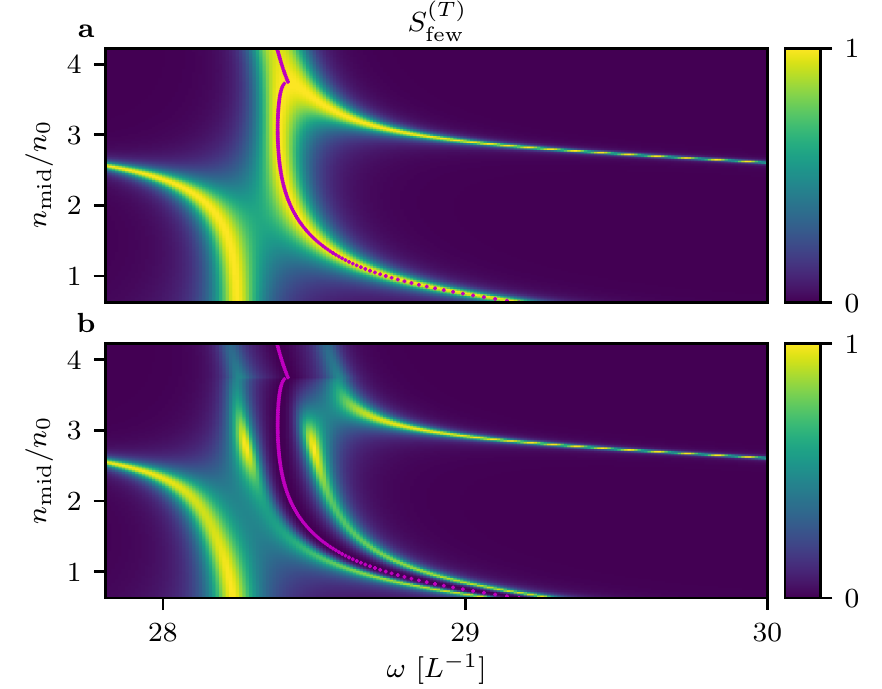}
		\caption{Linear transmission spectra for the double Fabry-Perot cavity in Fig.~\ref{fig::maxwell_doubleFabryPerot} as a function of the central mirror refractive index, calculated using ab initio few-mode theory with a single system mode ($\chi_9$). (a) shows the empty cavity without atom. A transition between different regimes featuring an avoided crossing is observed. In (b), results of the interacting system with an atom at the center of the first cavity are shown. The atom's resonance frequency is chosen resonant with the empty cavity spectral peak, indicated by the magenta dots. A splitting of the resonant cavity mode is observed, with interesting features in the transition regions. The shown few-mode results are found to agree well with linear dispersion theory in both cases.}\label{fig::Ovlap_Spectra}
	\end{figure}
	
	Interestingly, a dip in $\Delta_{\textrm{few}}$ is found at a single mode, which is related to the above observation that the spectral peak first shifts away from its final position with increasing mode number, before it converges in reverse direction to the correct result. 
	This effect can be understood from the fact that the modes $\chi_\lambda$ with $\lambda<9$ cause a positive spectral shift of the peak, whereas the modes with higher mode number induce an opposite frequency shift. 
	To verify this interpretation, we also study the convergence using an alternative ordering where the modes are included starting with $\chi_1$ and simply counting up. For example, $N_\textrm{modes}=2$ corresponds to $\Lambda_Q \in \{\chi_1, \chi_3\}$, $N_\textrm{modes}=3$ corresponds to $\Lambda_Q \in \{\chi_1, \chi_3, \chi_{5}\}$ and so on. With this labeling order, a monotonic convergence is found (green crosses), because now the competition between opposite shifts of red-detuned and blue-detuned modes is avoided. These results demonstrate that in particular in the multi-mode strong coupling regime, the convergence properties are affected by the choice of the system basis.

	Numerically, it is impossible to show whether the expansion indeed converges in a mathematical sense. In particular for the multi-mode strong coupling results, one may object that Fig.~\ref{fig::maxwell_convergence_new} does not exclude the possibility that the deviation oscillates very slowly or even diverges in the limit of infinitely many modes. In order to provide justification that this is not the case, we explicitly evaluate the relevant series terms of the few-mode expansion in Appendix \ref{sec::app_AnalytConvergence} and show that the final result is convergent as for the free cavity. The calculation is restricted to a specific geometry (see Appendix \ref{sec::app_AnalytConvergence} for details), but applies at any coupling strength within the validity range of the rotating wave approximation. The series expansion is therefore non-perturbative in the atom-cavity coupling strength, in the sense that it does not require a small parameter.
	
	We note that multi-mode convergence of light-matter models has been discussed extensively in the literature, for example in the context of the Rabi and related models at ultra-strong coupling \cite{DeLiberato2014,Gely2017,SanchezMunoz2018,Malekakhlagh2017} as well as in the context of time-dependent problems extending the Wigner-Weisskopf theory of spontaneous emission \cite{Krimer2014,Malekakhlagh2016}. Divergences can be handled by cutoffs (see, for example, \cite{Krimer2014} for an explicit account), cutoff-free methods have only been developed more recently by fully accounting for gauge invariance \cite{Malekakhlagh2017}. We note that these discussions are mainly concerned with the complete treatment of the light-matter interaction, which is particularly important at ultra-strong coupling. Ab initio few-mode theory is based on a complete treatment of the cavity-bath interaction in few-mode theory. In our present analysis, we avoided issues arising at ultra-strong coupling by applying the rotating wave approximation. It will therefore be interesting to see if ab initio few-mode theory can be married with ultra-strong coupling theory, where phenomenological few-mode models are also a valuable tool (see, for example, \cite{FriskKockum2019,Forn-Diaz2019,Kirton2018}). An outlook in this direction is given in Sec.~\ref{sec::outlook}. 

	\subsubsection{Quantum optical properties in overlapping modes cavities}

	\begin{figure*}
		\includegraphics[scale=1.0]{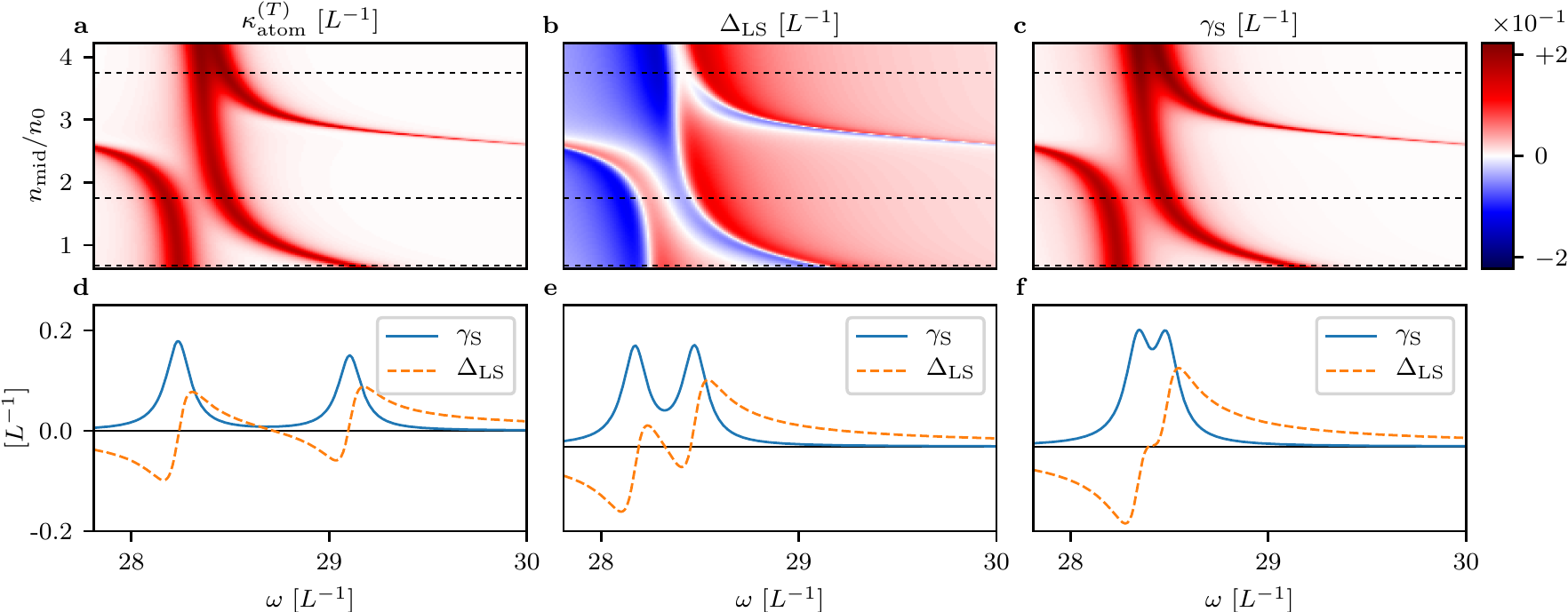}
		\caption{Quantum optical quantities corresponding to the atom-cavity spectra in Fig.~\ref{fig::Ovlap_Spectra}. The Purcell enhanced line width $\gamma_\textrm{S}$ of the atom, its cavity modified Lamb shift $\Delta_\textrm{LS}$ as well as $\kappa_\textrm{atom}^{(T)}=2\pi|[\doubleunderline{\mathcal{W}}^\dagger(\omega) \doubleunderline{\mathcal{D}}^{-1}(\omega) \underline{g}^* \underline{g}^T \doubleunderline{\mathcal{D}}^{-1}(\omega) \doubleunderline{\mathcal{W}}(\omega)]_{1,0}|$ are shown. These quantities are the atomic line analogues of the empty cavity couplings shown in Fig.~\ref{fig::maxwell_doubleDelta2D_couplings} and have been computed using ab initio few-mode theory with a single system mode ($\chi_9$). The lower panels (d-f) show slices at $n_\textrm{mid}=2.7,~7.0,~15.0$, respectively, demonstrating the transition between different regimes and that even the single mode theory is able to include effects beyond the isolated resonance approximation. \label{fig::Ovlap_Const}} 
	\end{figure*}

	One remaining advantage of the ab initio method that we have not demonstrated yet is the effect of frequency-dependent quantum optical couplings in interacting few-mode theory. To this end, we consider the double Fabry-Perot cavity depicted in Fig.~\ref{fig::maxwell_doubleFabryPerot}, with a varying refractive index $n_\textrm{mid}$ of the central mirror. Unlike before, we choose a finite mirror thickness $t=0.01$ and outer mirror refractive index $n_0=4.0$ in order to demonstrate that our approach is not limited to Ley-Loudon type \cite{Ley1987} potentials with its unrealistic infinitely thin mirrors.
	
	Such a double cavity geometry is ideal to investigate the transition from an isolated resonance to an overlapping modes regime that still features strong confinement. In this case, the overlap results from the near-degeneracy of modes in each respective cavity, which couple via leakage through the central mirror.
	Fig.~\ref{fig::Ovlap_Spectra} shows linear transmission spectra for this cavity as a function of $n_\textrm{mid}$, calculated using ab initio few-mode theory with a single system mode $\chi_9$. The top panel (a) shows the empty cavity spectrum which already displays interesting overlapping modes features, including an avoided crossing at intermediate $n_\textrm{mid}$ as well as a merging of spectral lines at high $n_\textrm{mid}$.
	
	The spectral structure can be understood as follows. At $n_\textrm{mid} \ll n_0$, the separating barrier is insignificant such that the cavity behaves like the single Fabry-Perot case from before with approximately twice the cavity length, showing isolated resonances.
	
	As the central barrier increases, a transition from a single to a double cavity structure occurs. In the latter, a more useful intuitive picture is to think of the system modes in each cavity. Due to their interaction via leakage through the central barrier, the avoided crossing occurs. In this context it is interesting to point out the connection of ab initio few-mode theory to non-Hermitian Hamiltonian formulations \cite{Rotter2009,Moiseyev2011,El-Ganainy2018} that are often used to interpret such scenarios and have also been considered in the context of potential scattering \cite{Savin2003,Shamshutdinova2008}.
	
	At higher $n_{\rm{mid}}$, the central barrier becomes very reflective such that after the avoided crossing two lines merge together. We have thus observed the transition from a single cavity behavior via a double cavity structure with an avoided crossing to a third regime with a strong separating barrier.
	
	The lower spectra include the effect of an atom at the center of the first cavity with $d=0.03$ and $\omega_\textrm{a}$ chosen resonant with the cavity peak (magenta dots). We see that when adding the atomic interaction, vacuum Rabi-splitting of the spectral peak is observed, featuring modulations of peak intensity in the regimes of overlapping modes. One of the peaks even disappears completely at high $n_\textrm{mid}$. An interesting aspect is that panel (b) has been computed using only a single ``single-cavity'' mode ($\chi_9$ in the first cavity, see Fig.~\ref{fig::maxwell_doubleFabryPerot}). Nevertheless, this ab initio single mode theory correctly predicts the interacting spectrum across the whole shown range of $n_\textrm{mid}$, with multi-mode deviations from the linear dispersion theory result of less than a few percent at $n_\textrm{mid}/n_0<1$ and less than a few permille at $n_\textrm{mid}/n_0>1$. In this case, the single mode theory thus provides a good description of the interacting system even in a seemingly multi-mode regime, when the spectral peaks of the empty cavity overlap significantly. In comparison, phenomenological few-mode theory could only reproduce these results by including at least two modes, e.g. $\chi_9$ in each cavity.
	
	To investigate the reason behind this unexpected quality of the ab initio single-mode results in more detail, Fig.~\ref{fig::Ovlap_Const} shows the atom's cavity modified Lamb shift $\delta_\textrm{LS}$ and Purcell enhanced decay width $\gamma_\textrm{S}$ \cite{Purcell1946,Heeg2013b,Kristensen2014} (panels (b) and (c), respectively), along with slices at $n_\textrm{mid}=2.7,~7.0,~15.0
	$ (panels (d-f), respectively). These quantities are directly computed from the cavity geometry (see Appendix \ref{sec::app_fewEOM_linScatt} for details) and we observe a varying frequency dependence. 
	At low $n_\textrm{mid}=2.7$ (panel (d)), we observe two isolated resonance features, each having a frequency dependence as expected from the Lorentzian single-mode contributions in phenomenological few-mode models~\cite{Walls2008}. 
	However, beyond that, the modes overlap significantly at $n_\textrm{mid}=7.0$ (panel (e)) and $n_\textrm{mid}=15.0$ (panel (f)). Still, our ab initio approach is able to account for these non-trivial bath effects in the overlapping modes case as shown in panels (e,f). In the latter cases, the standard result from single mode phenomenological few-mode theory breaks down and the advantage of the frequency dependent couplings in ab initio few-mode theory can be seen. 
	For this reason, we conclude that ab initio few-mode theory can extend the validity range of a single mode description to new regimes by incorporating non-trivial bath effects beyond the isolated resonance approximation into the frequency dependent couplings.

	\subsection{Non-linear phenomena}\label{sec::nonlinear}
	In the previous sections, we have demonstrated that ab initio few-mode theory establishes a powerful expansion scheme for problems involving interactions. But so far, we have only considered the linear limit of the interacting system, which has allowed us to systematically investigate various features of the expansion scheme. In the following, we show that the approach can also be applied in the non-linear regime.
	
	As in the linear case, we will again exploit that ab initio few-mode theory gives rise to Hamiltonians of a form similar to those used in phenomenological few-mode and system-bath theory \cite{Gardiner2004,Carmichael2008}. This central feature allows us to make use of many existing methods to tackle non-linear open-system dynamics, and to promote these methods to new regimes, by basing them on an ab initio few-mode Hamiltonian instead of on a phenomenological model.

	As a concrete example, we extend the previous model of a two-level atom in a cavity to stronger external driving fields. We employ the semi-classical monochromatic drive approximation, which is a textbook example that enables an explicit analytical computation of scattering observables, and which is of significance, for example, in spectroscopy~\cite{Walls2008,Carmichael2008}. 
	This model serves to demonstrate the applicability of ab initio few-mode theory in a scattering regime where the precise frequency dependence of the coupling constants matters, and that the resulting few-mode equations can indeed be solved using appropriate methods. In the process, we obtain analytic solutions for non-linear spectra of a two-level atom in an overlapping modes cavity in the form of a few-mode expansion.

	\subsubsection{Few-mode equations of motion with a semi-classical driving field}
	
	In order to show that ab initio few-mode theory can describe non-linear phenomena beyond the isolated resonance case, we employ the semi-classical assumption which is that the operator $\hat{b}_m^{\mathrm{(in)}}(\omega)$ can be treated as a commuting classical variable $b_m^{\mathrm{(in)}}(\omega) = b_m^{\mathrm{(in)}} 2\pi\delta(\omega-\omega_\mathrm{in})$, where $\omega_\mathrm{in}$ and $b_m^{\mathrm{(in)}}$ are the driving frequency and amplitude, respectively. In the time domain, one can alternatively write $b_m^{\mathrm{(in)}}(t) =  b_m^{\mathrm{(in)}}e^{-i\omega_\mathrm{in}t}$. Physically, this scenario corresponds to the steady-state response of the atom-cavity system for a monochromatic laser input when quantum fluctuations are neglected. The approximation has become a standard tool in quantum optics \cite{Cohen-Tannoudji1998b,Walls2008,Scully1997,Carmichael1999}, and can also be interpreted as the calculation of coherent state scattering probabilities \cite{Fan2010}.
	
	With this driving term, the solution of the few-mode equations of motion is given by the cavity operators and atomic operators all oscillating at a constant frequency, $\hat{a}_\lambda(t) =  \hat{a}_\lambda e^{-i\omega_\mathrm{in}t}$, $\hat{\sigma}^-(t) =  \hat{\sigma}^- e^{-i\omega_\mathrm{in}t}$, $\hat{\sigma}^+(t) =  \hat{\sigma}^+ e^{+i\omega_\mathrm{in}t}$ and $\hat{\sigma}^z(t) =  \hat{\sigma}^z$. Substituting into the equations of motion gives
	\begin{subequations}\label{eq::fm-nonlinear-eom}
	\begin{align}
		0 &= i(\omega^{\mathstrut}_a - \omega^{\mathstrut}_\mathrm{in}) \hat{\sigma}^+ - i \hat{\sigma}^z\sum_{\lambda} \hat{a}^\dagger_\lambda g^*_\lambda,
		\\
		0 &= i(\omega^{\mathstrut}_\mathrm{in}-\omega^{\mathstrut}_a) \hat{\sigma}^- + i \hat{\sigma}^z\sum_{\lambda} \hat{a}^{\mathstrut}_\lambda g^{\mathstrut}_\lambda,
		\\
		0 &= -i \hat{\sigma}^+ \sum_{\lambda} \hat{a}^{\mathstrut}_\lambda g^{\mathstrut}_\lambda + i \hat{\sigma}^- \sum_{\lambda} \hat{a}^{\dagger}_\lambda g^*_\lambda,
		\\
		\hat{a}_\lambda &= \sum_{\lambda'} \mathcal{D}^{-1}_{\lambda \lambda'}(\omega_\mathrm{in}) \big[2\pi \sum_m \mathcal{W}^{\mathstrut}_{\lambda'm}(\omega_\mathrm{in}) b^{\mathrm{(in)}}_m \nonumber
		\\
		&\qquad\qquad\qquad\quad + g^*_{\lambda'} \hat{\sigma}^-\big].
	\end{align}
	\end{subequations}
	
	\subsubsection{Steady-state non-linear spectra}
	\begin{figure*}
		\includegraphics[scale=1.0]{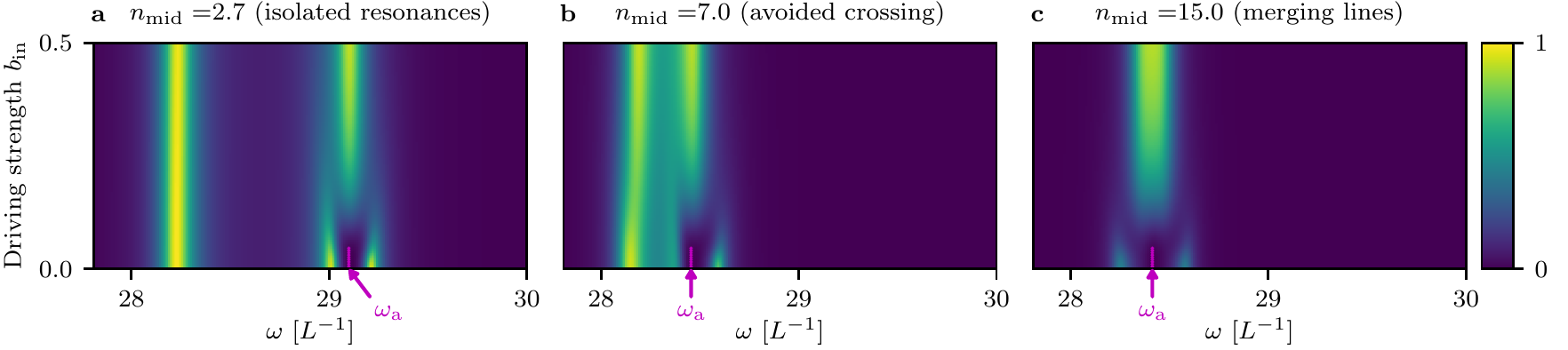}
		\caption{Non-linear spectra as a function of driving strength, corresponding to each of the regimes in Fig.~\ref{fig::Ovlap_Const}. Panel a-c show results for $n_\textrm{mid}=2.7,~7.0,~15.0$, respectively. The cavity is driven from one side, such that $\underline{b}^{\mathrm{(in)}} = b_\textrm{in} (\begin{smallmatrix}1\\0\end{smallmatrix})$, and the corresponding transmission spectra are shown. At $b_\textrm{in}=0$ the spectrum coincides with the linear interaction spectra shown in Fig.~\ref{fig::Ovlap_Spectra}b at each $n_\textrm{mid}$. At large $b_\textrm{in}$ the corresponding empty cavity spectra in Fig.~\ref{fig::Ovlap_Spectra}a are approached due to saturation of the atoms excitation. The atoms resonance frequency for each spectrum has been chosen as before and is indicated by the magenta arrows. \label{fig::Ovlap_NonLin}}
	\end{figure*}
	
	Eliminating the cavity mode operator in Eqs.~(\ref{eq::fm-nonlinear-eom}), we obtain closed equations for the atomic averages in terms of the given semi-classical drive amplitude
	\begin{subequations}
	\begin{align}
		0 &= -i\Delta\langle\hat{\sigma}^+\rangle - i\Omega^*\langle\hat{\sigma}^z\rangle + i\delta^*\langle\hat{\sigma}^+\rangle\,,
		\\
		0 &= i\Delta\langle\hat{\sigma}^-\rangle + i\Omega\langle\hat{\sigma}^z\rangle - i\delta\langle\hat{\sigma}^-\rangle\,,
		\\
		0 &= -i\Omega\langle\hat{\sigma}^+\rangle + i\Omega^*\langle\hat{\sigma}^-\rangle + \mathrm{Im}[\delta] (\langle\hat{\sigma}^z\rangle + 1)\,,
	\end{align}
	\end{subequations}
	with the parameters
	\begin{subequations}
	\begin{align}
	\Delta &= \omega_\mathrm{in} -\omega_\textrm{a}\,,
	\\
	\Omega &= 2\pi \underline{g}^T \doubleunderline{\mathcal{D}}^{-1}(\omega_\mathrm{in}) \doubleunderline{\mathcal{W}}(\omega_\mathrm{in}) \underline{b}_\mathrm{in}\,,
	\\
	\delta &= \underline{g}^T\doubleunderline{\mathcal{D}}^{-1}(\omega_\mathrm{in})\underline{g}^*\,.
	\end{align}
	\end{subequations}
	The solution for the atomic operators is given by
	\begin{subequations}
	\begin{align}
		\langle\hat{\sigma}^- \rangle &= \frac{\Omega}{\Delta - \delta + \frac{2|\Omega|^2}{\Delta-\delta^*}}\,,
		\\
		\langle\hat{\sigma}^z \rangle &= -\frac{\Delta-\delta}{\Delta - \delta + \frac{2|\Omega|^2}{\Delta-\delta^*}}\,,
	\end{align}
	\end{subequations}
	such that the expectation value of the output field $\langle\hat{b}^{\mathrm{(out)}}_m(\omega)\rangle = \langle\hat{b}^{\mathrm{(out)}}_m\rangle \delta(\omega-\omega_\mathrm{in})$ is 
	\begin{align}
	\langle\underline{\hat{b}}^{\mathrm{(out)}}\rangle =  \underline{b}^{\mathrm{(in)}} -i \doubleunderline{\mathcal{W}}^\dagger(\omega_\mathrm{in}) \doubleunderline{\mathcal{D}}^{-1}(\omega_\mathrm{in}) \big[&2\pi \doubleunderline{\mathcal{W}}^{\mathstrut}(\omega_\mathrm{in}) \underline{b}^{\mathrm{(in)}} \nonumber
	\\
	&+ \underline{g}^* \langle\hat{\sigma}^-\rangle\big].
	\end{align}
	
	To illustrate the results that can be obtained with this approach, Fig.~\ref{fig::Ovlap_NonLin} shows non-linear transmission spectra as a function of driving strength for each of the slices that were shown in panels (d-f) of Fig.~\ref{fig::Ovlap_Const}. That is the central barrier height is now held fixed at $n_\textrm{mid}=2.7,~7.0,~15.0$ corresponding to panels (a-c), respectively, such that non-linear effects in each of the three regimes investigated before can be seen. The atoms resonance frequency is also chosen as before and indicated in the figure for clarity.
	
	At weak driving, that is when $b_\textrm{in}=0$, the spectra coincide with the linear interaction spectra shown in Fig.~\ref{fig::Ovlap_Spectra}b for each of the depicted $n_\textrm{mid}$. At large $b_\textrm{in}$ on the other hand the corresponding empty cavity spectra in Fig.~\ref{fig::Ovlap_Spectra}a are approached, since the atom saturates at high driving strength in steady state. The transition region between these two extremes features rich behavior in the three regimes, all of which are now captured by ab initio few-mode theory with a single system mode.

	We have thus demonstrated that ab initio few-mode theory interfaces with an existing method and in particular, that the aspects of the theory which are usually neglected in its phenomenological counterpart, such as frequency dependent couplings and background scattering, can be incorporated fully in the semi-classical monochromatic drive approximation.

	\subsubsection{Other few-mode solution methods}
	In general, to what extent the novel aspects of ab initio few-mode theory can be incorporated into other existing methods will depend on the specific regime and appropriate approximations. A common example is the derivation of a Markovian Master equation for the cavity-atom part of the system by tracing out the bath modes. This method can easily be combined with ab initio few-mode theory in the regime of isolated cavity resonances, where phenomenological few-mode models are expected to apply as well (see also Section \ref{sec::maxwell_FabryPerot}). However, in regimes where frequency dependent cavity-bath couplings can not be eliminated by an appropriate choice of system states, the Markov approximation breaks down \cite{Gardiner2004}. On the other hand, non-Markovian Master equations and alternative methods to solve such systems have also been studied extensively in the literature \cite{Carmichael1999,Breuer2002_BOOK,Gardiner2004,DeVega2017}, and even non-Markovian input-output theory has been developed \cite{Diosi2012,Zhang2013}. This case demonstrates that while ab initio few-mode theory provides a new foundation for existing few-mode methods and allows the underlying Hamiltonian to be applied in extreme regimes, the precise application of each method for computing observables on this basis is regime dependent and should be revisited separately.

	\section{Outlook and generalizations}\label{sec::outlook}
	We have presented a number of advantages that ab initio few-mode theory provides, as well as interesting regimes that it already applies to in its current form. The concept of ab initio few-mode theory, namely extracting dominant degrees of freedom from a continuum, is very general. We therefore envision a number of additional possible applications, some of which will require extensions of the method.

	A natural application of ab initio few-mode theory would be quantum plasmonics \cite{Savage2012,Tame2013,Stockman2018}, which is a promising experimental platform due to its unique and extreme physical properties \cite{Alpeggiani2012,Delga2014,Chikkaraddy2016}. On the theoretical side, there are still various challenges \cite{Esteban2012,Fernandez-Dominguez2018,Stockman2018}. One of them is the high openness of these systems, which has cast doubts on the applicability of input-output models for quantum plasmonics \cite{Barnett1988a,Dutra2000,Khanbekyan2005,Franke2019}. Consequently, much effort has been invested into developing alternative quantum mechanical descriptions \cite{Esteban2012}, for example by quantizing quasi-modes, which have had much success in the semi-classical domain \cite{Ching1998,Kristensen2014,Lalanne2018}. While much progress has been made \cite{Hughes2018,Franke2019}, no direct alternative to the input-output formalism or another solution to the quantum scattering problem has been found yet. This is the exact feature that our method provides, as we showed in detail in this paper. Ab initio few-mode theory would thus allow the input-output formalism and its associated toolbox to be utilized in quantum plasmonics even at high leakage. On the other hand, plasmonic cavities usually also feature significant material absorption and losses \cite{Tame2013}, or even quantum effects of the resonator material \cite{Esteban2012,Savage2012}, which we have not accounted for in this paper. In order to include such effects in the formalism, ab initio few-mode theory could be applied to more complete quantization schemes such as macroscopic QED \cite{Knoell1987,Scheel2008} or microscopic Hamiltonians \cite{Huttner1992a}, which may require a generalized method to choose the relevant system modes appropriately. Alternatively, the phenomenological inclusion of absorptive baths, an approach frequently used for random media \cite{Viviescas2003,Beenakker1998}, while treating scattering and leakage via ab initio few-mode theory may be sufficient in certain scenarios \cite{Waks2010,Franke2019}, which is a straightforward application of our formalism.

	Another regime that has recently received much attention is extreme light-matter coupling, including ultra-strong \cite{Carusotto2013,FriskKockum2019,Forn-Diaz2019} as well as multi-mode strong coupling \cite{Krimer2014,Sundaresan2015}, where phenomenological few-mode models are also used extensively. Ab initio few-mode theory would be useful in these regimes providing the advantages outlined in Sec.~\ref{sec::eff_advantages}. In the context of applying the phenomenological input-output formalism at ultra-strong coupling, previous works \cite{Ciuti2006,Ridolfo2013} have shown how to modify the input-output relation in the presence of light-matter hybridization and counter-rotating cavity-bath coupling terms. Both of these approaches should combine straightforwardly with ab initio few-mode theory, up to the numerical computation of a relevant contour integral as outlined in Sec.~\ref{sec::Maxwell_beyondRot}. We note further work in this direction \cite{Bamba2013,DeLiberato2014_comment,Bamba2014} and alternative master equation methods to tackle open quantum systems at ultra-strong coupling \cite{DeLiberato2009,Beaudoin2011}, as reviewed in \cite{FriskKockum2019,Forn-Diaz2019}. In the context of multi-mode convergence at ultra-strong coupling, the proper treatment of gauge invariance and counter-rotating terms have been found to be crucial \cite{DeLiberato2014,Gely2017,SanchezMunoz2018,Malekakhlagh2017} (see also Sec.~\ref{sec::int_convergence}). It would therefore be interesting to see if ultra-strong coupling theory can be combined with ab initio few-mode theory to develop theoretical tools for highly open systems at ultra-strong coupling, complementing recent advances from circuit QED in this direction \cite{Malekakhlagh2016,Malekakhlagh2017}.

	A platform where ab initio few-mode theory applies almost directly and where it could serve as a useful tool is non-Hermitian photonics \cite{Peng2014,El-Ganainy2018,Miri2019,Ozdemir2019}, where loss and other open system effects are exploited and engineered. While many interesting phenomena of this kind are already observed in classical systems (see for example \cite{El-Ganainy2018,Ozdemir2019,Miri2019} for recent reviews), extensions to the quantum domain are imminent \cite{Quijandria2018,Ozdemir2019}. For example, spontaneous emission in an environment featuring exceptional points has been discussed in \cite{Pick2017} and interesting enhancement effects of the Petermann factor \cite{Petermann1979} have been found. These quantum effects are precisely due overlapping modes and complex bath structures. We therefore expect ab initio few-mode theory to be useful particularly for studying quantum dynamical effects in the presence of such exotic environments.

	The above are three concrete examples where ab initio few-mode theory can be applied. Beyond a direct application, it is also valuable that the connection of few-mode theory to other ab initio methods is now clear. To name a few, examples include SALT (steady state ab initio laser theory) \cite{Tureci2006,Tureci2008a,Cerjan2016} from laser theory, which employs constant flux states, which in turn have also found applications for example in circuit QED \cite{Malekakhlagh2016}, and quasi-modes \cite{Ching1998,Tureci2005,Kristensen2014,Lalanne2018}, which are mostly used to describe complex wave scattering phenomena \cite{Rotter2017} but have recently also entered the quantum domain \cite{Franke2019,Hughes2018}. Beyond light scattering, analogies to resonance and scattering theory in atomic \cite{Burke1965,Smith1966} and nuclear physics \cite{Mahaux1969,Mitchell2010}, mesoscopic systems \cite{Blanter2000,Rotter2017}, electronic transport \cite{Datta1995}, and even relativistic scenarios such as the Dirac equation \cite{Thaller1992,DiPiazza2012} can be found. Our formalism now allows few-mode theory to be treated on an equal footing as these well understood theories. This result may advance the exchange of methods and concepts \cite{Rotter2017} between currently separated fields.
	
	\section{Summary and conclusion}\label{sec::conclusions}
	In this work, we introduce ab initio few-mode theory, a method to describe quantum dynamics in open and scattering systems. The method and results presented can be understood from different perspectives.

	From a general point of view, we presented an approach to extract relevant degrees of freedom from a quantum field continuum. The concept exploits that in many physical systems and models, the quantum dynamics are often dominated by resonances or other meta-stable states, such that not the whole continuum participates in the dynamics. For non-interacting theories, we have presented an exact formalism that allows to rewrite the continuum in terms of a chosen set of relevant states. In the presence of interactions, this provides the option of simplifying the dynamics by only considering the interaction with these states. We have presented a systematic way to construct an effective few-mode expansion on this basis.

	More specifically in the theory of light-matter interactions, our method closes a gap in the current theoretical description by linking a large existing toolbox based on phenomenological few-mode and input-output models to ab initio theory. This connection is provided by a set of results. We have firstly presented a systematic approach to derive ab initio few-mode Hamiltonians. As a main result, we have demonstrated how to rigorously reconstruct the entire scattering matrix from such Hamiltonians using an input-output formalism, and have shown its equivalence to standard scattering theory. In the process, we have found crucial differences to phenomenological few-mode theory, such as a previously unknown background scattering contribution. In the presence of interactions such as atoms coupling to the light field, a systematic expansion scheme has been obtained providing a number of advantages, which are inherited from the exact treatment of the non-interacting theory in the ab initio approach. We have demonstrated each of the advantages explicitly using the paradigmatic situation of a two-level atom in a cavity as an example. In the process, we have shown that ab initio few-mode theory applies in extreme regimes and can be used to compute various observables for linear and non-linear systems.

	In conclusion, we have shown that ab initio few-mode theory provides a useful tool for describing a number of physical scenarios in quantum dynamics including extreme regimes, and that due to the generality of its concept, a broad class of systems, ranging from cavity QED to even relativistic scenarios, may be accessible through extensions of the method.

	\section*{Acknowledgements}
	We gratefully acknowledge fruitful discussions with G.~Amato, R.~Bennett, A.~Buchleitner, S.~Buhmann, H.~S.~Dhar, K.~P.~Heeg, Q.~Z.~Lv, S.~Rotter, F.~Salihbegovic, C.~Viviescas and M.~Zens. This work is part of and supported by the DFG Collaborative Research Centre ``SFB 1225 (ISOQUANT).''

	\appendix
	\section{Canonical quantization of the Schr\"odinger equation}\label{sec::app_schroed_CanQuant}
	
	Eq.~\eqref{equ::schrodinger_eom} is an example of a wave equation that can be quantized using the standard canonical quantization procedure \cite{Schiff1968,Imamoglu1999}, which we recapitulate in the following. The Lagrangian for the system reads
	\begin{align}
	L &= \int dr [i \psi^\dag(r,t) \dot{\psi}(r,t) - \frac{1}{2}\frac{\partial}{\partial r} \psi^\dag(r,t) \frac{\partial}{\partial r}\psi(r,t) \nonumber
	\\
	&- V(r)\psi^\dag(r,t)\psi(r,t)]\,, \label{equ::schrodinger_lagrangian}
	\end{align}
	such that the Euler-Lagrange equations yield Eq.~\eqref{equ::schrodinger_eom}. The conjugate momentum of $\psi(r,t)$ is then obtained as
	\begin{align}\label{equ::schrodinger_conjugateMomentum}
	\pi(r,t) & =  \frac{\partial}{\partial[ \dot{\psi}(r,t)]} L  =  i \psi^\dag(r,t)\,.
	\end{align}
	For quantization, we promote $\psi(r,t)$ [$\pi(r,t)$] to operators $\hat{\psi}(r,t)$ [$\hat{\pi}(r,t)$] and impose the bosonic commutation relations Eq.~\eqref{equ::schrodinger_fieldCommutator}.
	
	The second quantized Hamiltonian is then obtained from the Lagrangian as Eq.~\eqref{equ::schrodinger_hamiltonian}. Together with the commutation relations, the Heisenberg equations of motion can be verified to give Eq.~\eqref{equ::schrodinger_eom}.
	
	\section{Mode normalization and orthogonality}\label{sec::app_modeNorm}
	We choose the normalization of the normal modes such that the orthogonality condition reads
	\begin{align}\label{equ::schrodinger_orthoFull}
	\int dr\: \phi^*_{m'}(r,k') \phi^{\mathstrut}_m(r,k) & =   \delta^{\mathstrut}_{mm'}\:  \delta(E(k)-E(k'))\,.
	\end{align}
	The normal modes form a complete set in the sense that
	\begin{align}\label{equ::schrodinger_completenessFullCoord}
	\sum_m \int dE(k) \:\phi^*_{m}(r',k) \phi^{\mathstrut}_m(r,k) = \delta(r-r') \,.
	\end{align}
	We note that $k$ is simply a relabeling of the energy eigenstates, which we find convenient to introduce. A natural choice is $E(k)=k^2/2$, since $k$ then has a physical interpretation as the wave number.
	
	Similarly, hermicity of the subspace Hamiltonians \cite{Domcke1983} implies orthogonality of their corresponding eigenstates, the system and bath states. We choose their normalization such that
	\begin{equation}\label{equ::proj_Q_ortho}
	\braket{\chi_\lambda}{\chi_{\lambda'}} = \delta_{\lambda\lambda'}
	\end{equation}
	and
	\begin{equation}\label{equ::proj_PHP_ortho}
	\braket{\tilde{\psi}_{m}(k)}{\tilde{\psi}_{m'}(k')}= \delta^{\mathstrut}_{mm'}\: \delta\left(E(k) - E(k')\right)\,.
	\end{equation}
	
	Analogously to Eq.~\eqref{equ::proj_Q_statesDef}, the bath modes diagonalize the $P$-space projector via
	\begin{equation}\label{equ::proj_PProjector}
	P=\sum_{m}\int dE(k) \: \ket{\tilde{\psi}_{m}(k)}\bra{\tilde{\psi}_{m}(k)}\,.
	\end{equation}
	
	Since $P = 1-Q$, the system modes can furthermore be chosen orthogonal to the bath modes \cite{Domcke1983}
	\begin{equation}\label{equ::proj_QP_ortho}
	\braket{\chi^{\mathstrut}_\lambda}{\tilde{\psi}^{\mathstrut}_{m}(k)} = 0\,.
	\end{equation}
	We note that these orthogonality conditions are crucial for quantization.
	
	\section{Subspace expansion}\label{appSub}
	Here we summarise the calculation of the matrix elements in the expansion of the full eigenstates Eq.~\eqref{equ::proj_modeSep_b} as presented by Domcke \cite{Domcke1983}.
	
	We start by writing the Schr\"odinger equation Eq.~\eqref{equ::schrodinger_eom} in Dirac notation
	\begin{align}\label{equ::proj_DiracSchrodingerEq}
		H\ket{\phi_m(k)} = E(k)\ket{\phi_m(k)}\,,
	\end{align}
	which can be expressed as a pair of coupled equations in the two subspaces \cite{Feshbach1958}
	\begin{subequations}
	\label{equ:appSub_both} 
	\begin{align}
		H_{PP}\ket{\phi_m(k)} + H_{PQ}\ket{\phi_m(k)} &= E(k) P\ket{\phi_m(k)}\,, \label{equ::appSub_Pspace_Schroedinger}
		\\
		H_{QP}\ket{\phi_m(k)} + H_{QQ}\ket{\phi_m(k)} &= E(k) Q\ket{\phi_m(k)}\,.  \label{equ::appSub_Qspace_Schroedinger}
	\end{align}
	\end{subequations}
	The Lippmann-Schwinger equation for the P-space part Eq.~\eqref{equ::appSub_Pspace_Schroedinger} reads
	\begin{align}
		P\ket{\phi_m(k)} &= \ket{\tilde{\psi}_{m}^{(+)}(k)} \nonumber
		\\
		&+ \left(E(k) - H_{PP} + i\eta \right)^{-1} H_{PQ} \ket{\phi_m(k)}\,, \label{LS_P}
	\end{align}
	where we chose the incoming solution for the homogeneous part. Substitution into Eq.~\eqref{equ::appSub_Qspace_Schroedinger} and solving for $Q\ket{\phi_m(k)}$ gives
	\begin{align}\label{Q-expansion}
		Q\ket{\phi_m(k)} & =   G_{QQ} H_{QP} \ket{\tilde{\psi}_{m}^{(+)}(k)}\,,
	\end{align}
	where we have defined
	\begin{subequations}
	\begin{align}
	G_{QQ}&=(E(k) - H_{QQ} - H_{QP}\tilde{G}^{(+)} H_{PQ})^{-1}\,,\label{equ::app::subExp_GQQ}
	\\
	\tilde{G}^{(+)}&=\left(E(k) - H_{PP} + i\eta\right)^{-1}\,.\label{fullPGreen}
	\end{align}
	\end{subequations}
	Substitution into Eq.~\eqref{LS_P} gives Eq.~(2.11) from \cite{Domcke1983}
	\begin{align}\label{P-expansion}
		P\ket{\phi_m(k)} & =  \ket{\tilde{\psi}_{m}^{(+)}(k)}\nonumber
		\\
		&+ \tilde{G}^{(+)} H_{PQ}  G_{QQ} H_{QP}\ket{\tilde{\psi}_{m}^{(+)}(k)}\,.
	\end{align}
	Adding Eqs.~\eqref{Q-expansion} and \eqref{P-expansion} we obtain an expansion for the full eigenstates in terms of the subspace eigenstates
	\begin{align}\label{equ::app_mode_separationOp}
		\ket{\phi_m(k)} & =  Q\ket{\phi_m(k)} + P\ket{\phi_m(k)}
		\\
		&= G_{QQ} H_{QP} \ket{\tilde{\psi}_{m}^{(+)}(k)} \nonumber
		\\
		& + [1 + \tilde{G}^{(+)} H_{PQ}  G_{QQ} H_{QP}]\ket{\tilde{\psi}_{m}^{(+)}(k)}\,.
	\end{align}
	Note that this constitutes a generalization of results obtained in \cite{Viviescas2003} to a finite number of system modes and to the Schr\"odinger equation.
	
	The expansion coefficients in Eq.~\eqref{equ::proj_modeSep_b} are therefore
	\begin{align}\label{equ::alpha}
		\braket{\chi_{\lambda}}{\phi_m(k)} = \bra{\vphantom{\hat{k}^-}\chi_{\lambda}}G_{QQ} H_{QP} \ket{\tilde{\psi}_{m}^{(+)}(k)} 
	\end{align}
	and
	\begin{align}\label{equ::beta}
		&\braket{\tilde{\psi}_{m'}^{(+)}(k')}{\phi^{\mathstrut}_m(k)} = \bra{\tilde{\psi}_{m'}^{(+)}(k')}1 +  \nonumber\\
		&\qquad \tilde{G}^{(+)} H_{PQ}  G_{QQ} H_{QP}\ket{\tilde{\psi}_{m}^{(+)}(k)}\,. 
	\end{align}
	These expressions can be conveniently evaluated for a certain class of potentials using so-called separable expansions, as has been shown in detail by Domcke \cite{Domcke1983}.
	
	\section{System-and-bath operators}\label{sec::app_SBoperators}
	Here we derive the system-bath expansion and show that the operators associated with the subspace eigenstates naturally fulfill the desired conditions for bosonic system and bath operators as they are used in quantum noise theory \cite{Gardiner1985,Gardiner2004,Viviescas2003}.
	
	Eq.~\eqref{equ::proj_modeSep_b} can be used to write the field operator Eq.~\eqref{equ::schrodinger_fieldOpModeExpansion} as
	\begin{align}
	\hat{\psi}(r,t) &= \sum_{\lambda \in \Lambda_Q} \hat{a}^{\mathstrut}_\lambda \chi^{\mathstrut}_\lambda(r) \nonumber \\
	&+ \sum_{m'}\int dE(k') \:\hat{b}^{\mathstrut}_{m'}(k') \tilde{\psi}_{m'}^{\mathstrut}(r,k')\label{equ::proj_mode-SB-Expansion}
	\end{align}
	where we have defined~\cite{Viviescas2003}
	\begin{subequations}\label{equ::SB-operators} 
	\begin{align}
	\hat{a}_\lambda &= \sum_m \int dE(k) \hat{c}_m(k,t) \alpha_{\lambda m}(k)\,, \label{equ::proj_aOperatorsDef}
	\\
	\hat{b}_{m'}(k') &= \sum_m \int dE(k) \hat{c}_m(k,t) \beta_{m m'}(k, k') \label{equ::proj_bOperatorsDef}
	\end{align}
	\end{subequations}
	as the system and bath operators, respectively. Inverting Eqs.~\eqref{equ::SB-operators} by using the coefficient identities in Appendix \ref{app_coeffIds} gives Eq.~\eqref{equ::schrodinger_cExpansion2} \cite{Viviescas2003}.
	
	Using Eqs.~\eqref{equ::SB-operators} and the coefficient identities in Appendix \ref{app_coeffIds}, the commutation relations for the system-bath operators are obtained as
	\begin{subequations}\begin{align}\label{equ::proj_Q_Op_commutators}
	&\big[\hat{a}^{\mathstrut}_\lambda, \hat{a}^\dag_{\lambda'}\big] = \delta_{\lambda \lambda'}\,,
	\\
	&\big[\hat{b}^{\mathstrut}_{m}(k), \hat{b}^\dag_{m'}(k')\big] = \delta_{m m'} \delta\left(E(k)-E(k')\right)\,,
	\\
	&\big[\hat{a}^{\mathstrut}_{\lambda}, \hat{b}^\dag_{m}(k)\big] = 0\,,
	\\
	&\big[\hat{a}^{\mathstrut}_{\lambda}, \hat{a}^{\mathstrut}_{\lambda'} \big] = \big[\hat{b}^{\mathstrut}_{m}(k), \hat{b}^{\mathstrut}_{m'}(k')\big] = 0\,,\\ 
	&\big[\hat{a}^{\dag}_{\lambda}, \hat{a}^{\dag}_{\lambda'} \big] = \big[\hat{b}^{\dag}_{m}(k), \hat{b}^{\dag}_{m'}(k')\big] = 0\,,
	\end{align}
	\end{subequations}
	which are indeed the desired bosonic commutation relations \cite{Viviescas2003}.
	
	We note that due to the few-mode projection, the system states do not necessarily form a complete set in the region of the system modes. It is thus necessary to account for the bath state contribution in Eq.~\eqref{equ::proj_mode-SB-Expansion}, even when $r$ lies inside this region. This feature is relevant when field-matter interactions are included in the theory, which we discuss in Sec.~\ref{sec::eff}.
	
	\section{Expansion coefficient identities}\label{app_coeffIds}
	Using the completeness relation in full space
	\begin{equation}
	\mathbb{I} = \sum_m \int dE(k) \: \ket{\phi_m(k)} \bra{\phi_m(k)}
	\end{equation}
	and the orthogonality relations in the subspaces Eqs.~(\ref{equ::proj_Q_ortho}, \ref{equ::proj_PHP_ortho}, \ref{equ::proj_QP_ortho}) we obtain the coefficient identities
	\begin{subequations}
	\begin{align}
	&\int dE(k) \sum_m \alpha^{\mathstrut}_{\lambda m}(k) \alpha^{*\mathstrut}_{\lambda' m}(k) = \braket{\chi_{\lambda}}{\chi_{\lambda'}} = \delta_{\lambda \lambda'}\,, \label{equ::app_coeffProps1}
	\\[1ex]
	&\int dE(k') \sum_{m'}\alpha^{\mathstrut}_{\lambda m'}(k') \beta^{*\mathstrut}_{m' m}(k',k) \nonumber \\
	&\qquad =  \braket{\chi^{\mathstrut}_{\lambda}}{\tilde{\psi}_{m}^{(+)}(k)} =  0 \,,\label{equ::app_coeffProps3}
	\\[1ex]
	&\int dE(k'') \sum_{m''} \beta^{\mathstrut}_{m'' m}(k'',k) \beta^{*\mathstrut}_{m'' m'}(k'',k')
	\nonumber
	\\
	&\qquad = \braket{\tilde{\psi}_{m}^{(+)}(k)}{\tilde{\psi}_{m'}^{(+)}(k')} \nonumber \\
	&\qquad =  \delta^{\mathstrut}_{mm'} \delta(E(k)-E(k'))\,.
	\end{align}
	\end{subequations}
	Similarly,
	\begin{subequations}
	\begin{align}\label{equ::app_coeff_properties2.1} 
	&\int dE(k) E(k) \sum_m \alpha^{\mathstrut}_{\lambda m}(k) \alpha^{*\mathstrut}_{\lambda' m}(k) = E^{\mathstrut}_{\lambda}\delta^{\mathstrut}_{\lambda \lambda'}\,,
	\\[1ex]
	&\int dE(k'') \sum_{m''} E(k'') \beta^{\mathstrut}_{m'' m}(k'',k) \beta^{*\mathstrut}_{m'' m'}(k'',k')
	\nonumber\\
	&\quad =  E(k) \delta_{mm'} \delta(E(k)-E(k'))\,, \label{equ::app_coeff_properties2.2} 
	\\[1ex]
	&\int dE(k') \sum_{m'} E(k') \alpha^{\mathstrut}_{\lambda m'}(k') \beta^{*\mathstrut}_{m' m}(k',k)
	 \nonumber\\
	& \quad = \bra{\chi^{\mathstrut}_{\lambda}\vphantom{\tilde{\psi}_{k m}^{(+)}}} H_{QP} \ket{\tilde{\psi}_{m}^{(+)}(k)} =: W^{\mathstrut}_{\lambda m}(k)\,. \label{equ::app_coeff_properties2.3}
	\end{align}
	\end{subequations}
	Note that these relations are analogous to expressions obtained in \cite{Viviescas2003} for the dielectric Maxwell equations, but refer to different modes since our few-mode projection scheme differs.
	
	\section{Scattering matrix in Viviescas-Hackenbroich quantization}\label{sec::app_ViviescasScattering}
	
	In this Appendix, we calculate the scattering matrix for an example cavity using Viviescas\&Hackenbroich's Feshbach projection scheme \cite{Viviescas2003,Viviescas2004}.
	
	From Eq.~(68) in \cite{Viviescas2003} their scattering matrix reads
	\begin{align}\label{equ::app_Viviescas_Smatrix}
	S(\omega)
	&= 1 - 2\pi i \sum_{\lambda, \lambda' = 1}^{\infty}\mathcal{W}_{\lambda}^\dagger(\omega) \left(D^{-1}(\omega)\right)_{\lambda \lambda'} \mathcal{W}^{\mathstrut}_{\lambda'}(\omega)\,.
	\end{align}
	Here the matrix $D$ is defined by Eqs.~(65, 66) in \cite{Viviescas2003} as
	\begin{equation}\label{equ::app_Viviescas_Dmatrix}
	\left(D(\omega)\right)_{\lambda \lambda'} = (\omega-\omega_\lambda)\delta_{\lambda \lambda'} + \Gamma_{\lambda \lambda'}(\omega)\,,
	\end{equation}
	with
	\begin{align}
	\Gamma_{\lambda \lambda'}(\tilde{\omega}) &= \lim\limits_{\epsilon\rightarrow 0^+}\int d\omega'\: \frac{\mathcal{W}^{\mathstrut}_{\lambda}(\omega') \mathcal{W}^*_{\lambda'}(\omega')}{\omega'-\omega - i\epsilon}\,.
	\end{align}
	These expressions are similar to our input-output scattering matrix Eq.~\eqref{equ::max_scattIO}, except for the different projection scheme and the infinite number of system modes.
	
	In \cite{Viviescas2004}, the authors have demonstrated their formalism on the example of a one dimensional cavity with a single homogeneous dielectric layer of thickness $d$ and refractive index $n$ terminated by a perfectly reflecting mirror. In the following we attempt a calculation of the corresponding scattering matrix using the input-output result Eq.~\eqref{equ::app_Viviescas_Smatrix} from their method. The coupling coefficients for Neumann basis states are given by Eq.~(46) in \cite{Viviescas2004} as
	\begin{equation}
	\mathcal{W}_\lambda(\omega) = (-1)^\lambda\: \sqrt{\frac{\omega_\lambda}{\pi \omega d}}\,,
	\end{equation}
	where the cavity mode frequencies are
	\begin{equation}\label{equ::app_Viviescas_resFrequ}
	\omega_\lambda = \frac{c \pi \lambda}{n d}
	\end{equation}
	with $\lambda \in \{1,2,\dots\}$. We can simply plug this into the Eq.~\eqref{equ::app_Viviescas_Dmatrix} above to get
	\begin{align}
	\left(D(\omega)\right)_{\lambda \lambda'} &= \left(\omega-\omega_\lambda \right)\delta_{\lambda \lambda'} \nonumber \\
	&+ \tilde{\Gamma}(\omega) (-1)^{\lambda + \lambda'} \sqrt{\omega_\lambda \omega_{\lambda'}} 
	\end{align}
	where
	\begin{equation}
	\tilde{\Gamma}(\omega)=\int d\omega' \: \frac{1}{\omega} \frac{1}{\omega'-\omega - i\epsilon}\,.
	\end{equation}
	The inverse of this $D$-matrix can be calculated exactly using the Sherman-Morrison formula \cite{Domcke1983,Golub1996}. Substitution into Eq.~\eqref{equ::app_Viviescas_Smatrix} yields, after a short calculation,
	\begin{align}\label{Equ::app_Viviescas_Smatrix2}
	S(\omega) &=  1 - \frac{2i}{\omega d} \frac{K(\omega)}{1 - \tilde{\Gamma}(\omega) K(\omega)},
	\end{align}
	where
	\begin{equation}
	K(\omega) = \sum_{\lambda=1}^{\infty} \frac{\omega_\lambda}{\omega - \omega_\lambda}\,.
	\end{equation}
	Substitution of the resonance frequencies Eq.~\eqref{equ::app_Viviescas_resFrequ} gives
	\begin{align}
	K(\omega)
	&= \sum_{\lambda=1}^{\infty} \frac{\lambda}{\frac{\omega n d}{c \pi} - \lambda}\,.
	\end{align}
	This sum indeed diverges. There is also no well defined notion of taking a limit of $\lambda$ to infinity, since the projection is performed directly onto infinitely many modes. Similar behavior is observed for other one dimensional examples in \cite{Viviescas2004}, including a one-sided Ley-Loudon cavity.
	
	We conclude that in Viviescas\&Hackenbroich's formalism \cite{Viviescas2003}, there is no straightforward way to calculate scattering matrices from the input-output formalism due to the convergence behavior of the  infinitely many modes included in their projection scheme. For the example cavity investigated above, we have further observed that truncation approximations or cut-off schemes can be used to approximate the spectra around a single resonance for good cavities. For multiple or overlapping modes, however, such approximations fail. In these regimes it is thus crucial to understand how to precisely reconstruct the scattering information in system-bath theory. By using a different projection scheme and few-mode Hamiltonians, the approach presented in this work addresses this topic.
	
	\section{Domcke's Feshbach projection formalism for potential scattering}\label{sec::app_DomckeScatt}
	In Sec.~\ref{sec::potScatt_Domcke} we have focused on defining and interpreting the background and resonant scattering matrices. We further have shown how the former corresponds to an asymptotic basis transformation. In this Appendix we extract the relevant parts of Domcke's \cite{Domcke1983} derivation of this separation based on Lippmann-Schwinger equations and give his formulae for the $T$-matrices.
	
	The goal is to expand the $P$-space projection of the full eigenstate, which contains all the scattering information, in terms of the various subspace eigenstates. During the quantization procedure, we already derived Eq.~\eqref{P-expansion}, in which we now only have to expand the homogeneous part in terms of free states.
	
	We first write down the Lippmann-Schwinger equation for the bath eigenstates
	\begin{equation}
		\ket{\tilde{\psi}_{m}^{(\pm)}(k)} = \ket{k^{\mathstrut}_{m}} + G_0^{(\pm)}H_{PP} \ket{\tilde{\psi}_{m}^{(\pm)}(k)}\,,
	\end{equation}
	where we have defined the free Green function in full space
	\begin{equation}
		G_0^{(\pm)} = (E(k) - K \pm i\epsilon)^{-1}
	\end{equation}
	and the free eigenstates
	\begin{equation}
		K\ket{k_m}=E(k)\ket{k_m}.
	\end{equation}
	Upon substitution into Eq.~\eqref{P-expansion} we obtain \cite{Domcke1983}
	\begin{align}
		P\ket{\phi_m(k)} & =  \ket{k_m} +  G_0^{(\pm)} (H_{PP} - K) \ket{\tilde{\psi}_{m}^{(\pm)}(k)} \nonumber \\
		&\quad + \tilde{G}^{(+)} H_{PQ}  G_{QQ} H_{QP}\ket{\tilde{\psi}_{m}^{(+)}(k)}\,. 
	\end{align}
	From there we obtain the separation of the $T$-matrix \cite{Domcke1983}
	\begin{equation}
		T(k) = T_{\textrm{bg}}(k) + T^{(F)}_{\textrm{res}}(k),
	\end{equation}
	where, omitting subscripts for brevity,
	\begin{align}
		T_{\textrm{bg}}(k) &\equiv \bra{k}T_{\textrm{bg}}\ket{k} \nonumber
		\\
		&= \bra{k} (H_{PP}-K) \ket{\tilde{\psi}^{(+)}(k)}
	\end{align}
	and
	\begin{align}
		T^{(F)}_{\textrm{res}}(k) &\equiv \bra{\tilde{\psi}^{(-)}(k)}T^{\mathstrut}_{\textrm{res}}\ket{\tilde{\psi}^{(+)}(k)} \nonumber
		\\
		&= \bra{\tilde{\psi}^{(-)}(k)} H_{PQ}  G_{QQ} H_{QP}\ket{\tilde{\psi}^{(+)}(k)}\,.
	\end{align}
	The matrix element from the main text giving the resonant scattering matrix is
	\begin{align}
		T^{\mathstrut}_{\textrm{res}}(k) &\equiv \bra{\tilde{\psi}^{(+)}(k)}T^{\mathstrut}_{\textrm{res}}\ket{\tilde{\psi}^{(+)}(k)} \nonumber\\
		&= S_\mathrm{bg}^{-1}\: T_\mathrm{res}^{(F)}\,. \label{equ::app_DomckeScatt_Tres}
	\end{align}
	Consequently one obtains \cite{Domcke1983}
	\begin{equation}
		S(k) = \mathbb{I} - 2\pi iT(k) = S_{\textrm{bg}}(k)S_{\textrm{res}}(k)
	\end{equation}
	as expected.

	\section{The operator scattering matrix in second quantized potential scattering}\label{sec::app_opScatt}
	In this Appendix we derive the result used in Sec.~\ref{sec::potScatt_opScatt}, that the operator scattering matrix relating asymptotically free in- and out-operators is the same as the conventional on-shell scattering matrix for the corresponding states~\cite{Newton1982}. We proceed by solving the operator equations of motions for appropriately defined asymptotically free operators, following Glauber\&Lewenstein's method \cite{Glauber1991}.
	
	To define the asymptotically free operators one has to adiabatically turn off the interaction in the infinite past and future, such that these operators are actually evolving freely in the corresponding limits. To do so we replace the potential $V(r)$ by a potential $V(r,t)$ slowly varying in time such that
	\begin{equation}
		\lim\limits_{t\rightarrow \pm \infty} V(r,t) \rightarrow 0
	\end{equation}
	and
	\begin{equation}
		V(r,0) = V(r).
	\end{equation}
	Consequently, the normal modes also become time-dependent. In general, they fulfill an explicitly time-dependent form of the wave equation, however in the adiabatic limit they correspond to the time-independent normal modes at each time slice, such that Eq.~\eqref{equ::schrodinger_timeIndep} becomes
	\begin{align}\label{equ::app_opScatt_timeIndep}
		\left(-\frac{1}{2}\frac{\partial^2}{\partial r^2} + V(r,t)\right) \phi_m(r,k,t) = E(k,t) \phi_m(r,k,t)\,.
	\end{align}
	
	The in [out] operators are then defined as the corresponding free interaction picture operators in the infinite past [future], that is
	\begin{equation}\label{equ::app_opScatt_dIn}
		\hat{d}^{(\textrm{in})}_{m}(k) = \lim\limits_{t\rightarrow -\infty} e^{iE(k)t} \: \hat{d}^{\mathstrut}_{m}(k,t)
	\end{equation}
	and
	\begin{equation}\label{equ::app_opScatt_dOut}
		\hat{d}^{(\textrm{out})}_{m}(k) = \lim\limits_{t\rightarrow +\infty} e^{iE(k)t}\: \hat{d}^{\mathstrut}_m(k,t).
	\end{equation}
	In Eq.~\eqref{equ::schrodinger_fieldOpModeExpansion} and Eq.~\eqref{equ::potScatt_freeFieldExpansion}, two separate expansions of the quantum field have been introduced, one in terms for normal modes and one in terms of free states
	\begin{align}
		\hat{\psi}(r,t) &= \sum_m \int dE(k)\: \phi_m(r,k)\hat{c}_m(k,t) \nonumber
		\\
		&=\sum_m \int dE(k) \: \phi^{(\textrm{free})}_{m}(r,k)\hat{d}^{\mathstrut}_m(k,t)\,.\label{equ::app_opScatt_freeExpansion_general}
	\end{align}
	Using the orthogonality properties of these states one obtains a linear relation between the two operator bases
	\begin{equation}
		\hat{d}^{\mathstrut}_m(k,t) = \sum_{m'} \int dE(k') \braket{\phi^{(\textrm{free})}_{m}(k)}{\phi^{\mathstrut}_{m'}(k')} \hat{c}^{\mathstrut}_{m'}(k',t).
	\end{equation}
	
	The construction of the basis transformation between asymptotically free in- and out-operators proceeds similarly via comparing asymptotic expansions. Let us first asymptotically expand the field in the infinite past in terms of the in-operators using Eq.~\eqref{equ::app_opScatt_freeExpansion_general} and Eq.~\eqref{equ::app_opScatt_dIn} to get
	\begin{equation}\label{equ::app_opScatt_+freeExpansion}
		\hat{\psi}(r,t\rightarrow-\infty) = \sum_m \int dE(k) \phi^{(\textrm{free})}_{m}(r,k)\hat{d}^{(\textrm{in})}_m(k)e^{-iE(k)t}.
	\end{equation}
	To obtain a second expansion to compare to let us note that the normal modes are not uniquely defined since we have not specified their boundary conditions. The choice that is relevant in the infinite past are the states with a controlled incoming state $\ket{\phi^{(+)}_m(k,t)}$ \cite{Newton1982}. The corresponding expansion reads
	\begin{equation}\label{equ::app_opScatt_+Expansion}
		\hat{\psi}(r,t) = \sum_m \int dE(k) \:\phi^{(+)}_{m}(r,k,t)\hat{\tilde{c}}^{\mathstrut}_m(k)e^{-iE(k)t}\,,
	\end{equation}
	where $\hat{\tilde{O}}_m(k,t)=\hat{O}_m(k,t)e^{iE(k)t}$ is the relevant interaction picture operator \cite{Glauber1991}, which is independent of $t$ for the normal modes operators. These states by construction have the property that \begin{equation}
		\lim\limits_{t\rightarrow -\infty}\ket{\phi^{(+)}_m(k,t)} = \ket{\phi^{(\textrm{free})}_m(k)}\,.
	\end{equation}
	Comparing Eqs.~\eqref{equ::app_opScatt_+freeExpansion} and \eqref{equ::app_opScatt_+Expansion}, we thus find that
	\begin{equation}
		\hat{\tilde{c}}^{\mathstrut}_m(k) = \hat{d}^{(\textrm{in})}_m(k)\,.
	\end{equation}
	Consequently, since Eq.~\eqref{equ::app_opScatt_+Expansion} applies at all times, there are now two ways to express the field at the time slice $t=0$,
	\begin{align}
		\hat{\psi}(r,t=0) &= \sum_m \int dE(k)\: \phi^{(+)}_{m}(r,k,t=0)\hat{d}^{(\textrm{in})}_m(k)  \label{equ::app_opScatt_inExp_at0}
		\\
		&= \sum_m \int dE(k) \:\phi^{(\textrm{free})}_{m}(r,k)\hat{\tilde{d}}^{\mathstrut}_m(k,t=0)\,. \label{equ::app_opScatt_freeExp_at0}
	\end{align}
	At $t=0$ our potential has the desired physical value such that $\phi^{(+)}_{m}(r,k,t=0)=\phi^{(+)}_{m}(r,k)$, where the latter solves the original mode equation Eq.~\eqref{equ::schrodinger_timeIndep}.
	
	From Eqs.~(\ref{equ::app_opScatt_inExp_at0}, \ref{equ::app_opScatt_freeExp_at0}) we can obtain the transformation between asymptotically free operators in the infinite past and free operators at the time slice $t=0$ as
	\begin{align}
		\hat{\tilde{d}}_m(k,t=0) &= \sum_{m'} \int dE(k')\nonumber
		\\
		&\braket{\phi^{(\textrm{free})}_{m}(k)}{\phi^{(+)}_{m'}(k')}\: \hat{d}^{(\textrm{in})}_{m'}(k')\,. \label{equ::app_opScatt_in-0}
	\end{align}
	Analogously, by expanding in the $\ket{\phi^{(-)}_m(k,t)}$ basis and performing an asymptotic expansion in the infinite future, we obtain a second expansion
	\begin{align}
		\hat{\tilde{d}}_m(k,t=0) &= \sum_{m'} \int dE(k')\nonumber
		\\
		&\braket{\phi^{(\textrm{free})}_{m}(k)}{\phi^{(-)}_{m'}(k')} \:\hat{d}^{(\textrm{out})}_{m'}(k')\,.  \label{equ::app_opScatt_out-0}
	\end{align}
	Upon combining Eqs.~(\ref{equ::app_opScatt_in-0}, \ref{equ::app_opScatt_out-0}) and using that the matrix elements vanish off the energy shell, we obtain the operator scattering relation
	\begin{align}
		\hat{d}^{(\textrm{out})}_{m}(k) &= \sum_{m'} \int dE(k')\nonumber
		\\
		& \braket{\phi^{(-)}_{m}(k)}{\phi^{(+)}_{m'}(k')} \: \hat{d}^{(\textrm{in})}_{m'}(k')\,.
	\end{align}
	Indeed, the matrix element in this expression is the scattering matrix \cite{Newton1982}
	\begin{equation}
		S^{\mathstrut}_{mm'}(k,k') = \braket{\phi^{(-)}_{m}(k)}{\phi^{(+)}_{m'}(k')}\,,
	\end{equation}
	which is related to the on-shell scattering matrix $S_{mm'}(k)$ used in the main text by \cite{Newton1982}
	\begin{equation}
		S_{mm'}(k,k') = S_{mm'}(k) \: \delta(E(k)-E(k'))\,.
	\end{equation}
	We thus obtain the result Eq.~\eqref{equ::potScatt_free_IO} as
	\begin{equation}
		\hat{d}^{(\textrm{out})}_m(k) = \sum_{m'}S^{\mathstrut}_{mm'}(k)\: \hat{d}_{m'}^{(\textrm{in})}(k).
	\end{equation}

	\section{Regularization of Fourier integrals in the input-output formalism}\label{sec::app_FourReg}
	In this Appendix we provide a derivation of Eq.~\eqref{equ::IO_modeSol}. In the process we show how the Fourier integrals are regularized in the input-output formalism and how this relates to time-independent scattering theory \cite{Newton1982}.
	
	We start by Fourier transforming Eq.~\eqref{equ::IO_Heisenberg1} to get
	\begin{align}\label{equ::app_FourReg_FTHeisenberg1}
		0
		&= i(E(\tilde{\omega})-E_\lambda) \hat{a}_\lambda(\tilde{\omega}) \nonumber
		\\
		& -i\sum_m \int dE(k) W_{\lambda m}(k) \int_{-\infty}^{\infty} dt e^{iE(\tilde{\omega})t} \hat{b}_m(k,t)\,. 
	\end{align}
	Substitution of Eq.~\eqref{equ::IO_formalT0} gives
	\begin{align}
		0
		&= i(E(\tilde{\omega})-E_\lambda) \hat{a}_\lambda(\tilde{\omega}) \nonumber
		\\
		& -i\sum_m \int dE(k) W_{\lambda m}(k) \int_{-\infty}^{\infty} dt e^{iE(\tilde{\omega})t} \times \nonumber
		\\
		& \qquad e^{-iE(k)(t-t_0)} \hat{b}_m(k,t_0)\nonumber
		\\[1ex]
		& -\sum_m \int dE(k) W_{\lambda m}(k) \int_{-\infty}^{\infty} dt e^{iE(\tilde{\omega})t} \times \nonumber
		\\
		&\qquad\sum_{\lambda'\in\Lambda_Q} W_{\lambda' m}^*(k) \int_{t_0}^{t} dt' e^{-iE(k) (t-t')} \times \nonumber
		\\
		& \qquad\qquad \int_{-\infty}^{\infty} dE(\tilde{\omega}')\frac{1}{2\pi} e^{-iE(\tilde{\omega'})t'}\hat{a}^{\mathstrut}_{\lambda'}(\tilde{\omega}')\,. \label{equ::app_FourReg_FTsubst1}
	\end{align}
	We note that the integration over energies from negative to positive infinity enters via the inverse Fourier transform of Eq.~\eqref{equ::IO_FTdef}, where the energy definition range has to be suitably extended beyond the physical spectrum for the inverse Fourier transform to be defined. This does not constitute an approximation, but rather a definition of an energy dispersion beyond the physical spectrum, such that inverse Fourier transforms can be used as a mathematical tool.
	
	The first of the three terms in this Eq.~\eqref{equ::app_FourReg_FTsubst1} is simple enough already, the second can be reduced using the definition of the input operator and the Fourier identity
	\begin{equation}
		\int_{-\infty}^{\infty} dt e^{i(E(\tilde{\omega})-E(k))t} = 2\pi\delta(E(\tilde{\omega})-E(k)).
	\end{equation}
	The third term can be simplified in the scattering limit $t_0\rightarrow-\infty$. However, we notice that the integral is in fact divergent in this limit. This is a well known feature of time-independent scattering theory and can be dealt with through regularization \cite{Newton1982}. In our case we require a substitution
	\begin{equation}
		E(k)\rightarrow E(k)-i\epsilon
	\end{equation}
	and taking the limit ${\epsilon\rightarrow 0^+}$ at the end, which regularizes the integral as $t_0 \rightarrow -\infty$. Physically this corresponds to solving an initial value problem \cite{Newton1982}.
	
	Evaluation of the integrals in Eq.~\eqref{equ::app_FourReg_FTsubst1} then yields Eqs.~(\ref{equ::IO_modeSol}-\ref{equ::IO_decayMatrix}).
	\section{Few-mode Hamiltonian for the scalar Maxwell wave equation}
	In this Appendix we provide details on the application of our formalism to the dielectric Maxwell wave equation Eq.~\eqref{equ::max_scalarEOM}, which constitutes a combination of the system-bath formalism by Viviescas\&Hackenbroich \cite{Viviescas2003}, the projection scheme by Domcke \cite{Domcke1983} and the relation of the input-output formalism to scattering theory presented in the main text for the Schr\"odinger equation.
	
	\subsection{Canonical quantization}\label{sec::app_maxwell_quant}
	The quantization of the vectorial dielectric Maxwell equation has been presented by Glauber\&Lewenstein \cite{Glauber1991}. Here we follow their approach, simplifying the results to the scalar wave equation Eq.~\eqref{equ::max_scalarEOM}. For simplicity we work with $\hbar=c=1$.
	The Lagrangian for the system is \cite{Glauber1991}
	\begin{equation}
	L = \frac{1}{2} \int dr \left(\varepsilon(r)\dot{A}^2(r) - \left(\frac{\partial A(r)}{\partial r}\right)^2\right)\,,
	\end{equation}
	such that the resulting Euler-Lagrange equations can be checked to give Eq.~\eqref{equ::max_scalarEOM}. The conjugate momentum of $A(r)$ can then be obtained as \cite{Glauber1991}
	\begin{equation}
	\Pi(r) = \frac{\delta L}{\delta \dot{A}(r)} = \varepsilon(r)\dot{A}(r)\,.
	\end{equation}
	Therefore, the Hamiltonian reads \cite{Glauber1991}
	\begin{align}
	H[A,\Pi] &= \int dr\: \Pi(r,t) \dot{A}(r,t) - L \nonumber
	\\
	&= \frac{1}{2} \int dr \left[\frac{\Pi^2(r)}{\varepsilon(r)} + \left(\frac{\partial A(r)}{\partial r}\right)^2\right]\label{equ::max_H_0_fields}\,.
	\end{align}
	This Hamiltonian can now be expressed in its normal mode basis \cite{Glauber1991}, as we have done for the Schr\"odinger equation earlier. To do so we expand the $A$-field as \cite{Glauber1991}
	\begin{equation}\label{equ::max_A-expansion}
	A(r,t) = \sum_m \int d\omega \: \hat{q}_m(\omega,t) f_m(r,\omega)\,,
	\end{equation}
	and similarly the conjugate momentum via
	\begin{equation}\label{equ::max_Pi-expansion}
	\Pi(r,t) = \sum_m \int d\omega \:\varepsilon(r) \hat{p}_m(\omega,t) f^*_m(r,\omega)\,.
	\end{equation}
	Here, $\hat{q}_m(\omega,t)$ are coordinate operators and $\hat{p}_m(\omega,t)$ the corresponding momentum operators \cite{Glauber1991}, both associated with the normal modes $f_m(r,\omega)$ defined as eigenfunctions of the Fourier transformed equations of motion Eq.~\eqref{equ::max_timeIndep}.

	The electric field is given by
	\begin{equation}\label{equ::max_E-expansion}
		E(r,t) = -\frac{\Pi(r,t)}{\varepsilon(r)}= -\sum_m \int d\omega \: \hat{p}_m(\omega,t) f^*_m(r,\omega).
	\end{equation}
	
	We choose the mode normalization such that
	\begin{equation}
	\int dr \:\epsilon(r)\, f_{m'}^{*\mathstrut}(r,\omega')f^{\mathstrut}_m(r,\omega) = \delta^{\mathstrut}_{mm'}\delta(\omega-\omega')\,.
	\end{equation}
	We note that the normalization and energy labeling we have chosen here differ from the choice for the Schr\"odinger equation, in order to stay close to conventions usually adopted in quantum optics. As a result, care has to be taken to translate between the two cases. Specifically, to go from a Maxwell mode $f_m(r,\omega)$ to a Schr\"odinger mode $\phi_m(r,k)$ does  not only require the substitution of the energy dependent potential  $\tilde{V}(r,\omega) \rightarrow V(r)$, but also $\omega \rightarrow \sqrt{2E(k)}$ and $f_m(r, \omega)/\sqrt{\omega} \rightarrow \phi_m(r,k)$. Additionally, we note that unlike the Schr\"odinger equation, the kinetic term in the scalar Maxwell equation does not have a factor of $1/2$, such that effectively $H\rightarrow 2H$. The normalization of the system and bath states as well as their associated operators is modified correspondingly.
	
	Applying the normal mode expansions Eqs.~(\ref{equ::max_A-expansion}, \ref{equ::max_Pi-expansion}) to the Hamiltonian Eq.~\eqref{equ::max_H_0_fields} gives \cite{Glauber1991}
	\begin{align}
	\hat{H} &= \frac{1}{2} \sum_m \int d\omega \big[\hat{p}_m^\dagger(\omega,t) \hat{p}_m^{\mathstrut}(\omega,t) \nonumber
	\\
	& \qquad\qquad + \omega^2 \hat{q}_m^\dagger(\omega,t) \hat{q}_m^{\mathstrut}(\omega,t) \big]. \label{equ::max_H_1_pq-ops}
	\end{align}
	The operators fulfill the equal-time commutation relations \cite{Glauber1991,Viviescas2003}
	\begin{subequations}
	\begin{align}
	\big[\hat{q}^{\mathstrut}_m(\omega,t),\hat{q}^{\mathstrut}_{m'}(\omega',t)\big] &= \big[\hat{q}^{\mathstrut}_m(\omega,t),\hat{q}^\dagger_{m'}(\omega',t)\big] = 0 \,,\\
	\big[\hat{p}^{\mathstrut}_m(\omega,t),\hat{p}^{\mathstrut}_{m'}(\omega',t)\big] &= \big[\hat{p}^{\mathstrut}_m(\omega,t),\hat{p}^\dagger_{m'}(\omega',t)\big] = 0 \,,\\
	\big[\hat{q}_m(\omega,t),\hat{p}_{m'}(\omega',t)\big] &= i \delta_{mn}\delta(\omega-\omega')\,,\\
	\big[\hat{q}^{\mathstrut}_m(\omega,t),\hat{p}^\dagger_{m'}(\omega',t)\big] &= i \mathcal{M}^*_{mm'}(\omega,\omega')\,,
	\end{align}
	\end{subequations}
	where $\mathcal{M}_{mn}(\omega,\omega')$ is defined by
	\begin{align}
	\mathcal{M}_{mm'}(\omega,\omega') &= \braket{f_{m}^{*}(\omega)}{f^{\mathstrut}_{m'}(\omega')} \nonumber
	\\
	&=\int dr \varepsilon(r) f_m(r,\omega) f_{m'}(r,\omega')\,.
	\end{align}
	
	We see that the main difference to the single time derivative case is that the Hamiltonian Eq.~\eqref{equ::max_H_0_fields} contains momentum operators and therefore the coordinate operators have different commutation relations. One can introduce bosonic normal mode ladder operators $\hat{c}^{\vphantom{\dagger}}_m(\omega,t)$ and $\hat{c}_m^\dagger(\omega,t)$ via an operator rotation \cite{Glauber1991,Viviescas2003}
	\begin{align}
	\hat{q}^{\mathstrut}_m(\omega) &=  \sqrt{\frac{1}{2\omega}}[\hat{c}^{\mathstrut}_m(\omega) \nonumber
	\\
	&\quad + \sum_{m'} \int d\omega' \mathcal{M}^*_{mm'}(\omega,\omega')\hat{c}^\dagger_{m'}(\omega')]\,,
	\end{align}
	\begin{align}
	\hat{p}^{\mathstrut}_m(\omega) &=  i\sqrt{\frac{\omega}{2}}[\hat{c}^{\dagger}_m(\omega) \nonumber
	\\
	&\quad - \sum_{m'} \int d\omega' \mathcal{M}_{mm'}(\omega,\omega')\hat{c}^{\mathstrut}_{m'}(\omega')]\,,
	\end{align}
	where we have omitted each operators time-dependence for brevity. In this basis, the Hamiltonian Eq.~\eqref{equ::max_H_1_pq-ops} can then be written as~\cite{Glauber1991,Viviescas2003}
	\begin{equation}\label{equ::max_NormH}
	\hat{H} = \sum_m \int d\omega \:\omega \,\hat{c}^{\dagger}_m(\omega) \hat{c}_m^{\mathstrut}(\omega) + \textrm{const.}
	\end{equation}
	and  thus is again diagonal. We note, however, the difference in energy dependence to Eq.~\eqref{equ::kDep_Hdiag}, which is a result of the double time derivative. In addition the field expansions Eqs.~(\ref{equ::max_A-expansion}-\ref{equ::max_E-expansion}) now contain the coordinate operators instead of the ladder operators. If expanded in terms of ladder operators, the expansions then contain both raising and lowering operators. For example for the electric field we have, dropping time-dependences for brevity,
	\begin{equation}\label{equ::max_E_normLadderExpansion}
		E(r) = i \sum_m \int d\omega \sqrt{\frac{\omega}{2}} [\hat{c}^{\mathstrut}_m(\omega)f^{\mathstrut}_m(r, \omega) - \hat{c}^\dagger_m(\omega)f^*_m(r, \omega)]\,,
	\end{equation} 
	and for the $A$-field
	\begin{equation}\label{equ::max_A_normLadderExpansion}
		A(r) = \sum_m \int d\omega \sqrt{\frac{1}{2\omega}} [\hat{c}^{\mathstrut}_m(\omega)f^{\mathstrut}_m(r, \omega) + \hat{c}^\dagger_m(\omega)f^*_m(r, \omega)]\,.
	\end{equation}
	We further note that this canonical quantization scheme works explicitly in the Coulomb gauge \cite{Glauber1991}, which is relevant to obtain the correct coupling term in the presence of light-matter interactions (see also Sec.~\ref{sec::int_lightMatter}).
	
	\subsection{Feshbach projection}\label{sec::app_maxwell_Feshbach}
	Since the mode equation Eq.~\eqref{equ::max_timeIndep} features a wave operator that is Hermitian under the modified inner product Eq.~\eqref{equ::kDep_innerProduct}, we can apply the projection operator formalism analogously to the Schr\"odinger equation.
	In particular, we can write similarly to Eq.~\eqref{equ::proj_modeSep_b}, adapting the energy normalization,
	\begin{align}
	\ket{f_m(\omega)}
	&= \sum_{\lambda\in\Lambda_Q} \ket{\chi_\lambda} \alpha_{\lambda m}(\omega) \nonumber
	\\
	& + \int d\omega' \ket{\tilde{\psi}^{\mathstrut}_{m'}(\omega')}\beta^{\mathstrut}_{m m'}(\omega, \omega')\,,  \label{equ::max_modeSep_b}
	\end{align}
	where the coefficients are now 
	\begin{subequations}
	 \begin{align}
	  \alpha_{\lambda m}(\omega)&=\braket{\chi_\lambda}{f_m(\omega)}\,,\\
	  \beta_{m m'}(\omega, \omega')&=\braket{\tilde{\psi}_{m'}^{\mathstrut}(\omega')}{f^{\mathstrut}_m(\omega)}\,.
	 \end{align}
	\end{subequations}
	The system and bath states each fulfill eigenvalue equations with the energy dependent potential.
	
	To obtain a few-mode Hamiltonian, we now have to apply the resulting operator basis transformation to a different normal mode Hamiltonian given by Eq.~\eqref{equ::max_H_1_pq-ops}. A related expansion of this form has already been performed by Viviescas\&Hackenbroich \cite{Viviescas2003}. There are two differences to our case that have to be considered. Firstly, we have a finite number of system modes $\ket{\chi_\lambda}$, while in \cite{Viviescas2003} an infinite set of modes has been defined by imposing boundary conditions on a spatial region. Secondly, we use the energy-dependent potential form of the wave equation, while Viviescas\&Hackenbroich have performed a variable substitution to obtain a wave equation that is Hermitian under the ordinary inner product. These modifications result in the input-output scattering matrices being well defined, convergent and numerically calculable (see example calculations in Sec.~\ref{sec::exSyst} and Sec.~\ref{sec::int_convergence} for convergence considerations in the interacting case). The reason is that the infinite mode limit has to be taken with care due to certain coupling contributions that vanish in this limit, but still contribute to the scattering, as has already been noted by Domcke \cite{Domcke1983}.
	
	Apart from these differences, the derivation (see Appendix \ref{appMax} for details) of the Gardiner-Collett Hamiltonian follows analogously to \cite{Viviescas2003}, yielding the Hamiltonian Eq.~\eqref{equ::max_GCH},
	where
	\begin{subequations}
	\begin{align}
	\mathcal{W}_{\lambda m}(\omega) &= \frac{1}{2\sqrt{\omega_\lambda \omega}} \bra{\chi_\lambda\vphantom{\tilde{\psi}_{km}^{(+)}}} H_{QP} \ket{\tilde{\psi}_{m}^{\mathstrut}(\omega)}\,,
	\\
	\mathcal{V}_{\lambda m}(\omega) &= \frac{1}{2\sqrt{\omega_\lambda \omega}} \bra{\chi^*_\lambda\vphantom{\tilde{\psi}_{m}^{}(\omega)}} H_{QP} \ket{\tilde{\psi}_{m}^{\mathstrut}(\omega)}\,.
	\end{align}
	\end{subequations}
	The system and bath operators fulfill the equal time commutation relations
	\begin{subequations}\label{equ::max_ladderComm}
	\begin{align}
	&\big[\hat{a}^\dagger_{\lambda}, \hat{a}^{\mathstrut}_{\lambda'}\big] = \delta^{\mathstrut}_{\lambda \lambda'}\,,
	\\
	&\big[\hat{b}^\dagger_{m}(\omega), \hat{b}^{\mathstrut}_{m'}(\omega')\big] = \delta^{\mathstrut}_{mm'} \delta(\omega-\omega')\,,
	\\
	&\big[\hat{a}^\dagger_{\lambda}, \hat{b}^{\mathstrut}_{m}(\omega)\big] = 0\,.
	\end{align}
	\end{subequations}
	The electric field operator Eq.~(\ref{equ::max_E-expansion}) can be expanded in this basis as
	\begin{align}
	E(r,t) &= i \sum_\lambda \sqrt{\frac{\omega_ \lambda}{2}} [\hat{a}^{\mathstrut}_\lambda(t) \chi^{\mathstrut}_\lambda(r) - \hat{a}^{\dagger}_\lambda(t) \chi^{*}_\lambda(r)] \nonumber
	\\
	& +i \sum_m \int d\omega \sqrt{\frac{\omega}{2}} \times \nonumber
	\\
	& \quad[\hat{b}^{\mathstrut}_m(\omega,t) \tilde{\psi}_{m}^{}(\omega,r) - \hat{b}^{\dagger}_m(\omega,t) \tilde{\psi}_{m}^{*}(\omega,r)]\,.\label{equ::max_E-expansion_ladderOps}
	\end{align}
	
	We note that, unlike in Viviescas\&Hackenbroich's approach \cite{Viviescas2003}, the external modes contribute to the field even inside the cavity, as has already been noted in Appendix \ref{sec::app_SBoperators}. This feature is crucial when light-matter interactions are included in the theory, which we discuss in Sec.~\ref{sec::eff}.
	
	\subsection{Details on the system-bath expansion of the Maxwell Hamiltonian}\label{appMax}

	We can apply the system-bath expansion for the normal modes Eq.~\eqref{equ::max_modeSep_b} to the Maxwell fields given by Eqs.~(\ref{equ::max_A-expansion}, \ref{equ::max_Pi-expansion}), to get \cite{Viviescas2003}
	\begin{align}\label{equ::appMax_A-expansion}
		A(r,t) &= \sum_{\lambda} \hat{Q}_\lambda \chi_\lambda(r) \nonumber
		\\
		&+ \sum_m \int d\omega\:\hat{Q}_m(\omega) \tilde{\psi}_m(r,\omega)\,,  
	\end{align}
	and similarly  the conjugate momentum \cite{Viviescas2003}
	\begin{align}\label{equ::appMax_Pi-expansion}
		\Pi(r,t) &= \sum_{\lambda} \varepsilon(r)\hat{P}^{\mathstrut}_\lambda \chi^*_\lambda(r) \nonumber
		\\
		&+ \sum_m \int d\omega\: \varepsilon(r)\hat{P}^{\mathstrut}_m(\omega) \tilde{\psi}^{*}_m(r,\omega)\,.  
	\end{align}
	Here we defined the position operators in system space
	\begin{equation}
		\hat{Q}_\lambda = \int d\omega\: \hat{q}(\omega) \alpha_\lambda(\omega)\,,
	\end{equation}
	in bath space
	\begin{equation}
		\hat{Q}_m(k) = \sum_{m'}\int d\omega' \: \hat{q}_{m'}(\omega') \beta_{mm'}(\omega,\omega')\,,
	\end{equation}
	as well as the momentum operators in system space
	\begin{equation}
	\hat{P}_\lambda = \sum_m\int d\omega \:\hat{p}^{\mathstrut}_m(\omega) \alpha^\dagger_{\lambda m}(\omega)\,,
	\end{equation}
	and in bath space
	\begin{equation}
	\hat{P}_m(k) = \sum_{m'}\int d\omega'\: \hat{p}^{\mathstrut}_{m'}(\omega') \beta^\dagger_{mm'}(\omega,\omega')\,.
	\end{equation}
	These relations can again be inverted (cf. Sec.~\ref{sec::schroed_systBathBasis}) to give \cite{Viviescas2003}
	\begin{equation}\label{equ::appMax_cExpansion2_coord}
		\hat{q}_m(\omega) = \sum_{\lambda \in Q} \hat{Q}^{\mathstrut}_\lambda \alpha^{*\mathstrut}_{\lambda m}(\omega) + \sum_{m'} \int d\omega' \hat{Q}^{\mathstrut}_{m'}(\omega') \beta^*_{m m'}(\omega, \omega')\,,
	\end{equation}
	and
	\begin{equation}\label{equ::appMax_cExpansion2_momentum}
		\hat{p}_m(\omega) = \sum_{\lambda \in Q} \hat{P}^{\mathstrut}_\lambda \alpha^{\mathstrut}_{\lambda m}(\omega) + \sum_{m'} \int d\omega' \hat{P}^{\mathstrut}_{m'}(\omega') \beta^{\mathstrut}_{m m'}(\omega, \omega')\,,
	\end{equation}
	similarly to Eq.~\eqref{equ::schrodinger_cExpansion2}.
	
	Applying these two expansions to the Maxwell normal mode Hamiltonian Eq.~\eqref{equ::max_H_1_pq-ops} and using the coefficient identities analogous to Appendix \ref{app_coeffIds} gives the system-bath Hamiltonian \cite{Viviescas2003}
	\begin{align}
		\hat{H} &= \frac{1}{2}\sum_\lambda [\hat{P}^\dagger_\lambda \hat{P}^{\mathstrut}_\lambda + E^{\mathstrut}_\lambda\hat{Q}^\dagger_\lambda \hat{Q}^{\mathstrut}_\lambda]\nonumber
		\\
		&+ \frac{1}{2}\sum_m \int d\omega [\hat{P}^\dagger_m(\omega) \hat{P}^{\mathstrut}_m(\omega) + \omega^2\hat{Q}^\dagger_m(\omega) \hat{Q}^{\mathstrut}_m(\omega)]\nonumber
		\\
		&+\frac{1}{2}\sum_{\lambda, m} \int d\omega [\tilde{W}^{\mathstrut}_{\lambda m}(\omega) \hat{Q}_\lambda^\dagger \hat{Q}_m^{\mathstrut}(\omega) + h.c.]\,,\label{equ::appMax_pqHamiltonian}
	\end{align}
	with the coupling coefficients in Maxwell normalization
	\begin{equation}
		\tilde{W}_{\lambda m}(\omega) = \bra{\chi^{\mathstrut}_\lambda}H\ket{\tilde{\psi}^{\mathstrut}_{m}(\omega)}\,.
	\end{equation}
	
	As shown earlier, for the Schr\"odinger equation this point constitutes the final system-bath Hamiltonian and is of Gardiner-Collett form. However now the operators in the Hamiltonian are not ladder operators, instead the system operators fulfill the commutation relations \cite{Viviescas2003}
	\begin{subequations}
	\begin{align}\label{equ::appMax_comm_PQ_system}
		&\big[\hat{Q}^{\mathstrut}_{\lambda},\hat{Q}^{\mathstrut}_{\lambda'}\big] = [\hat{Q}^{\mathstrut}_{\lambda},\hat{Q}^{\dagger}_{\lambda'}] = 0\,,
		\\
		&\big[\hat{P}^{\mathstrut}_{\lambda},\hat{P}^{\mathstrut}_{\lambda'}\big] = \big[\hat{P}^{\mathstrut}_{\lambda},\hat{P}^{\dagger}_{\lambda'}\big] = 0\,,
		\\
		&\big[\hat{Q}^{\mathstrut}_{\lambda},\hat{P}^{\mathstrut}_{\lambda'}\big] = i \delta_{\lambda \lambda'}\,,
		\\
		&\big[\hat{Q}^{\mathstrut}_{\lambda},\hat{P}^{\dagger}_{\lambda'}\big] = i \mathcal{N}^*_{\lambda \lambda'}\,,
	\end{align}
	\end{subequations}
	and the bath operators fulfill
	\begin{subequations}\begin{align}\label{equ::appMax_comm_PQ_bath}
		&\big[\hat{Q}^{\mathstrut}_{m}(\omega),\hat{Q}^{\mathstrut}_{m'}(\omega')\big] = [\hat{Q}^{\mathstrut}_{m}(\omega),\hat{Q}^{\dagger}_{m'}(\omega')] = 0\,,
		\\
		&\big[\hat{P}^{\mathstrut}_{m}(\omega),\hat{P}^{\mathstrut}_{m'}(\omega')\big] = \big[\hat{P}^{\mathstrut}_{m}(\omega),\hat{P}^{\dagger}_{m'}(\omega')\big] = 0\,,
		\\
		&\big[\hat{Q}^{\mathstrut}_{m}(\omega),\hat{P}^{\mathstrut}_{m'}(\omega')\big] = i \delta_{m m'} \delta(\omega-\omega')\,,
		\\
		&\big[\hat{Q}^{\mathstrut}_{m}(\omega),\hat{P}^{\dagger}_{m'}(\omega')\big] = i \mathcal{N}^*_{m m'}(\omega,\omega')\,.
	\end{align}
	\end{subequations}
	To obtain a Gardiner-Collett Hamiltonian in terms of ladder operators, we have to perform an operator rotation on the system operators
	\begin{equation}
		\hat{Q}^{\mathstrut}_\lambda = \sqrt{\frac{1}{2\omega_\lambda}}[\hat{a}^{\mathstrut}_\lambda + \sum_{\lambda'} \mathcal{N}^*_{\lambda\lambda'} \hat{a}^\dagger_{\lambda'}]\,,
	\end{equation}
	\begin{equation}
		\hat{P}^{\mathstrut}_\lambda = i \sqrt{\frac{\omega_\lambda}{2}}[\hat{a}^{\dagger}_\lambda - \sum_{\lambda'} \mathcal{N}^{\mathstrut}_{\lambda\lambda'} \hat{a}^{\mathstrut}_{\lambda'}]\,,
	\end{equation}
	and on the bath operators
	\begin{align}
		\hat{Q}^{\mathstrut}_m(\omega) &= \sqrt{\frac{1}{2\omega}}[\hat{b}^{\mathstrut}_m(\omega)\nonumber
		\\
		&+ \sum_{m'}\int d\omega' \mathcal{N}^*_{mm'}(\omega,\omega') \hat{b}^\dagger_{m'}(\omega')]\,,
	\end{align}
	\begin{align}
		\hat{P}^{\mathstrut}_m(\omega) &= i \sqrt{\frac{\omega}{2}}[\hat{b}^{\dagger}_m(\omega)\nonumber
		\\
		&- \sum_{m'}\int d\omega' \mathcal{N}^{\mathstrut}_{mm'}(\omega,\omega') \hat{b}^{\mathstrut}_{m'}(\omega')]\,.
	\end{align}
	Here, we have defined the overlap matrices \cite{Glauber1991}
	\begin{equation}
		\mathcal{N}_{\lambda \lambda'} = \braket{\chi_\lambda^{\mathstrut*}}{\chi_{\lambda'}^{\mathstrut}} = \int dr\:\varepsilon(r)\, \chi_\lambda(r) \chi_{\lambda'}(r)\,,
	\end{equation}
	\begin{align}
		\mathcal{N}_{m m'}(\omega,\omega') &= \braket{\tilde{\psi}_{m}^{*}(\omega)}{\tilde{\psi}_{m'}^{}(\omega')} \nonumber
		\\
		&= \int dr\: \varepsilon(r)\, \tilde{\psi}_{m}^{}(r,\omega) \tilde{\psi}_{m'}^{}(r,\omega')\,.
	\end{align}
	Substitution into the Hamiltonian Eq.~\eqref{equ::appMax_pqHamiltonian} gives the Gardiner-Collett Hamiltonian Eq.~\eqref{equ::max_GCH} for Maxwell's equations. The associated ladder operator commutation relations Eq.~\eqref{equ::max_ladderComm} can be obtained from substitution of the operator rotation into Eqs.~(\ref{equ::appMax_comm_PQ_system}, \ref{equ::appMax_comm_PQ_bath}).
	
	\section{Maxwell scattering in the slowly varying envelope approximation}\label{sec::app_maxwell_scatt}

	The rotating wave approximation employed in Sec.~\ref{sec::Maxwell_rot} simplifies the second quantized Hamiltonian by omitting counter-rotating terms. Recognizing that these terms arise due to a double time derivative in the wave equation, we may consider a modified wave equation with a single time derivative
	\begin{align}\label{equ::kDep_eom}
	-\frac{1}{2}\frac{\partial^2}{\partial r^2} \psi(r,t) & =  i \varepsilon(r) \frac{\partial}{\partial t} \psi(r,t)\,.
	\end{align}
	This can be regarded as a variant of the slowly-varying envelope approximation of Eq.~\eqref{equ::max_scalarEOM}.
	
	\subsection{Canonical quantization}
	For this wave equation, the canonical quantization is completely analogous to the Schr\"odinger case in Sec.~\ref{sec::schroed_CanQuant}. We again get a Hamiltonian
	\begin{align}\label{equ::kDep_Hdiag}
	\hat{H} & =  \sum_m \int dE(k) \: E(k)\,\hat{c}^\dag_m(k,t) \hat{c}^{\mathstrut}_m(k,t)\,,
	\end{align}
	only now, the mode operators $\hat{c}_m(k,t)$ are associated with states $\phi_m(r,k)$ defined by the eigenvalue equation
	\begin{align}\label{equ::kDep_timeIndep}
	-\frac{1}{2}\frac{\partial^2}{\partial r^2} \phi_m(r,k) = \varepsilon(r) E(k) \phi_m(r,k)\,.
	\end{align}
	
	\subsection{Feshbach projection}
	To reveal its similarity with the Schr\"odinger equation \cite{Rotter2017}, we rewrite Eq.~\eqref{equ::kDep_timeIndep} in the form
	\begin{align}\label{equ::kDep_timeIndep2}
	[-\frac{1}{2}\frac{\partial^2}{\partial r^2} + \tilde{V}(r,k) ]\phi_m(r,k) = E(k)\phi_m(r,k)\,.
	\end{align}
	It is thus convenient to use the normalization and energy labeling that we used for the Schr\"odinger equation, such that the energy-dependent potential is given by
	\begin{equation}
		\tilde{V}(r,\omega) = \left [1-\varepsilon(r)\right] E(k)\,.
	\end{equation}
	
	Again accounting for the modified inner product, the system-bath separation via projection operators in Sec.~\ref{sec::FeshOps} can be performed identically to yield the same ab initio Gardiner-Collett Hamiltonian as in Sec.~\ref{sec::schroed_GC}, namely
	\begin{align}\label{equ::GC_H2}
	\hat{H} & =  \sum_{\lambda\in\Lambda_Q} E^{\mathstrut}_\lambda \hat{a}^\dag_\lambda \hat{a}^{\mathstrut}_\lambda + \sum_m \int dE(k) \: E(k) \hat{b}^\dag_m(k) \hat{b}^{\mathstrut}_m(k) \nonumber \\
	& + \sum_{\lambda\in\Lambda_Q} \sum_m \int dE(k)\:\left [W^{\mathstrut}_{\lambda m}(k) \hat{a}^\dag_\lambda \hat{b}^{\mathstrut}_m(k) + \mathrm{h.c.} \right ]\,, 
	\end{align}
	with
	\begin{equation}\label{equ::GC_H2_couplings}
	W_{\lambda m}(k) := \bra{\vphantom{\hat{k}^-}\chi^{}_\lambda} H \ket{\tilde{\psi}_{m}(k)^{}}\,.
	\end{equation}
	The only differences are now the changed inner product in the couplings definition and that the system and bath states are defined by the eigenvalue equation Eq.~\eqref{equ::kDep_timeIndep2} with an energy dependence of the potential.
	
	Therefore, the equivalence between input-output formalism and scattering theory follows analogously to Sec.~\ref{sec::scattEquiv}. This wave equation serves as a useful intermediate between the Schr\"odinger and Maxwell case, since it already features the modified inner product while no counter-rotating terms appear.
	
	\section{Linear dispersion theory}\label{sec::app_LinDisp}
	
	Linear dispersion theory is a method that allows to translate atoms whose transitions couple to the light field into a linear refractive index. The main assumption is that all of the transitions are weakly excited, such that a linear effective medium description can be used.
	
	Historically, this approach was developed in the early days of quantum mechanics (see \cite{Lax1951} for a review) and can be employed for a variety of systems (see for example \cite{Lax1951, Born1980, Rohlsberger2005}). Later it was realized, that even strong coupling effects such as vacuum Rabi-splitting can be described \cite{Zhu1990} by it. For us, linear dispersion theory can thus serve as an ideal practical benchmark, since it has been extensively tested experimentally and its limitations are well understood. In addition, due to the effective medium description, the concept of a mode is not necessary, that is linear dispersion theory can be understood as a basis-free method. This independence of a mode description makes linear dispersion theory also a perfect conceptual benchmark for our few-mode theory.
	
	For completeness, we provide a slightly unusual derivation for the single transition case in the following. In particular it is shown that, apart from the weak excitation as well as the dipole approximation, no further assumptions are necessary.
	
	The derivation is inspired by a similar account in \cite{Malekakhlagh2016b}, which focused on the consequences of the $A^2$-term in cavity and circuit QED.
	
	From Eq.~\eqref{equ::int_interactionHamiltonian}, the atom-field interaction Hamiltonian reads
	\begin{equation}
		\hat{H}_\mathrm{int} = -i\omega_\textrm{a}  (d\hat{\sigma}^+ - d^*\hat{\sigma}^-) A(r_a) + c_A A^2(r_a).
	\end{equation}
	We have now also included the $A^2$-term, introducing an additional constant $c_A$, which depends on the physical realization of the two-level system \cite{Malekakhlagh2016b}.
	
	The Heisenberg equations of motion for the atomic lowering operator then read, applying the weak excitation approximation $\hat{\sigma}^z(t) \approx -1$,
	\begin{equation}
		\dot{\hat{\sigma}}^-(t) = -i\omega_\textrm{a} \hat{\sigma}^-(t) - \omega_\textrm{a} d A(r_a, t).
	\end{equation}
	The solution of this equation can be written as
	\begin{equation}
		\hat{\sigma}^-(t) = \int_{-\infty}^{\infty} \frac{d\omega}{2\pi} e^{-i\omega t} \frac{i\omega_\textrm{a} d A(r_a, \omega)}{\omega_\textrm{a} - \omega},
	\end{equation}
	where $A(r_a, \omega)$ is the Fourier transformed field operator. Similarly, for the raising operator one obtains
	\begin{equation}
		\hat{\sigma}^+(t) = \int_{-\infty}^{\infty} \frac{d\omega}{2\pi} e^{-i\omega t} \frac{-i\omega_\textrm{a} d^* A(r_a, \omega)}{\omega_\textrm{a} + \omega}.
	\end{equation}
	The field equations of motion for the coupled system are
	\begin{align}
		\varepsilon(r) \frac{\partial^2}{\partial t^2} A(r, t) =& \frac{\partial^2}{\partial r^2} A(r, t) + c_A A(r, t)\delta(r-r_a)  \nonumber
		\\
		&+ i\omega_\textrm{a} \big(d\hat{\sigma}^+(t) - d^*\hat{\sigma}^-(t)\big) \delta(r-r_a).
	\end{align}
	Substituting the solutions for the atomic operators and moving to the frequency domain, we obtain an effective Maxwell equation
	\begin{align}
		 \frac{\partial^2}{\partial r^2} A(r, \omega) =& -\omega^2 \varepsilon(r) A(r, \omega) + c_A \delta(r-r_a) A(r,\omega) \nonumber
		 \\
		 &+ \frac{2\omega_\textrm{a}^3 |d|^2}{\omega^2 - \omega_\textrm{a}^2} \delta(r-r_a) A(r,\omega).
	\end{align}
	As a result, we can write an effective energy dependent permittivity for the two-level system as
	\begin{equation}\label{eq::app_effEpsilon}
		\varepsilon'(r) = \varepsilon(r) - \left(\frac{\omega_\textrm{a}^2}{\omega^2}\frac{2\omega_\textrm{a} |d|^2}{\omega^2 - \omega_\textrm{a}^2} + \frac{c_A}{\omega^2}\right) \delta(r-r_a).
	\end{equation}
	We note that for an atomic medium of number density $\rho$ sufficiently large compared to the wavelength, the more standard expression
	\begin{equation}
		\varepsilon'(r) = \varepsilon(r) - \frac{|d|^2}{\omega - \omega_\textrm{a}} \rho(r)
	\end{equation}
	can be obtained in the weak coupling regime, where the rotating wave approximation $\omega^2 - \omega_\textrm{a}^2\approx 2\omega_\textrm{a}(\omega - \omega_\textrm{a})$ as well as $\omega_\textrm{a}^2/\omega^2 \approx 1$, and $c_A \approx 0$ are assumed. Usually, a decay rate $\gamma$ is also included to account for additional decay channels. In our case, this contribution does not appear, since we only consider radiative losses which are already accounted for by the dipole coupling.
	
	We further note that in one dimension and for layered systems, the scattering solutions of these modified Maxwell equations can be found efficiently using a transfer matrix formalism \cite{Parratt1954,Rohlsberger2005}, which we employ to perform the calculations for the examples shown in the main text. For simplicity, we also neglect the $A^2$-term contribution by setting $c_A = 0$. The formula Eq.~\eqref{eq::app_effEpsilon} nevertheless includes the contribution from counter-rotating terms, such that the applicability of the rotating wave approximation in both the cavity-bath and the atom-cavity coupling that were performed for the linear scattering calculation in ab initio few-mode theory can be tested.

	\section{Analytical convergence}\label{sec::app_AnalytConvergence}
	In this Appendix we show the convergence of the effective few-mode expansion for an analytically solvable example cavity. We choose the cavity geometry depicted in Fig.~\ref{fig::maxwell_fabryPerot_exact} and the Dirichlet basis states for Q-space with $\omega_\lambda = \lambda \pi/L$. This configuration is particularly convenient since the exact solution for the Schr\"odinger potential analogue of its free theory has been given by Domcke \cite{Domcke1983}. Adapting the expression from \cite{Domcke1983} to the Maxwell case gives
	\begin{equation}
		\mathcal{W}_\lambda(\omega) = \frac{w}{\alpha \cot{\alpha} - s - i\beta} \sqrt{ \lambda} (-1)^\lambda,
	\end{equation}
	where $\alpha$, $\beta$ and $w$ are constants in the sense that they do not depend on the mode index or number of chosen modes, but they may for example depend on frequency. $s$ is the sum
	\begin{equation}
		s = \sum_{\lambda \in \Lambda_Q} \frac{2\lambda^2\pi^2}{\alpha^2 - \lambda^2\pi^2},
	\end{equation}
	it therefore does not explicitly depend on the mode index, however it does depend on which modes are included in the few-mode basis. The latter is important if we want to take the limit of infinitely many system modes in the end, where the effective few-mode expansion should converge and become exact. We note that $s$ by itself does not converge on its own in this limit, however any observables quantities will in the end. This non-trivial convergence behavior has already been pointed out by Domcke \cite{Domcke1983} and will be encountered again multiple times in the following. Note that one consequence is that all system-bath couplings approach zero in this limit. We note that this poses no problem for practical calculations since any relevant observables, such as the scattering matrices, should converge. Furthermore, the couplings are finite at any finite mode number, such that numerical calculations can be performed and yield the correct observables as shown in the main text.
	
	For the free scattering matrix this convergence has already been shown by Domcke \cite{Domcke1983} for the Schr\"odinger case. The same derivation in essence also applies to the Maxwell case by invoking the result from Sec.~\ref{sec::Maxwell_rot}.
	
	In the following we will show the convergence in the linear interacting case to confirm the validity of the effective few-mode expansion scheme.
	
	\begin{figure}[t]
		\includegraphics[width=\columnwidth]{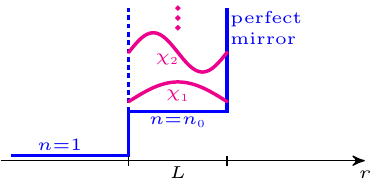}
		\caption{A cavity geometry whose Schr\"odinger analogue has been solved exactly by Domcke \cite{Domcke1983} using the Feshbach projection formalism.}\label{fig::maxwell_fabryPerot_exact}
	\end{figure}
	From Eq.~\eqref{equ::max_Dmatrix_rot} the $\mathcal{D}$-matrix is
	\begin{equation}
	\mathcal{D}^{\mathstrut}_{\lambda \lambda'}(\omega) = (\omega-\omega_\lambda)\delta_{\lambda \lambda'} + \Gamma'_{\lambda \lambda'}(\omega).
	\end{equation}
	For consistency with the rotating wave approximation, as outlined in Sec.~\ref{sec::Maxwell}, we employ
	\begin{align}
	 	\mathcal{D}_{\lambda \lambda'}(\omega)
		&\approx \frac{\omega^2 - \omega_\lambda^2}{2\omega_\lambda}\delta_{\lambda \lambda'} + \frac{\Gamma_{\lambda \lambda'}(\omega)}{\sqrt{\omega_\lambda \omega_{\lambda'}}} \,.
	\end{align}
	This approximation is also convenient for the contour integral in the level shift matrix to be computable via Domcke's separable expansion method \cite{Domcke1983}, resulting in the expression
	\begin{equation}
		\frac{\Gamma_{\lambda \lambda'}(\omega)}{\sqrt{\omega_\lambda \omega_{\lambda'}}} = \frac{\tilde{\gamma}}{\alpha \cot{\alpha} - s - i\beta} \sqrt{ \lambda\lambda'} (-1)^{\lambda+\lambda'}
	\end{equation}
	where $\tilde{\gamma} = \pi/L$ is a constant. We further note that it is crucial to approximate the diagonal and level shift term consistently within the rotating wave approximation in order to obtain a converging series expansion in the rotating wave approximation (see also Sec.~\ref{sec::Maxwell}).
	
	Inverting the $\mathcal{D}$-matrix via the Sherman-Morrison formula \cite{Domcke1983,Golub1996} gives
	\begin{align}
		\mathcal{D}^{-1}_{\lambda \lambda'}(\omega) &= \frac{2\omega_\lambda\delta_{\lambda\lambda'}}{\omega^2-\omega_\lambda^2} - \frac{\tilde{\gamma}}{\alpha \cot{\alpha} - i\beta +\tilde{\gamma}b - s}  \nonumber
		\\
		&\times\frac{4\omega_\lambda \omega_{\lambda'}\sqrt{\lambda\lambda'}(-1)^{\lambda+\lambda'}}{(\omega^2-\omega_\lambda^2)(\omega^2-\omega_{\lambda'}^2)},
	\end{align}
	where
	\begin{equation}
		b = \frac{L}{\pi}\sum_{\lambda \in \Lambda_Q} \frac{2\pi^2\lambda^2}{\alpha^2-\pi^2\lambda^2} = \frac{s}{\tilde{\gamma}}.
	\end{equation}
	We then have
	\begin{align}
		\mathcal{D}^{-1}_{\lambda \lambda'}(\omega) &= \frac{2\omega_\lambda\delta_{\lambda\lambda'}}{\omega^2-\omega_\lambda^2} \nonumber\\ &- \frac{1}{\alpha \cot{\alpha} - i\beta}  \frac{4\omega^{3/2}_\lambda (-1)^{\lambda} \omega^{3/2}_{\lambda'}(-1)^{\lambda'}}{(\omega^2-\omega_\lambda^2)(\omega^2-\omega_{\lambda'}^2)}.
	\end{align}
	The coupling constants in this basis are
	\begin{equation}
		g_\lambda = \tilde{g} \frac{\sin{\left(\frac{\pi \lambda}{L}r_\textrm{a}\right)}}{\sqrt{\lambda}},
	\end{equation}
	where $\tilde{g}$ is a constant containing $d$ and $\omega_\textrm{a}$. $r_\textrm{a}$ is the atom's position, which we take to be $r_\textrm{a}=\frac{L}{2}$, such that the atom is located at the cavity center. We are now in a position to check the convergence of the interaction sums appearing in Eq.~\eqref{equ::int_linearScatteringIO}. We can write
	\begin{align}
		\underline{g}^T \doubleunderline{\mathcal{D}}^{-1}(\omega) \underline{g}^* = G_1 - \frac{|\tilde{g}|^2 L \pi}{\alpha \cot{\alpha} - i\beta} (G_2)^2,
	\end{align}
	where the sums
	\begin{align}
		G_1 = 2|\tilde{g}|^2 L \pi \sum_{\lambda \in \Lambda_Q^{\textrm{odd}}} \frac{1}{\alpha^2 - \pi^2 \lambda^2}
	\end{align}
	and
	\begin{align}
		G_2 =  \sum_{\lambda \in \Lambda_Q} \sin\big(\frac{\pi\lambda}{2}\big)(-1)^\lambda\frac{\pi\lambda}{\alpha^2 - \pi^2 \lambda^2}.
	\end{align}
	In the limit of infinite number of system modes, that is $\Lambda_Q = \{\chi_1, \chi_2, ..., \chi_N\}$ with $N\rightarrow\infty$ one obtains
	\begin{equation}
		G_1 \rightarrow -\frac{|\tilde{g}|^2 L \pi}{2\alpha} \tan{\frac{\alpha}{2}},
	\end{equation}
	and $G_2$ can be expressed in terms of beta, gamma and hypergeometric functions. Therefore $\underline{g}^T \doubleunderline{\mathcal{D}}^{-1}(\omega) \underline{g}^*$ converges individually in this limit, only containing isolated poles at certain energies.
	
	Similarly,
	\begin{align}
		\underline{g}^T \doubleunderline{\mathcal{D}}^{-1}(\omega) \doubleunderline{\mathcal{W}}(\omega) &= \frac{w}{\alpha \cot{\alpha} - s - i\beta} \nonumber
		\\
		&\times \left[\tilde{g}L G_2 - \frac{\tilde{g} L}{\alpha \cot{\alpha} - i\beta} G_2 s \right],
	\end{align}
	and
	\begin{align}
		\doubleunderline{\mathcal{W}}^\dagger(\omega) \doubleunderline{\mathcal{D}}^{-1}(\omega) \underline{g}^* &= \left[\frac{w}{\alpha \cot{\alpha} - s - i\beta}\right]^* \nonumber
		\\
		&\times \left[\tilde{g}^*L G_2 - \frac{\tilde{g}^* L}{\alpha \cot{\alpha} - i\beta} G_2 s \right].
	\end{align}
	These terms contain the $s$-sum, which diverges in the limit $N\rightarrow \infty$. They can be understood by comparing to the empty cavity term
	\begin{align}
	\doubleunderline{\mathcal{W}}^\dagger(\omega) \doubleunderline{\mathcal{D}}^{-1}(\omega) \doubleunderline{\mathcal{W}}(\omega) &= \left|\frac{w}{\alpha \cot{\alpha} - s - i\beta}\right|^2  \nonumber
	\\
	&\times \left[\frac{L}{\pi} s - \frac{L/\pi}{\alpha \cot{\alpha} - i\beta} s^2 \right],
	\end{align}
	which, as already shown by Domcke \cite{Domcke1983}, is convergent and yields a well defined resonant scattering matrix. The non-convergent $s$-terms furthermore completely cancel when the result is multiplied by the background scattering matrix \cite{Domcke1983}.
	
	We have thus shown the convergence of the few-mode expansion in the infinite mode limit for a special case. We note that in order to obtain this converging series, there are two crucial factors. Firstly, for gauge consistency we require the $p\cdot A$ interaction term,  leading to a $1/\sqrt{\omega_\lambda}$ dependence in the couplings. Secondly, the rotating wave approximation in the system-bath coupling has to be applied consistently (see also Sec.~\ref{sec::Maxwell}). The latter is already necessary in the non-interacting case and thus a feature of the system-bath interaction, rather than of the light-matter coupling.
	
	\bibliographystyle{myprsty}
	\bibliography{library}
\end{document}